\documentclass[12pt,preprint]{aastex}

\shorttitle{M37 Transit Survey II}
\shortauthors{Hartman et al.}

\begin{document}

\title{Deep MMT\footnote{Observations reported here were obtained at the MMT Observatory, a joint facility of the Smithsonian Institution and the University of Arizona.} Transit Survey of the Open Cluster M37 II: Variable Stars}
\author{J.~D.~Hartman\altaffilmark{2}, B.~S.~Gaudi\altaffilmark{3}, M.~J.~Holman\altaffilmark{2}, B.~A.~McLeod\altaffilmark{2}, K.~Z.~Stanek\altaffilmark{3}, J.~A.~Barranco\altaffilmark{4}, M.~H.~Pinsonneault\altaffilmark{3} and J.~S.~Kalirai\altaffilmark{5,6}}
\altaffiltext{2}{Harvard-Smithsonian Center for Astrophysics, 60 Garden St., Cambridge, MA~02138, USA; jhartman@cfa.harvard.edu, mholman@cfa.harvard.edu, bmcleod@cfa.harvard.edu, jbarranc@cfa.harvard.edu}
\altaffiltext{3}{Department of Astronomy, The Ohio State University, Columbus, OH~43210, USA; gaudi@astronomy.ohio-state.edu, kstanek@astronomy.ohio-state.edu, pinsono@astronomy.ohio-state.edu}
\altaffiltext{4}{Department of Physics and Astronomy, San Francisco State University, 1600 Holloway Ave., San Francisco, CA~94132, USA; barranco@stars.sfsu.edu}
\altaffiltext{5}{University of California Observatories/Lick Observatory, University of California at Santa Cruz, Santa Cruz CA, 95060}
\altaffiltext{6}{Hubble Fellow}

\begin{abstract}
We have conducted a deep ($15 \la r \la 23$), 20 night survey for transiting planets in the intermediate age ($\sim 550~{\rm Myr}$) open cluster M37 (NGC 2099) using the Megacam wide-field mosaic CCD camera on the 6.5m Multiple Mirror Telescope (MMT). In this paper we present a catalog and light curves for 1445 variable stars; 1430 (99$\%$) of these are new discoveries. We have discovered 20 new eclipsing binaries and 31 new short period ($P < 1~{\rm day}$) pulsating stars. The bulk of the variables are most likely rapidly rotating young low-mass stars, including a substantial number ($\ga 500$) that are members of the cluster. We identify and analyze five particularly interesting individual variables including a previously identified variable which we suggest is probably a hybrid $\gamma$ Doradus/$\delta$ Scuti pulsator, two possible quiescent cataclysmic variables, a detached eclipsing binary (DEB) with at least one $\gamma$ Doradus pulsating component (only the second such variable found in an eclipsing binary), and a low mass ($M_{P} \sim M_{S} \sim 0.6 M_{\odot}$) DEB that is a possible cluster member. A preliminary determination of the physical parameters for the DEB+$\gamma$ Doradus system yields $M_{P} = 1.58 \pm 0.04 M_{\odot}$, $M_{S} = 1.58 \pm 0.04 M_{\odot}$, $R_{P} = 1.39 \pm 0.07 R_{\odot}$ and $R_{S} = 1.38 \pm 0.07 R_{\odot}$.
\end{abstract}


\section{Introduction}

This paper is the second in a series of papers on a deep survey for transiting planets in the open cluster M37 (NGC 2099) using the MMT telescope. In the first paper \citep[Paper I]{Hartman.07} we introduced the survey, described the spectroscopic and photometric observations, determined the fundamental parameters (age, metallicity, distance and reddening) of the cluster, and obtained its mass and luminosity functions and radial density profile. As a result, we have a good understanding of the population of stars that we have observed. This paper focuses on the variable stars in our survey. We will discuss stellar rotation and transiting planets in future papers.

As discussed in Paper~I, there have been a number of surveys for transiting planets in galactic open clusters. The rationale behind conducting such a survey is that since the properties of stars in clusters are easier to determine en masse than they are for field stars, and since cluster stars have more uniform properties than field stars, one can determine in a relatively straightforward fashion from the survey the frequency of planets in the cluster, or at least place a meaningful upper limit on it \citep[e.g.][]{Burke.06}. We were motivated in particular to conduct this survey by \citet{Pepper.05, Pepper.06a} who suggested that it may be possible to find Neptune sized planets transiting solar-like stars by surveying an open cluster with a large telescope. These authors developed a formalism for estimating the expected yield from an open cluster transit survey as a function of the cluster's parameters and the survey strategy. Using this formalism we found that based on its location on the sky, distance, reddening, and richness, M37 is one of the clusters with the highest expected yield of Neptune-sized planets for a transit survey conducted with the MMT.

While to date no confirmed transiting planet has been found in an open cluster, these surveys have produced a number of by-products unrelated to the study of planets. One of the primary by-products is the study of variable stars in the surveyed field. Studies of this sort have been published for the open clusters NGC~6791, NGC~2158, NGC~1245, M34, NGC~2516, and NGC~2547 as a by-product of transit surveys \citep{Mochejska.02, Mochejska.04, Pepper.06b, DeMarchi.07, Irwin.06, Irwin.07a, Irwin.07b}. Similarly, variable stars have also been studied in the globular clusters 47~Tucanae and $\omega$~Centauri as a by-product of transit surveys \citep{Weldrake.04, Weldrake.07}.

The study of variable stars in stellar clusters is particularly interesting since one can put constraints on the age, metallicity, distance and reddening of stars that are cluster members. This additional information allows for powerful tests of stellar pulsation models \citep[e.g.][]{Frandsen.98}, or stellar evolution theory in the case of eclipsing binaries \citep[e.g.][]{Southworth.04a}. By identifying variables in different clusters it is possible to study the evolution of stellar properties including rotation \citep[e.g.][]{Irwin.06, Irwin.07a, Irwin.07b} and pulsation \citep[e.g.][]{Krisciunas.98}. Even if most stars in a given survey are not members of a cluster, identifying the variables is still interesting as it is a good tool for finding rare or previously unknown phenomena that deserve detailed study. 

While we were primarily concerned with maximizing the expected planet yield when choosing to survey M37, this cluster is also an interesting object to survey for variable stars both because it is rich for an open cluster ($\sim 4000$ stars larger than $0.3M_{\odot}$, Paper~I) and because at an age of $550~{\rm Myr}$ (Paper~I) many of the stars are active enough to be detectable as variables with amplitudes greater than a few millimagnitudes. M37 has been the target of two previous surveys for variable stars \citep{Kiss.01,Kang.07} which have discovered 24 variables. The \citet{Kiss.01} survey consisted of 658 $R$-band observations taken over the course of $\sim 1$ month, and spanned a magnitude range of $12 < V < 18$, with a point-to-point precision of $0.01$ mag at the bright end and $0.03$ mag at the faint end. The \citet{Kang.07} survey consisted of 1220 $V$-band observations taken over 12 nights in 2004 and 3 nights in 2006, and was saturated for stars brighter than $V = 13$. Because our survey, consisting of $\sim 4000$ observations taken over $\sim 1$ month, and spanning a magnitude range of $15 < V < 23$, with a point-to-point precision of $\sim 0.002$ mag at the bright end and $0.05$ mag at the faint end, has more observations with higher precision for fainter targets than either of these surveys, we have identified 1445 variables, of which 1430 are new discoveries. 

In the following section we will describe the observations and the reduction of the time-series data. In \S 3 we describe the selection of variable stars and present the catalog. In \S 4 we match to the previously identified variables. In \S 5 we discuss the global properties of the variables noting that a substantial fraction are likely to be rapidly rotating spotted stars. In \S 6 we analyze five particularly interesting variables including a probable hybrid $\gamma$ Doradus/$\delta$ Scuti pulsator, two possible quiescent cataclysmic variables, a detached eclipsing binary with a $\gamma$ Doradus component and a low mass detached eclipsing binary with $\sim 0.6 M_{\odot}$ components that is a possible cluster member. Finally we summarize our results in \S 7.

\section{Observations and Data Reduction}

\subsection{Observations}

A detailed discussion of the observations was presented in Paper I, we provide only a brief overview here. The observations consist of both $gri$ photometry for $\sim 16000$ stars, and $r$ time-series photometry for $\sim 23000$ stars obtained with the Megacam mosaic imager \citep{McLeod.00} on the 6.5 m MMT and high-resolution spectroscopy of 127 stars obtained with the Hectochelle multi-fiber, high-dispersion spectrograph \citep{Szentgyorgyi.98} on the MMT. 

The primary time-series photometric observations were done using the $r$ filter and consist of $\sim 4000$ high quality images obtained over twenty four nights (including eight half nights) between December 21, 2005 and January 21, 2006. Where available we took $BV$ photometry for the sources from \citet{Kalirai.01}. We also transformed our $ri$ photometry into $I_{C}$ using the $I_{C}$ photometry from \citet{Nilakshi.02} to define the transformation. Finally we took $K_{S}$ photometry from 2MASS \citep{Skrutskie.06} where available.

The spectra were obtained on four separate nights between February 23, 2007 and March 12, 2007 and were used to measure $T_{eff}$, $[Fe/H]$, $v\sin i$ and the radial velocity (RV) via cross-correlation.

\subsection{Image Subtraction Photometry}

The CCD calibration, point-spread-function fitting photometry and photometric calibration routines were described in Paper I. To obtain light curves from the $r^{\prime}$ time series observations of the primary M37 field we used the image subtraction technique due to \citet{Alard.98} and \citet{Alard.00}, using a slightly modified version of Alard's {\scshape Isis} 2.1 package. Before performing photometry we construct a reference image for each chip from $\sim 100$ of the best seeing images. We choose one image as a template and register each image in the reference list to the template and then match the background and convolve it to the seeing and flux scale of the template. The matched images are then mean-combined and processed through PSF fitting routines as described in Paper I.

The reduction of each image in the time series involves the following steps:
\begin{enumerate}
\item Remove cosmic rays using the \emph{extract} program in {\scshape Isis}.
\item Identify stars in the image with the Source Extractor program (version 2.3.2) due to \citet{Bertin.96}. The resulting star list is matched to the star list for the reference image and a third order polynomial transformation from the reference coordinates to the image coordinates is determined. The reference image is then transformed to the image coordinate system using a cubic spline interpolation with the {\scshape ISIS} program \emph{interp}.
\item If the image has seeing better than $1\arcsec$ convolve the image with a Gaussian kernel to broaden the seeing up to $1\arcsec$.
\item Adjust the FWHM values of the Gaussians used in the image subtraction kernel according to the value expected, assuming Gaussian PSFs, from the difference in seeing between the (broadened) image and the reference. We use the expected FWHM, half the expected FWHM and twice the expected FWHM.
\item Use the \emph{mrj\_phot} program in {\scshape Isis} to convolve the reference into the image, saving the residual image which is set to the flux scale of the reference.
\item Transform the reference star list to the image coordinates and perform aperture photometry on the residual image at the location of the transformed stars. To perform the aperture photometry we sum all pixels within a fixed aperture radius and divide by the sum of the reference PSF convolved with the kernel from \emph{mrj\_phot} over the same aperture radius. Here we use the PSF from the PSF fitting routines discussed in Paper I as the reference PSF to ensure proper flux scaling in converting the light curves from differential flux to magnitudes.
\end{enumerate}

The differential flux light curves obtained from the above pipeline are then converted to magnitudes using the reference PSF magnitude. We obtained differential light curves for a total of 23790 sources detected in the stacked $r^{\prime}$ images, of which 14750 are detected in $g$, $r$ and $i$, 7245 are detected only in $r$ and $i$, 141 are detected only in $g$ and $r$ and 1654 are only detected in $r$. 

The light curves are passed through two cleaning procedures to remove unreliable measurements. The first procedure is to remove $5\sigma$ outliers from the light curves, where $\sigma$ is the measured, rather than the expected, $RMS$ of the light curve. This step is necessary to remove unidentified cosmic ray hits that fall within the aperture radius of a source, satellite trails or transient bad rows (which were seen to affect a number of images). All of these artifacts might affect only a few sources on a given image, but over the course of the run will affect almost every source at least a few times. Variable stars with sharp features, such as deep eclipses, may have reliable extremum measurements clipped by this procedure. Therefore, we investigate all identified variable star light curves on a case by case basis to recover reliable measurements that have been removed.

The second cleaning procedure is to remove points from images that are outliers in a large number of light curves. Letting $x$ be the fraction of light curves for which a given image is a $3\sigma$ outlier, we inspected the histogram of $x$ and chose its cutoff value for each chip. The typical cutoff was between $x = 5\% - 10\%$. After both cleaning procedures, typical light curves contained $\sim 4000$ points.

\begin{figure}[p]
\plotone{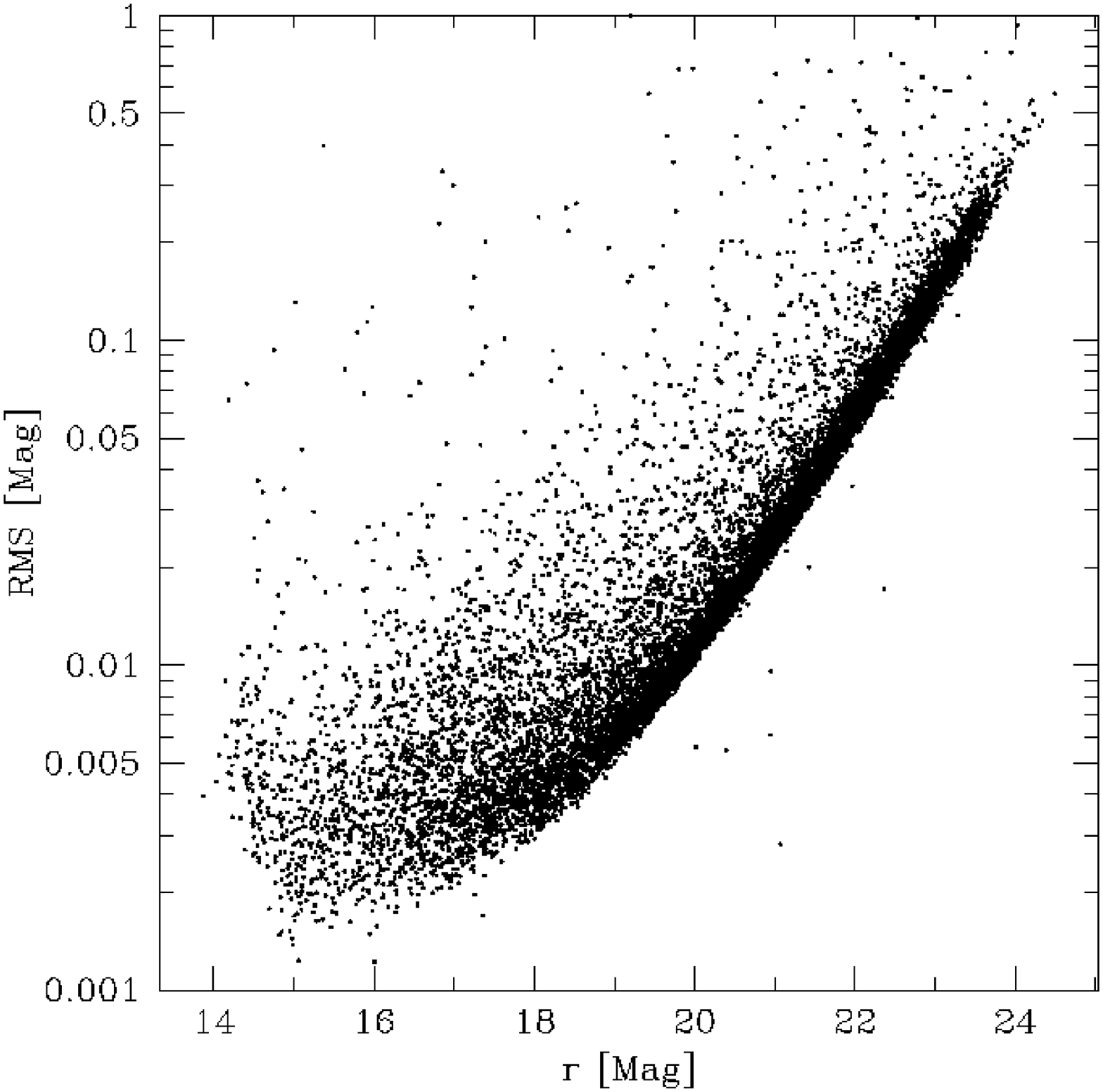}
\caption{RMS of the light curves as a function of $r$ magnitude. The point-to-point precision of the light curves approaches 1 mmag at the bright end.}
\label{fig:lcstat}
\end{figure}

Figure~\ref{fig:lcstat} shows the Root Mean Square (RMS) of the light curves as a function of the $r$ magnitude. The point-to-point precision of the light curves approaches 1 mmag at the bright end. This is approximately a factor of 5 poorer than one would expect based on the photon noise. Achieving better precision would require correcting for additional systematic variations in the light curves. We will correct for these systematic variations before performing the search for transit candidates.

\placefigure{fig:lcstat}

\section{Variable Star Selection}

We have used the following three different methods to identify periodic variable stars from our time series observations:
\begin{enumerate}
\item A search for periodic, sinusoidal variations with the Lomb-Scargle (L-S) algorithm \citep{Lomb.76, Scargle.82, Press.89, Press.92}.
\item A search for general periodic variability using the analysis of variance (AoV) periodogram due to \citet{Schwarzenberg-Czerny.89}, as implemented by \citet{Devor.05}.
\item A search for eclipses using the Box-fitting Least Squares (BLS) algorithm due to \citet{Kovacs.02}.
\end{enumerate}

Note that the version of AoV that we implement here uses phase binning. \citet{Schwarzenberg-Czerny.96} has also developed a version of AoV that uses triganometric orthogonal polynomials. This alternative version reduces to a form that is analogous to L-S when only one harmonic is used. Besides the difference in the variable used in the periodogram (AoV as opposed to power in L-S, note that there is a one-to-one correspondence between the two), there is a slight improvement in this version of AoV over L-S in that AoV also fits a constant term to the light curve rather than simply subtracting off the mean as is done in L-S.

While there will be overlap between the results of the methods that we used, each algorithm tends to be sensitive to different types of variables. The more general tool, AoV with phase binning, should in principle identify variables found by the more restrictive L-S and BLS algorithms, but it may also be more sensitive to non-astrophysical variability due to uncalibrated instrumental or atmospheric effects. As a result, AoV with phase binning cannot be used to search for sinusoidal or transit-like variability with amplitudes as low as that accessible to L-S and BLS. In the next three subsections we describe the results from using each selection tool, and in the last subsection we present the composite catalog of variables.

\subsection{L-S search}

\begin{figure}[p]
\plotone{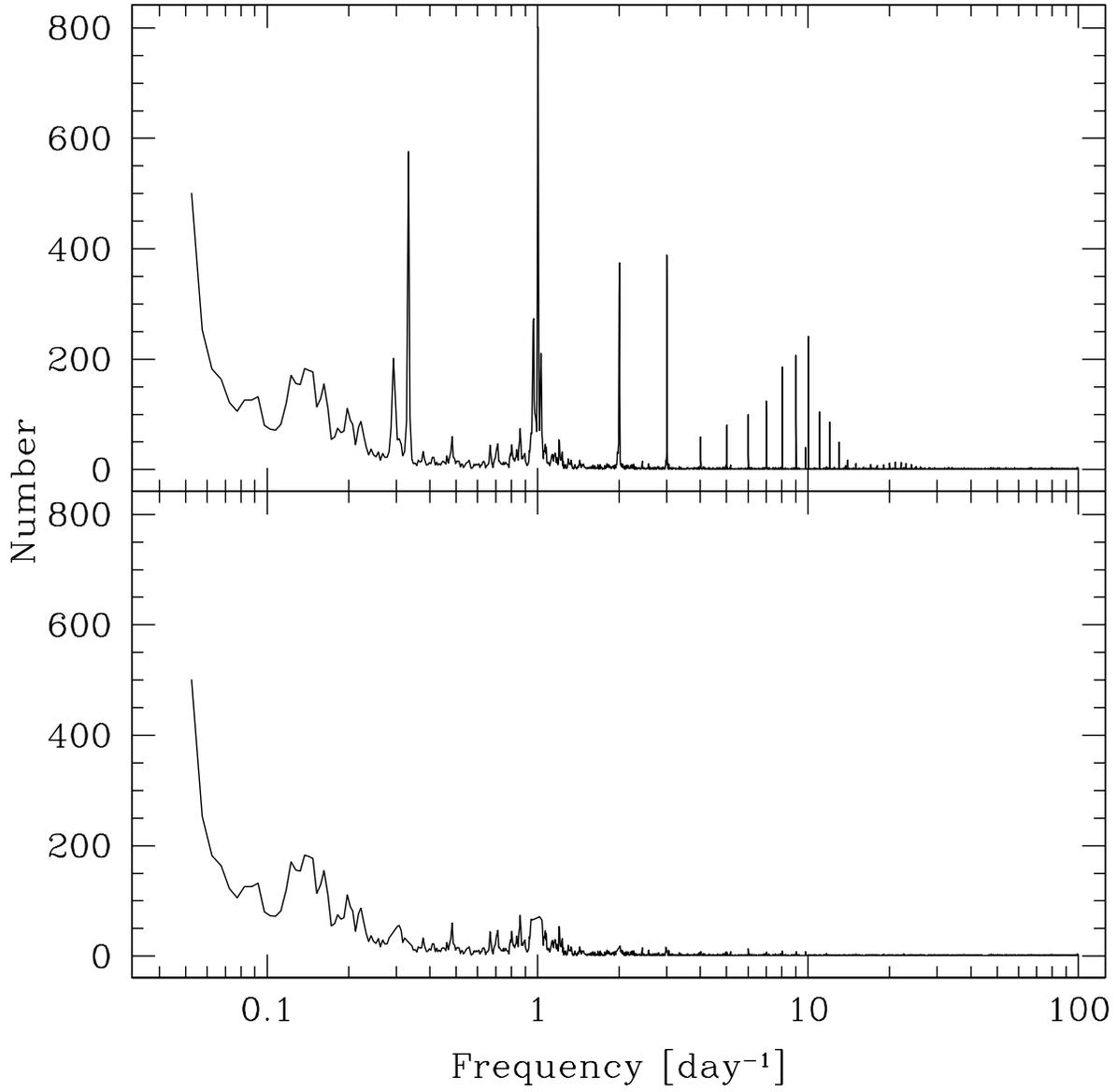}
\caption{Histogram of peak frequencies in the Lomb-Scargle periodograms of all light curves. The top panel shows the histogram before removing bad frequencies (seen as spikes in the periodogram), the bottom shows the histogram after removing them.}
\label{fig:LSPeriodHist}
\end{figure}

We used the L-S algorithm to search for sinusoidal periodic variables with periods between 0.01 and 20 days; the search was conducted with a sampling of 0.005 times the Nyquist sampling. Figure~\ref{fig:LSPeriodHist} shows the histogram of peak frequencies for all of our 23,790 light curves. Spikes in the histogram result from instrumental or atmospheric artifacts that systematically affect numerous light curves. Many of these occur at aliases of 1 sidereal day and result from artifacts such as the rotation of diffraction spikes from nearby stars or color dependent extinction. Other artifacts, such as a series of observations where the image subtraction procedure did not work optimally (due, for example, to coma or other non-symmetric distortions of the PSF) may be phased at periods that are not aliases of a sidereal day. Since the vast majority of candidate variables identified at these frequencies are spurious detections, we reject any detection at one of these spikes. The rejection removes 8791 of the 23,790 light curves. The lower panel of figure~\ref{fig:LSPeriodHist} shows the histogram after removing these frequencies. The rise in the number of detections towards short frequencies results both from real variables with periods between 5 and 10 days and from $\sim 1/f$ red noise \citep[e.g.][]{Pont.06} in the light curves.

\placefigure{fig:LSPeriodHist}

For each light curve we calculate the false alarm probability (FAP) of the period corresponding to the peak in the periodogram using equation 13.8.7 from \citet{Press.92}:
\begin{equation}
{\rm FAP}(z) = 1 - (1 - e^{-z})^{M},
\end{equation}
where $z$ is the peak value in the periodogram and $M$ is the number of independent frequencies sampled, and which we approximate by the number of points times the ratio of the maximum frequency searched to the Nyquist frequency. For convenience we henceforth refer to $\log(FAP)$ as FAP. The FAP provides a relative ordering of detection confidence among the light curves. How the formal FAP relates to the real FAP depends on the specific window function of the observations as well as on the noise distribution in the light curves. As the noise in our light curves is typically not frequency independent we do not expect the formal FAP to correspond to the real FAP, thus the selection threshhold in formal FAP is set by inspecting the light curves at a range of FAP values.

\begin{figure}[p]
\plotone{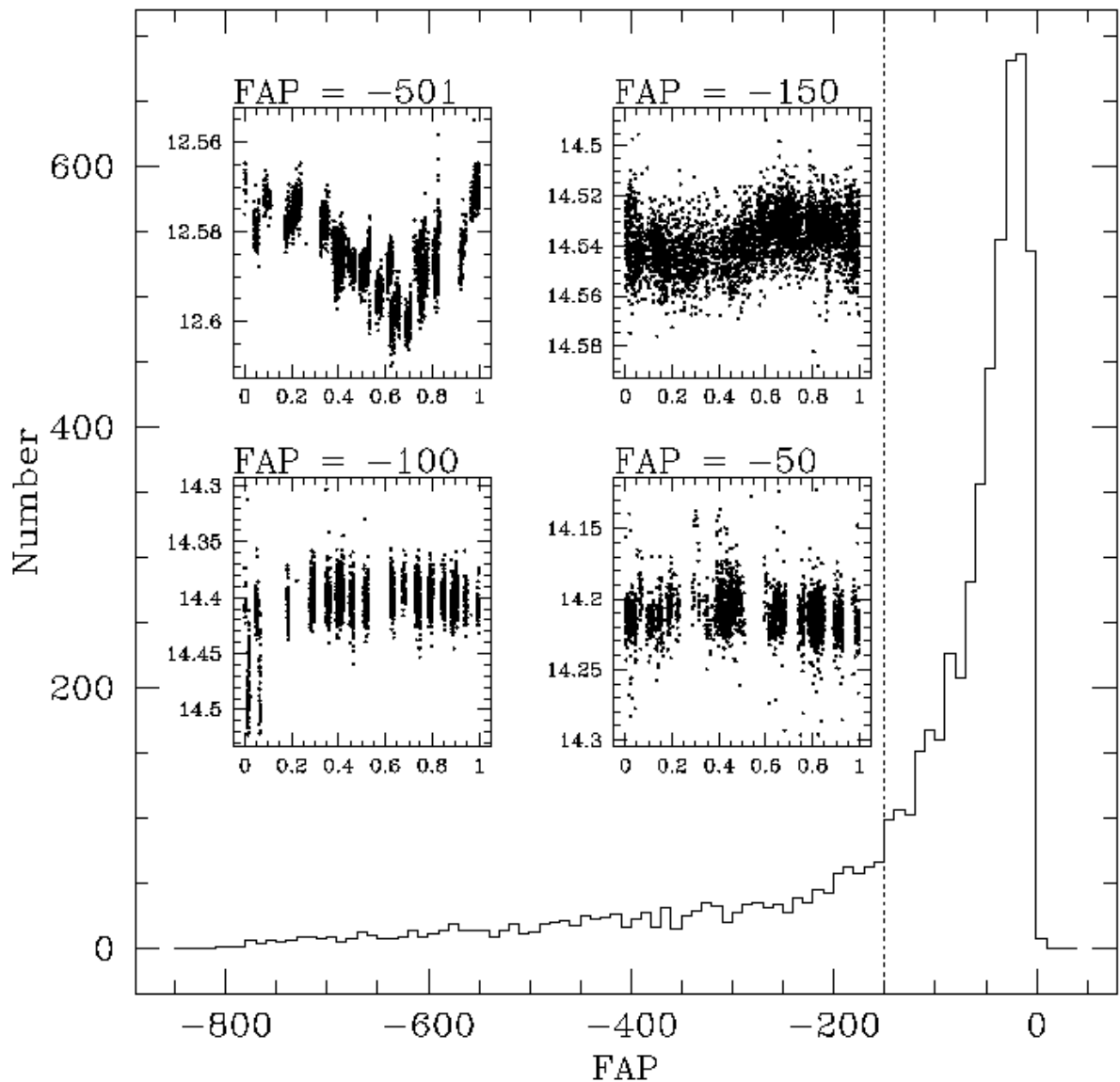}
\caption{Histogram of the logarithm of the formal false alarm probability values from the Lomb-Scargle period finding algorithm. Example phased light curves at a range of FAP values are shown. The units on the inset plots are phase for the x-axis and $r$ magnitude for the y-axis. We select stars with $FAP < -150$ and $P \ge 0.1~{\rm days}$ or $FAP < -10$ and $P < 0.1~{\rm days}$ as variables. The dotted line shows the selection for the longer period variables.}
\label{fig:LSfapdist}
\end{figure}

\begin{figure}[p]
\plotone{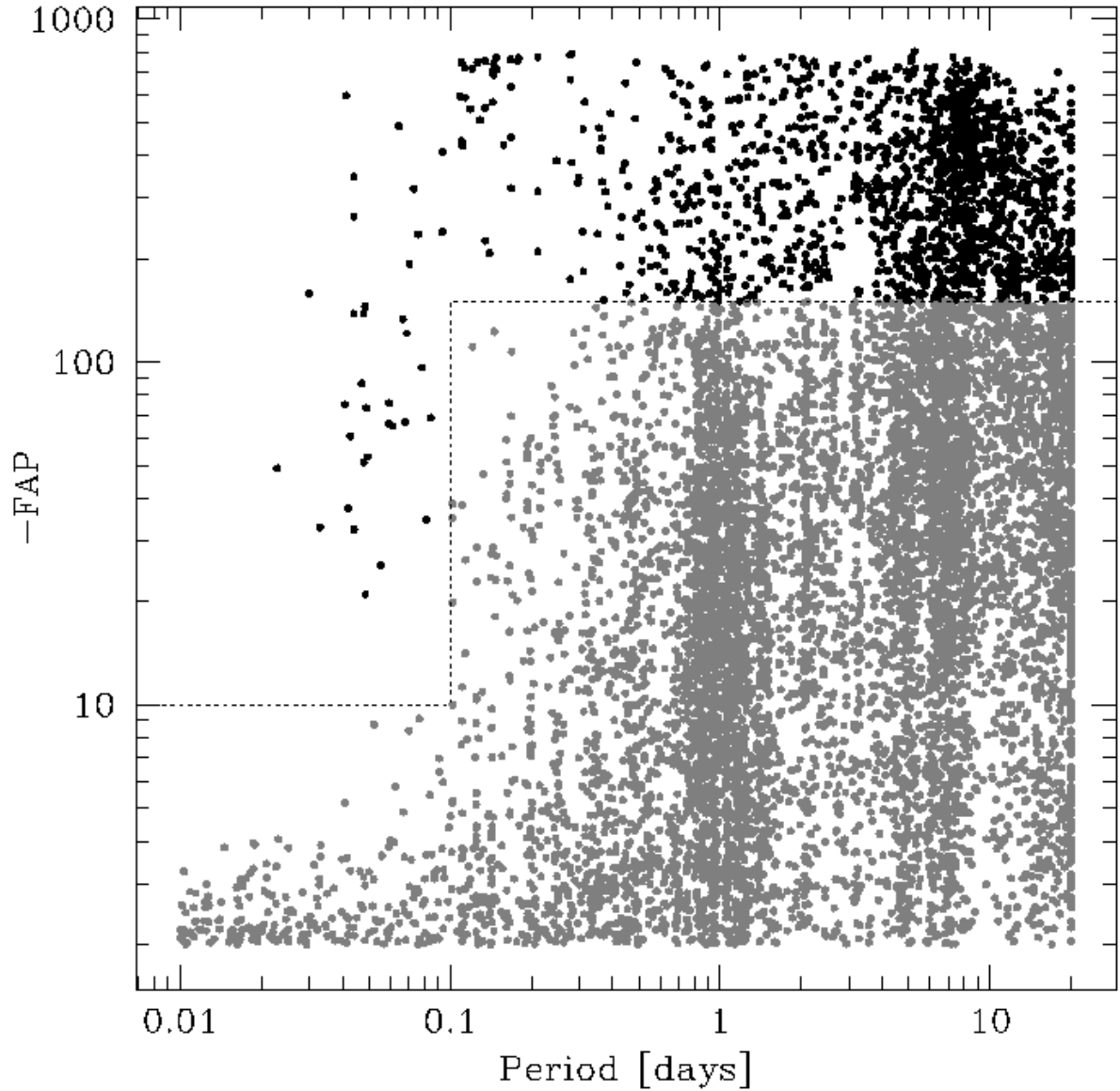}
\caption{Period vs. the negative logarithm of the formal false alarm probability values from the Lomb-Scargle period finding algorithm. We only plot points the pass the removal of bad frequencies described in the text and that have $FAP < -2$. The dotted line shows the selection. Dark points show selected variable candidates, light points show rejected stars.}
\label{fig:LSFAPvsPeriod}
\end{figure}

In figure~\ref{fig:LSfapdist} we plot the histogram of formal FAP values together with example light curves at a range of values. Only stars detected in all three filters ($gri$), with more than 500 points in their light curves, and with $r < 20$ are included in the histogram. As seen in figure~\ref{fig:LSFAPvsPeriod}, we adopt a selection criterion of $FAP < -150$ for stars with $P \geq 0.1~{\rm days}$ and $FAP < -10$ for stars with $P < 0.1~{\rm days}$. We use a less severe selection threshold for short period stars since they are less affected by red noise. This selects 1533 candidate variables. After applying the selections described below in \S 3.5, 1369 of these candidates are included as variables in the final catalog.

\placefigure{fig:LSfapdist}

\placefigure{fig:LSFAPvsPeriod}

Note that while the formal false alarm probability for the long period detections is an astonishingly small $10^{-150}$, the actual false alarm probability is much higher due to red noise. While methods such as the trend filtering algorithm \citep[TFA;][]{Kovacs.05} or SYSREM \citep{Tamuz.05} may filter red noise from an ensemble of light curves, they are known to distort the signal of large amplitude variables and may reduce the sensitivity to these variables as well. Since we are primarily interested in large amplitude variables, we do not apply either of these filtering methods to our light curves. A filtering method will be applied to the light curves before searching for planetary transits and other low amplitude variables; the results from this will be presented in a separate paper.

\subsection{AoV search}

\begin{figure}[p]
\plotone{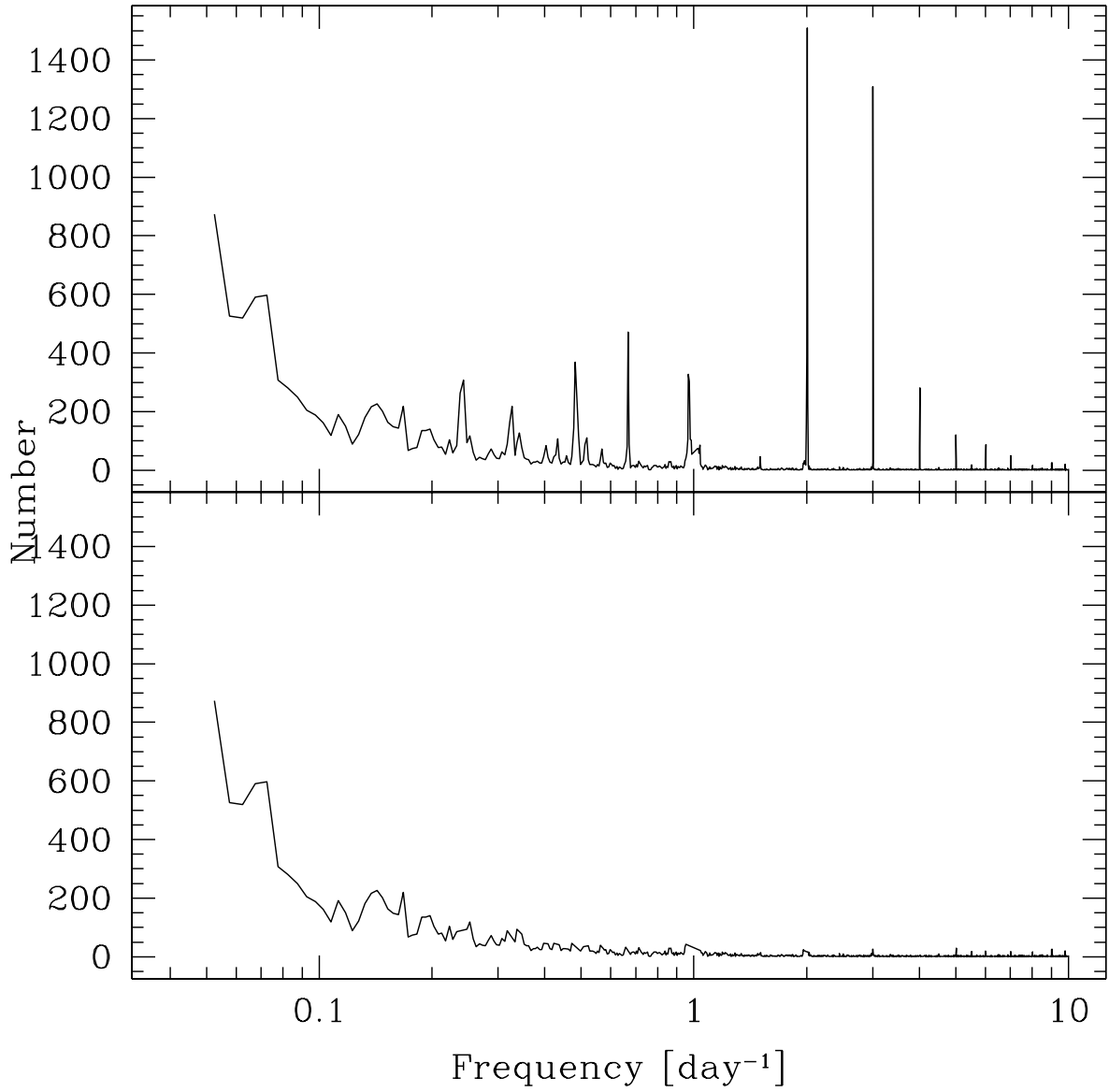}
\caption{Same as figure~\ref{fig:LSPeriodHist}, this time shown for AoV.}
\label{fig:AOVPeriodHist}
\end{figure}

\begin{figure}[p]
\plotone{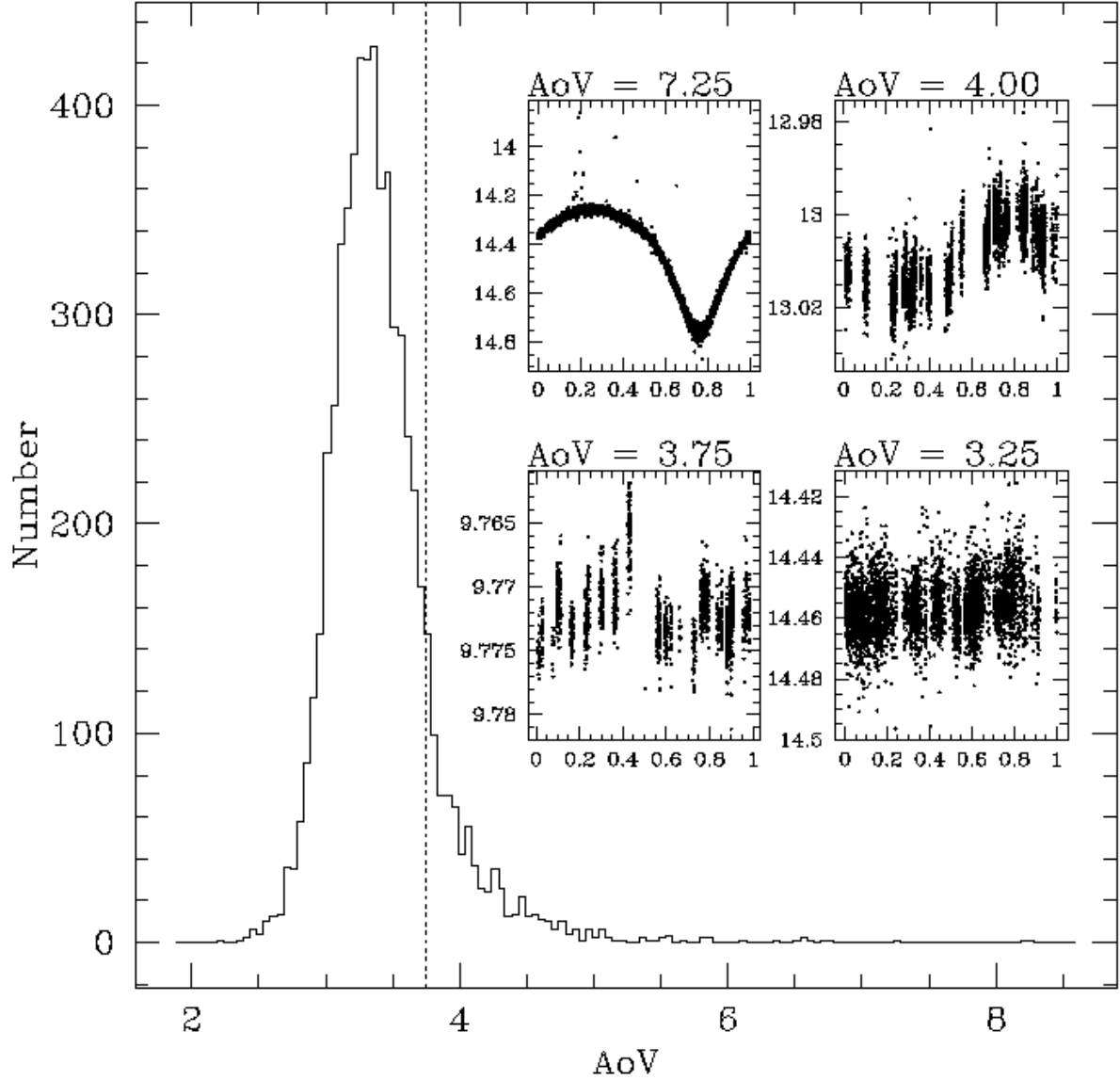}
\caption{Same as figure~\ref{fig:LSfapdist}, here we plot the histogram of AoV. The selection criterion is $AoV > 3.75$.}
\label{fig:AoVdist}
\end{figure}

We used the AoV algorithm with phase binning to search for variables with periods between 0.1 and 20 days; a coarse scan was conducted at 0.01 times the Nyquist sampling with a fine tuning search around the highest peaks at 0.0005 times the Nyquist sampling. Following the \citet{Devor.05} implementation, we normalize the AoV periodogram of each light curve to have zero mean and unit variance. As for L-S, we first reject spikes in the frequency histogram (removing 8968 light curves, figure~\ref{fig:AOVPeriodHist}) and set the selection criterion by visual inspection of light curves. Figure~\ref{fig:AoVdist} shows the histogram of normalized AoV values together with example light curves. We adopt a selection criterion of $AoV > 3.75$ which selects 982 candidate variables, 615 of these are included in the final catalog.

\placefigure{fig:AOVPeriodHist}

\placefigure{fig:AoVdist}

\subsection{BLS search}

\begin{figure}[p]
\plotone{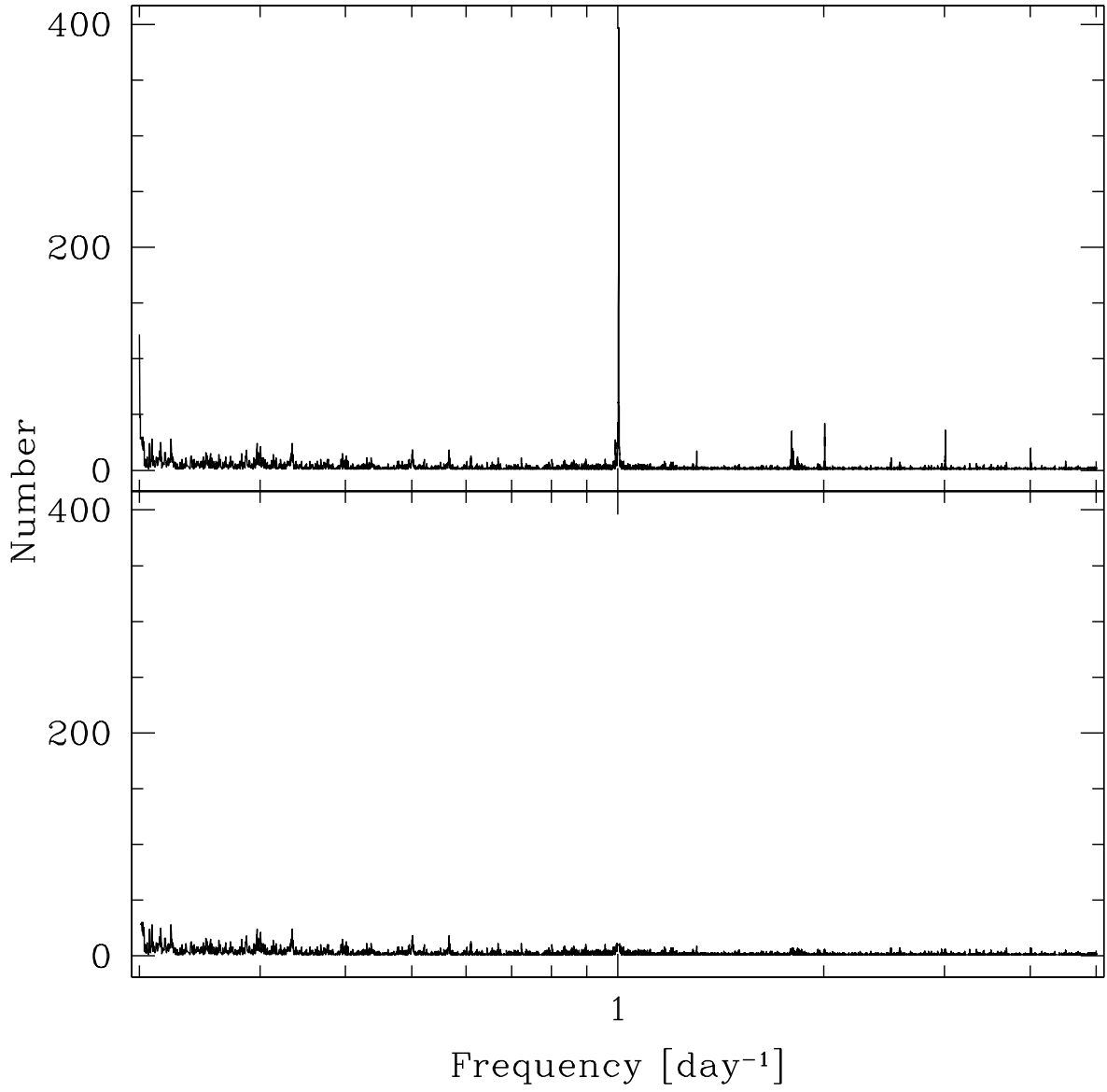}
\caption{Same as figure~\ref{fig:LSPeriodHist}, this time shown for BLS.}
\label{fig:BLSPeriodHist}
\end{figure}

We search for eclipsing variables using the BLS algorithm. The search is restricted to periods between 0.2 and 5.0 days as the detection efficiency drops for longer periods. As for L-S and AoV we first reject bad periods by searching for spikes in the histogram of detected periods (5621 light curves are rejected; see figure~\ref{fig:BLSPeriodHist}).

\placefigure{fig:BLSPeriodHist}

Following \citet{Pont.06} we use as our selection statistic the signal to pink noise:
\begin{equation}
SN^{2} = \frac{\delta^{2}}{\sigma_{w}^{2}/n_{t} + \sigma_{r}^{2}/N_{t}}
\end{equation}
where $\delta$ is the depth of the eclipse, $n_{t}$ is the number of points in the eclipse, $N_{t}$ is the number of distinct eclipses sampled, $\sigma_{w}$ is the white noise, and $\sigma_{r}$ is the red noise at the time-scale of the eclipse. We subtract the best BLS model from each light curve and set $\sigma_{w}$ equal to the standard deviation of the residual. We then bin the residual light curve in time with a binsize equal to the duration of the eclipse and set $\sigma_{r}$ equal to its standard deviation. 

\begin{figure}[p]
\plotone{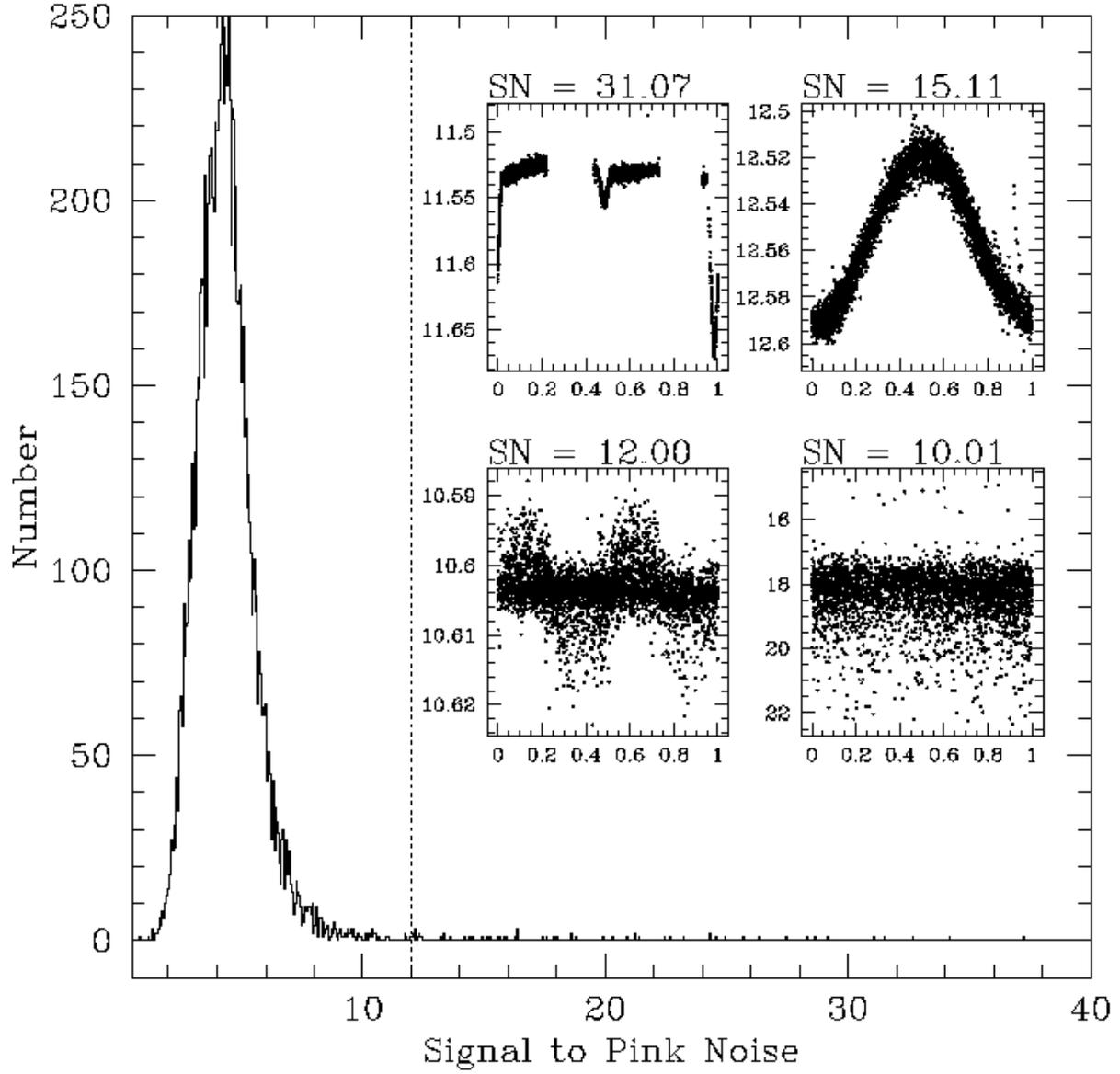}
\caption{Same as figure~\ref{fig:LSfapdist}, this time shown for BLS. The selection criterion is $SN > 12.0$.}
\label{fig:BLS_Hist}
\end{figure}

Figure~\ref{fig:BLS_Hist} shows a histogram of the resulting $SN$ values for all light curves, together with sample light curves at a range of $SN$ values. We select 90 light curves with $SN > 12.0$ as candidate variables, 43 of these are included in the final catalog.

\placefigure{fig:BLS_Hist}

\subsection{Estimation of the Period Errors}

\begin{figure}[p]
\plotone{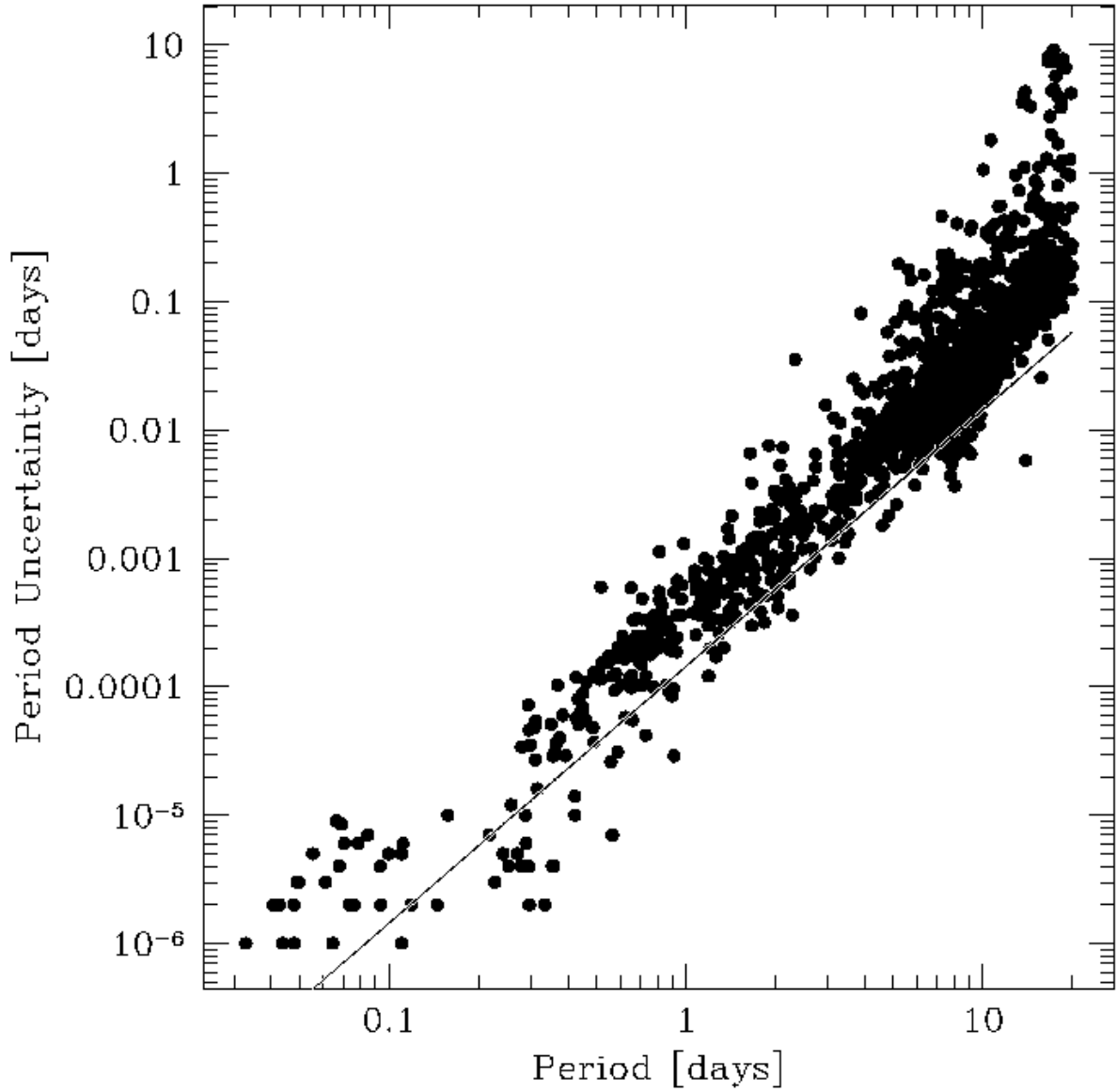}
\caption{The uncertainty on the variable star periods as a function of period. The solid lines shows the expected relation for a signal to noise of 10 and a residual correlation time-scale of 30 minutes.}
\label{fig:PeriodErr}
\end{figure}

For the variables with periods detected by L-S or AoV, we measured the period uncertainty by bootstrap. For each bootstrap iteration we draw a sample of points from the light curve with replacement and then search for the period within a frequency of $\pm 1/T$ from the period found for the original light curve, where $T$ is the time-baseline of the light curve. The period uncertainty for a light curve is given by the standard deviation of 100 bootstrap iterations. Figure~\ref{fig:PeriodErr} shows the period uncertainty of the variables as a function of their period. 

We can estimate the expected uncertainty on the period using equation 25 of \citet{Schwarzenberg-Czerny.91}:
\begin{equation}
\Delta P \sim P^{2} \sqrt{\frac{3 \sigma^{2} D}{T^3}}
\label{eqn:period_err}
\end{equation}
where $P$ is the period, $\sigma$ is the noise in units of the signal amplitude (i.e., the inverse of the signal to noise), $D$ is the average correlation time-scale for the residuals from the model periodic signal (equal to the sampling time for the case of pure gaussian white noise with a perfect sinusoidal signal), and $T$ is the time base-line. This relation holds for $T \gg P \gg D$. Note that the period error does not include uncertainties due to aliasing - it describes how well you can centroid the peak in a periodogram after choosing amongst aliases. We choose by eye the best period for each light curve from among the possible aliases.

We plot equation~\ref{eqn:period_err} on figure~\ref{fig:PeriodErr} for $T = 31.1$ days, assuming a typical signal to noise of 10 and a correlation time-scale of half an hour. Note that $D$ is effectively a free parameter, since it is difficult to know a priori how well the AoV ``model'' will fit the actual light curves, and it is adjusted to provide a good fit to the short period errors. The errors from the bootstrap appear to be in reasonable agreement with theoretical expections, though only if we allow for time-correlated noise. The disagreement for the longer period detections is due to the violation of the assumption $T \gg P$.

\placefigure{fig:PeriodErr}

\subsection{Combined Catalog of Variables}

The combined list of candidate variables contains 1963 sources. Of these, 38 were detected by all three methods, 543 were detected by AoV and L-S but not by BLS, 6 were detected by L-S and BLS but not by AoV, 16 were detected by AoV and BLS but not by L-S, 384 were detected only by AoV, 946 were detected only by L-S, and 30 were detected only by BLS.

Before proceeding we must first correct for the blending of variability between sources that results from performing aperture photometry on the subtracted images \citep[e.g.][]{Hartman.04}. To do this we identify groups of candidate variables that lie within $4\arcsec$ of each other (so that their PSFs may overlap significantly). We then choose from each group the source with the smallest average deviation of the centroid position measured on the subtracted images from the star's position measured on the reference image. After this procedure we are left with 1825 candidate variables. Requiring that the variables be detected in the $g$, $r$ and $i$ filters reduces this number to 1732.

\begin{figure}[p]
\plotone{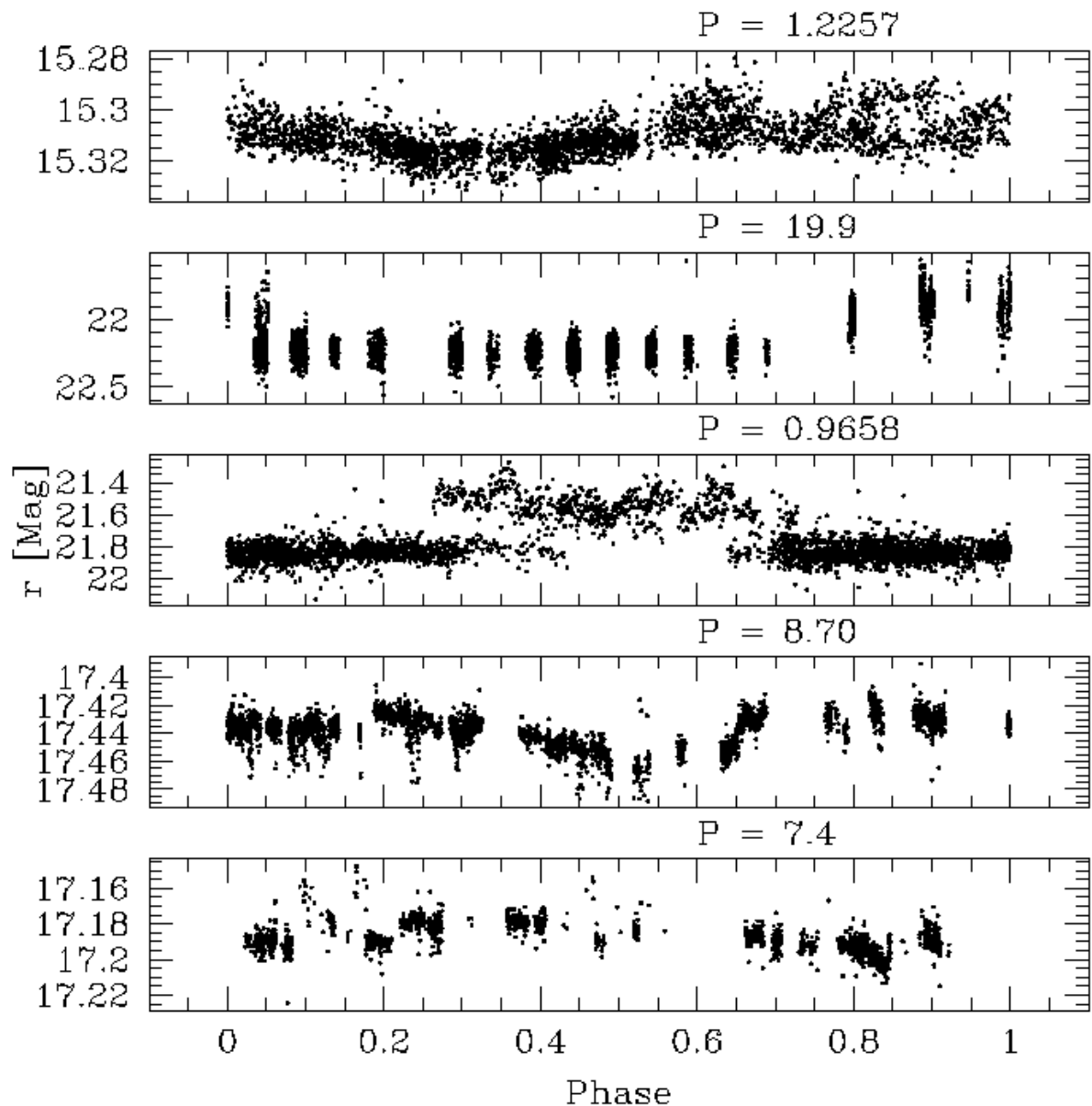}
\caption{Examples of candidate variable star light curves which were rejected by eye. The listed period is the value returned by LS and is in days. The top candidate may very well be real, we erred on the side of caution, however, and rejected it since it is bright enough for the light curve to be dominated by red noise, shows signficant scatter when phased and lies close to the edge of the chip where the image subtraction tended to be less reliable. The next two candidates are both located on the same chip. The apparent variation is due to the same five nights being brighter than average for both stars. This systematic variation was seen in the light curves of several stars on this chip and is due to the appearance of bad columns and rows on the chip that did not occur on other nights. The second candidate from the bottom may be real but was rejected due to its proximity to the edge of the chip. The bottom candidate was rejected because it lies in the wings of a much brighter saturated star.}
\label{fig:VarRej}
\end{figure}

We then examine all candidate light curves by eye, choosing among the possible periods the one that appears to phase the light curve most cleanly, and rejecting false positive detections. In addition to inspecting the light curves we also inspected the position of each star on the reference mosaic image rejecting stars on bad columns or in the wings of very bright stars. After this procedure we are left with 1445 variables, of which 1380 have periods (note that variables without periods correspond to light curves which showed believable variability but for which the period returned by the selection tools did not appear to be correct).  In figure~\ref{fig:VarRej} we show examples of some of the 287 light curves which were rejected by eye. Note that some of these may well be real variables, and unfortunately the selection process is fairly subjective, however it is very difficult to develop an automatic selection routine that is completely robust against false positives, but does not reject a significant number of true variables, when dealing with real data.

Table~\ref{tab:M37_varcatalog} gives the first few entries in our catalog of variable stars, the full catalog is available in the electronic edition of the journal. We continue the numbering system from \citet{Kiss.01}, relabeling KV1 through KV17 from \citet{Kang.07} as V8-V24 and labeling new variables begining with V25. The new variables are sorted by RA. For completeness we include unmatched known variables, though we caution that the coordinates for V8-10, V15-16, V18, and V23 are uncertain since we did not observe these stars and there appears to be an error in the table listing the variable star coordinates in \citet{Kang.07} (see next subsection). Light curves for all variable stars are available on request. Table~\ref{tab:M37_varsummary} provides a summary of the variable star catalog.

We also inspected the light curves of the 94 variable candidates that were not detected in all three filters, we selected 29 of these as real variables based on their light curves. Table~\ref{tab:M37_varcatalog_suspect} lists the positions, magnitudes and periods for these suspected variables.

\section{Match to Known Variables}

There are a total of twenty four variables that have been previously identified in the field of M37. Of these, seven were found by \citet{Kiss.01} and an additional seventeen were found by \citet{Kang.07}. Thirteen of these variables are eclipsing binaries, ten show $\delta$-Scuti type variations and one is an RRc variable.

We matched the previous variable star catalogs to the list of 23,790 sources that we obtained light curves for using an astrometric matching radius of $0\farcs5$. Fifteen of the known variables matched to sources in our input photometry list. Six of the unmatched variables (V1-2, V8-10 and V18) are saturated in our observations, the other three unmatched sources (V15-16, and V23) fell on chip gaps and were not observed. We independently select all fifteen matched sources as variables. In matching to the known variables we found that there appears to be a discrepancy between the coordinates for several of the KV variables listed in table~1 of \citet{Kang.07} and the light curves presented in that paper. We came to this conclusion by comparing the light curves and periods of the matched sources with the light curves in the paper. We revise the coordinates in our own catalog following the naming convention used for the light curves in \citet{Kang.07}.

\begin{figure}[p]
\plotone{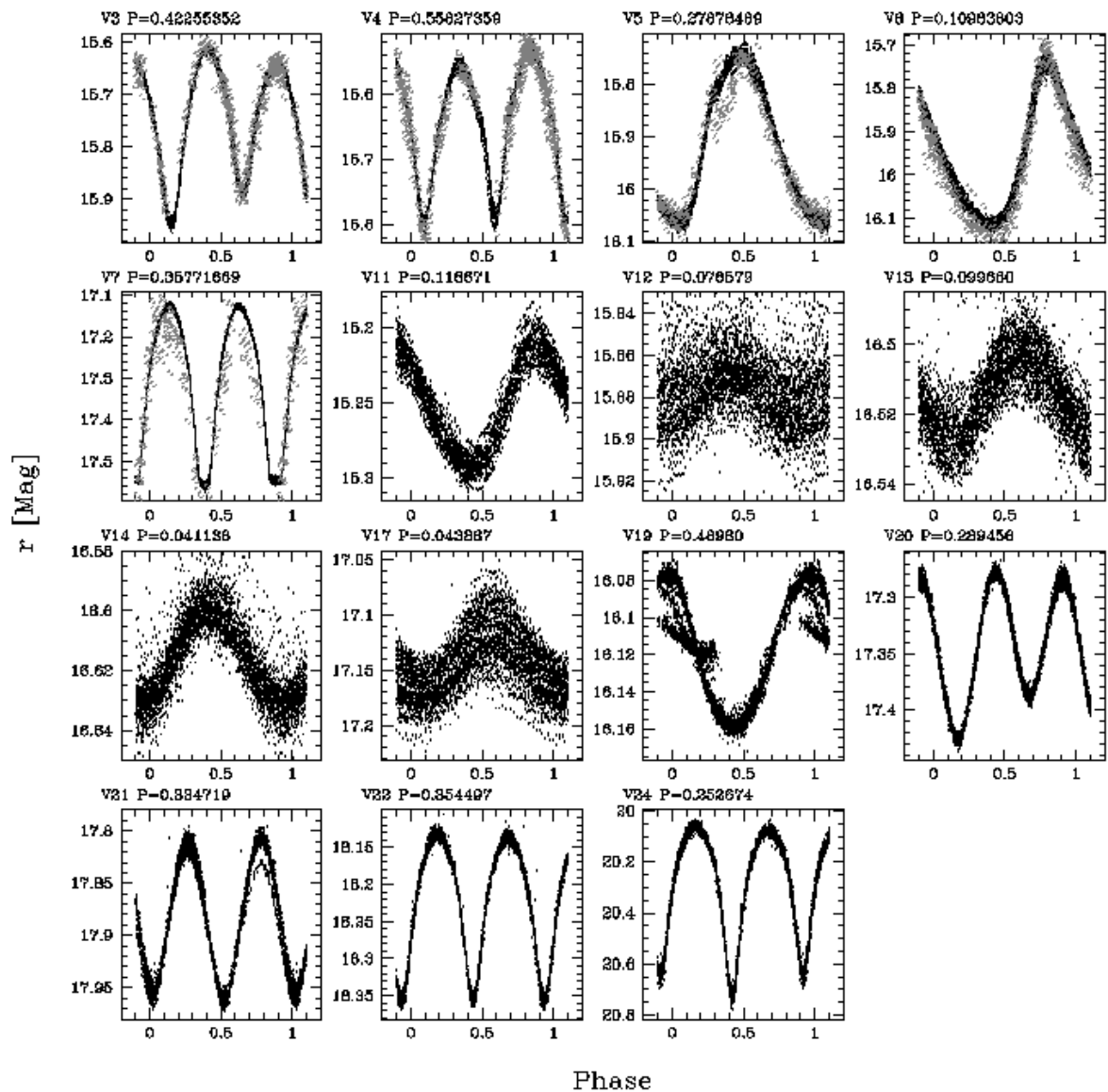}
\caption{Light curves of 15 previously identified variables that we match to. For variables V3-V7 we overplot the $R_{c}$ published photometry from \citet{Kiss.01} in gray. The \citet{Kiss.01} photometry has been shifted by an offset to match up with our data. Note that V4 exhibits a pronounced O'Connell effect and also shows evidence for variations of $\sim 0.01$ mag between cycles.}
\label{fig:matchvarLC}
\end{figure}

Figure~\ref{fig:matchvarLC} shows our light curves for the fifteen variables that we recover together with the published photometry from \citet{Kiss.01}. For the variables V3-V7 we combine our photometry with the \citet{Kiss.01} photometry which was taken between Dec. 1999, and Feb. 2000 to improve the precision of the periods. We will now discuss a few of these variables in detail, though we leave the very interesting star V17 for \S 6.

\placefigure{fig:matchvarLC}

\subsection{V3 and V4}

The contact eclipsing binaries V3 and V4 both exhibit a pronounced O'Connell effect \citep[unequal maxima;][]{OConnell.51}. The origin of this effect is still uncertain, though a number of theories have been put forward to explain it including stellar spot activity \citep{Binnendijk.60}, the formation of a hot spot on one of the stars due to mass exchange \citep{Shaw.94}, or an unequal interaction between the two components and any circumstellar material \citep{Liu.03}. In any of these theories, one would expect the effect to be variable. V4 does show some evidence for variation of $\sim 0.01$ mag between the 2000 and 2006 epochs as well as within the 2006 epoch as seen in figure~\ref{fig:matchvarLC}; V3 on the other hand, does not show evidence of variation. Note that four of the newly discovered contact binaries also exhibit the O'Connell effect (V37, V706, V936, and V1194, see figure~\ref{fig:EBLC}).

%\placefigure{fig:V4LC}
%
\subsection{V19}

\citet{Kang.07} classify this variable as a possible W UMa type eclipsing binary, though the classification is unsecure due to a lack of phase coverage. We find that the light curve shows substantial amplitude variations which make it most likely a fundamental mode RR Lyrae pulsator with the Blazhko effect.

\section{Properties of the Variables}

\begin{figure}[p]
\plotone{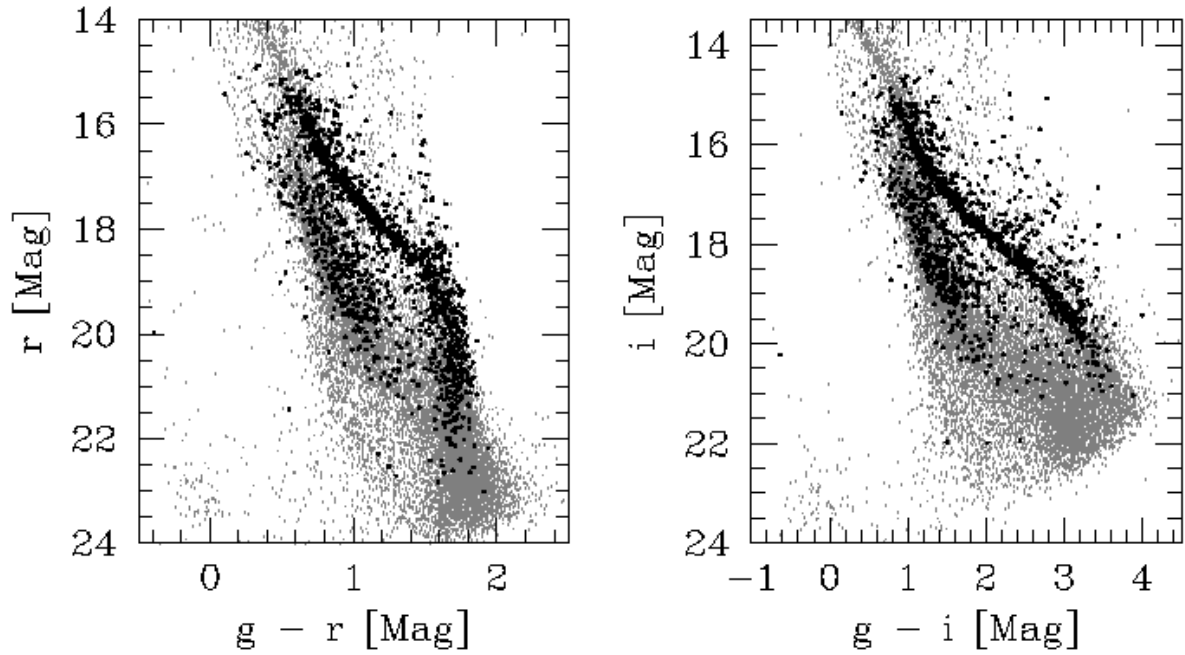}
\caption{Location of all variables identified by this survey on $gr$ and $gi$ CMDs. The dark points are the variables, the gray-scale is used to show all the point sources. The variables appear to lie preferentially along the cluster main sequence. The cut-off at the bright end is due to saturation of our primary time-series data. Note that the formal photometric error on the colors goes from $\sim 0.015$ mag at the bright end of each plot to $\sim 0.03$ mag at the faint end.}
\label{fig:VarsonCMD}
\end{figure}

In figure~\ref{fig:VarsonCMD} we show the location of all the variables identified by this survey on $gr$ and $gi$ CMDs. It is immediately apparent that the variables lie preferentially along the cluster main sequence. Most of these are rapidly rotating stars (relative to the Sun) with surface brightness inhomogeneities that are detectable at our level of photometric precision. We will discuss these variables in detail in a separate paper (Hartman et al. 2007, in preparation). Given the relatively young age of the cluster (550 Myr), the fact that we detect these variations for so many cluster stars is not surprising.

\placefigure{fig:VarsonCMD}

\begin{figure}[p]
\plotone{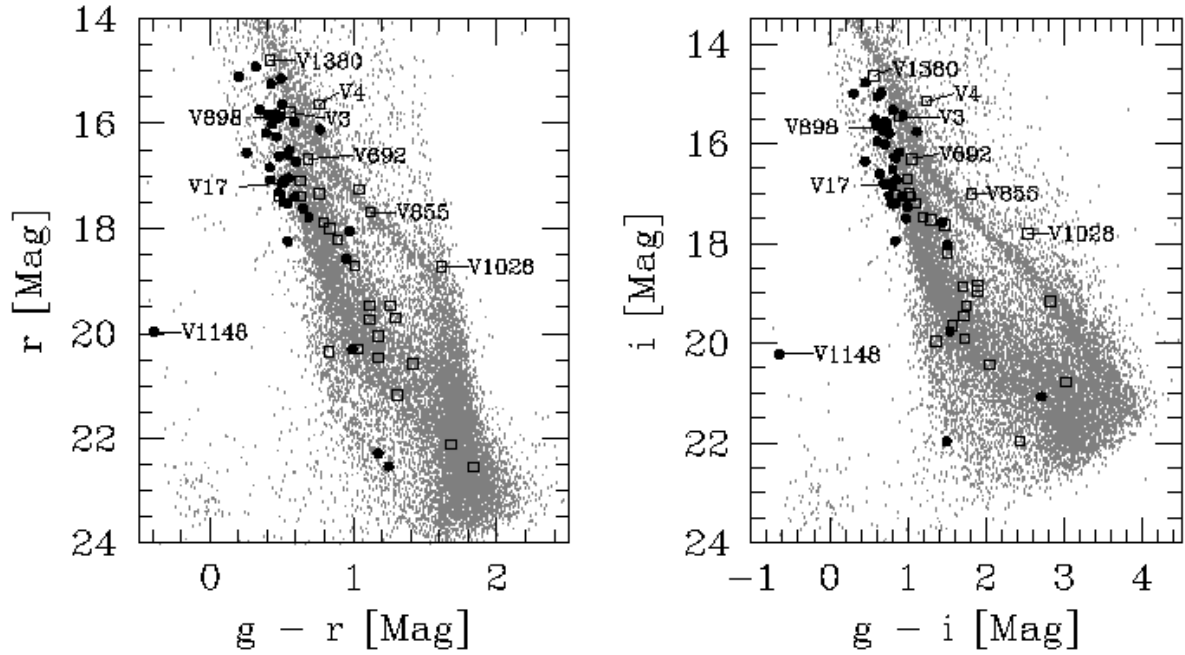}
\caption{Location of visually selected short-period pulsating variables (filled points) and eclipsing binaries (open squares) on $gr$ and $gi$ CMDs. Labels indicate some of the variables discussed in the text. Note that we do not use a symbol to plot V898 as it is neither an eclipsing binary nor a pulsating star, the label however shows its location on the CMDs.}
\label{fig:PulsandEBonCMD}
\end{figure}

From inspecting the light curves of all the variables identified by our survey, we classify 27 of the variables as eclipsing binaries (or ellipsoidal variables) and 39 of the variables as pulsating variables with periods less than 1 day. Including the 4 previously identified eclipsing binaries and 5 previously identified pulsating stars that we have not observed, there are a total of 31 eclipsing binaries and 44 pulsating stars known in the field of this cluster. We show the locations of these variables on the $gr$ and $gi$ CMDs in figure~\ref{fig:PulsandEBonCMD} and show the phased light curves for the newly discovered variables in figures~\ref{fig:PulsLC} and \ref{fig:EBLC}. As the red-edge of the $\delta$-Scuti instability strip intersects the M37 main sequence at $V \sim 14.4$ or $r \sim 14.2$ \citep{Kang.07}, we expect that any $\delta$-Scuti variables within this cluster will be saturated in our time series observations. The one or two pulsating variables in figure~\ref{fig:PulsandEBonCMD} that lie near the cluster main sequence are thus likely to be background disk stars.  There are six eclipsing binaries that lie near the cluster main sequence, these include V3, V4, V692, V855, V1028 and V1380. The binaries V3, V4 and V855 are contact systems, while V692, V1028 and V1380 are detached systems. We will discuss V1028 and V1380 in greater detail in \S 6.4 and \S 6.5. 

We have obtained spectra for the system V692 and detected it as a single lined spectroscopic binary. The cross-correlation routine gives a temperature for this system of $\sim 7000$ K which is too hot for it to be a member of the cluster given its position on the CMD, but consistent with it being a member of the background field. Based on the radial velocity curve, we estimate that the systemic RV is $\sim -2$ km/s compared with $9.4 \pm 1.4$ km/s for the cluster (Paper I). We conclude that this system is not a member of the cluster, and will not investigate it any further.

\placefigure{fig:PulsandEBonCMD}

\placefigure{fig:PulsLC}

\placefigure{fig:EBLC}

Figure~\ref{fig:PerMag} shows the period-$r$ relation for the variables. In two of the panels we split the variables into two groups, those which match to photometrically selected cluster members (see Paper I), and those which lie well away from the cluster main sequence. There is a strong correlation between period and magnitude for the cluster members with a break into two groups at $r \sim 18$. We are most likely seeing the relation between rotation period and stellar mass for the young cluster stars. The relation is quite similar to that seen, for example, in the Hyades by \citet{Radick.87}, and will be discussed in greater detail elsewhere. The non-cluster members appear to show no clear correlation between period and magnitude. This is not unexpected, though, since the field stars lie over a large range of distances from the Earth.

\placefigure{fig:PerMag}

We will now discuss a few of the populations of field variables.

\subsection{Short Period Pulsators}

The majority of the short-period pulsating variables appear to lie in the range $r < 19$, $0.2 < g - r < 1$. These stars have pulsation periods on the order of 0.1 days and have $0.45 < B - V < 1.18$, they are thus likely to be background $\delta$-Scuti or SX Phe (Population II $\delta$-Scuti) stars largely seen through $E(B-V) > 0.4$. 

Fundamental mode $\delta$-Scuti variables exhibit a well-known period-luminosity-color (PLC) relation that we can use to determine their distances. \citet{McNamara.00} give the following period-color and period-luminosity relations:
\begin{eqnarray}
(V-I_{C})_{0} &=& 0.195\log P + 0.514, \label{eqn:DeltaScutiPC}\\
M_{V} &=& -3.725\log P - 1.933. \label{eqn:DeltaScutiPL}
\end{eqnarray}
The first of these equations can be used to determine the $E(V-I_{C})$ reddening, then assuming $E(V-I_{C}) = 1.56E(B-V)$ (Paper I) and $A_{V} = 3.1E(B-V)$ we can determine the distance using equation~\ref{eqn:DeltaScutiPL}.

We identify four stars, based on their periods, amplitudes and light curve shapes, that are quite likely fundamental mode pulsators. We give the identifications, colors, periods and inferred distances in table~\ref{tab:DeltaScutidist}, the BV photometry for V6 comes from \citet{Kang.07} as this star was not observed by \citet{Kalirai.01}. We include V5 in this list, though we note that \citet{Kang.07} have argued that this star may be an RRc variable rather than a fundamental mode $\delta$-Scuti. Figure~\ref{fig:WUMAdist} shows the distance and extinction for these four stars as well as for a number of eclipsing binaries discussed below. One star lies well beyond the disk of the galaxy and is likely to be a halo SX Phe star. The total extinction in this direction appears to be $A_{V} \sim 2$ mag, or $E(B-V) = 0.65$ which is slightly larger than the value of $E(B-V) = 0.56$ from the \citet{Schlegel.98} extinction maps (note that the galactic latitude of the field is $\sim 3^{\circ}$, so the \citet{Schlegel.98} map is unreliable here).

\placetable{tab:DeltaScutidist}

\subsection{Eclipsing Binaries}

As seen in figure~\ref{fig:EBLC}, many of the eclipsing binaries are contact, W UMa type systems. Similar to $\delta$-Scuti variables, these binaries exhibit a PLC relation that we can use to determine their distances. We adopt the following PLC relation from \citet{Rucinski.97}:
\begin{equation}
M_{V} = -2.38 \log P + 4.26(B-V)_{0} + 0.28 \label{eqn:WUMAMvBV}
\end{equation}
where we assume solar metallicity, and note that the resulting uncertainty on the distance modulus for each source is $\sim 0.5$ mag. 

The reddening of each system can be estimated from the $B-V$-$V-I_{C}$ color-color plot assuming $E(V-I_{C}) = 1.56E(B-V)$ and using the fiducial relation for the cluster. Assuming $A_{V} = 3.1E(B-V)$, we can then find the distance to the system from equation~\ref{eqn:WUMAMvBV}. Table~\ref{tab:WUMAdist} gives the identifications, colors, periods and inferred distances and extinctions for 10 candidate contact binary systems. We do not include two systems which appear to have unreliable photometry (one has a negative reddening, the other yields $A_{V} > 4.0$ mag), and one star which has dereddened colors that are blueward of the cluster turnoff.

\placetable{tab:WUMAdist}

Figure~\ref{fig:WUMAdist} shows $A_{V}$ as a function of distance for the 10 systems listed in table~\ref{tab:WUMAdist}. The W UMa systems appear to show more scatter than the $\delta$-Scuti stars which may be attributed in part to the 0.5 mag uncertainty in the $P-L$ relation for these stars and in part to the greater uncertainty in their colors due to their larger amplitude variability. There is also a hint that the W UMa systems have systematically greater extinction at a given distance than what we infer from the $\delta$-Scuti stars. It is unclear what the source of this systematic discrepancy might be. 

We also show the extinction vs. distance from the Besan\c{c}on model \citep{Robin.03} for the galactic latitude/longitude of the field ($l = 177.6^{\circ}$, $b = 3.1^{\circ}$), assuming the default interstellar extinction of $0.7$ mag/kpc. While the value of $0.7$ mag/kpc appears to reproduce the extinction of the cluster, the model extinction/distance relation cuts off sharply at the scale-height of the galactic disc ($\sim 140$ pc) whereas the observed relation cuts off at about twice that distance. The model assumes that the dust follows a simple exponential density profile in both the radial and vertical directions; discrete clouds along the line of sight with higher than average density may cause the total extinction in this field to be greater than expected.

\placefigure{fig:WUMAdist}

\subsection{Rotating Field Stars}

There are $\sim 500$ variable stars that do not appear to be members of the cluster, are not eclipsing binaries, and are not $\delta$-Scuti or RR Lyrae type pulsators. We argue that most of these stars are likely to be rotating variable main sequence stars. Figure~\ref{fig:ExampleRotLC} shows light curves for 16 random variables of this type.

\placefigure{fig:ExampleRotLC}

In figure~\ref{fig:PerVI} we plot period against $V-I_{C}$ for the 481 of these variables that have period determinations and $V$ photometry. For stars with $1 < V-I_{C} < 2.3$ the periods appear to scatter between 0.1 and 20 days, with the highest density between 3 and 12 days, while stars with $V-I_{C} < 1$ or $V-I_{C} > 2.3$ tend to have shorter periods ($< 2$ days for stars with $V-I_{C} > 2.3$). The trend toward shorter periods among the bluer stars may either indicate a population of pulsators (e.g. $\gamma$ Doradus stars) or it may indicate shorter rotation periods for the hotter stars. From a visual inspection of the light curves alone it is impossible to distinguish between these two cases. For the redder stars, however, the trend toward shorter periods is probably due to the onset of short period rotation for cool stars. Note that for the cluster, $V-I_{C} = 2.3$ corresponds roughly to $r \sim 20$, which, as seen in the center panel of figure~\ref{fig:PerMag}, is where the stars appear to transition to a population of predominately short period rotators. The scatter between 3 and 12 days is also consistent with the range of rotation periods seen for brighter stars, keeping in mind that as the rotation periods increase the amplitudes decrease (thus making longer period rotators harder to detect) and that the rotation periods generally evolve over time from rapid rotation in the pre-main sequence stage to month long rotations for stars the age of the sun \citep[e.g.][]{Irwin.07a}.

\placefigure{fig:PerVI}

One way to test the consistency of our hypothesis that these variables are rapidly rotating, heavily spotted stars is to compare the amplitudes of their light curves to the Rossby number ($R_{o}$; the ratio of the rotation period to the convective turnover time-scale). Empirically there is a strong anti-correlation between the Rossby number and the x-ray activity of low-mass stars \citep{Noyes.84}, and a similar anti-correlation appears to hold for the light curve amplitudes of spotted rotating stars as well \citep[e.g.][]{Messina.01}. 

The turnover time-scale can be estimated from the empirical relation given by \citet{Noyes.84}. To apply this relation we must first de-redden our stars. As for the W UMa stars, this can be done using the $B-V$ vs. $V-I_{C}$ color-color plot together with the fiducial de-reddened relation for the cluster (figure~\ref{fig:BVVI_notincluster}). Because the reddening vector for stars with $B-V > 1.2$ or $V-I_{C} > 1.8$ may intersect the fiducial curve at more than one location, we select stars bluer than these limits. We also adopt a cut on the inferred $E(B-V)$ of $0 < E(B-V) < 0.7$ and a magnitude cut of $15 < V < 22$. To avoid including stars without surface convection zones we also only consider stars with $(B-V)_{0} > 0.5$. This last cut rejects 27 stars, many of which may be $\gamma$ Doradus variables.

\placefigure{fig:BVVI_notincluster}

In figure~\ref{fig:AmplitudeRossby_notincluster} we show the relation between $R_{o}$ and the peak-to-peak $r$-band amplitude ($A_{r}$) for the 232 stars which pass the above cuts (we also reject one star with $R_{o} \sim 2$ as it is an outlier from the rest of the observed stars). By eye there is a clear anti-correlation. To show that this is not due to a selection effect, we also plot for each star the estimated minimum amplitude that the light curve could have had and still have passed the variable star selection criteria. The methods for calculating the amplitude of the light curves and estimating the minimum detectable amplitudes will be described elsewhere (Paper III). To evaluate the signficance of the apparent correlation we calculate Kendell's normalized rank-based $\tau$ correlation coefficient, modified for the case of data suffering a one-sided truncation \citep{Tsai.90, Efron.92, Efron.99}. We find $\tau = -2316$ meaning the null hypothesis of no correlation is formally rejected at the $2316\sigma$ level. This correlation is stronger than what we find for the period-$A_{r}$ ($\tau = -1728$) or for the $(B-V)_{0}$-$A_{r}$ ($\tau = 565$) correlations. The anti-correlation between $R_{o}$ and $A_{r}$ is consistent with the hypothesis that rotation coupled with surface brightness inhomogeneities is the cause of the observed brightness variations for these stars.

\placefigure{fig:AmplitudeRossby_notincluster}

\section{Individual Variable Stars}

So far we have focused primarily on groups of variables. In this subsection we will investigate a few particularly interesting individual variables, including a hybrid $\gamma$ Doradus/$\delta$ Scuti pulsator (V17), a blue periodic variable star with $g-r < 0.$ (V1148), a blue flaring variable (V898), and two interesting detached eclipsing binaries (V1380 and V1028).

\subsection{V17}

\citet{Kang.07} emphasize the curious nature of the variable V17. They find two independent frequencies in its light curve: one at $f_{1} = 22.796$ day$^{-1}$ with a semi-amplitude of 32.9 mmag in $V$, and a second much lower frequency of $f_{2} = 2.396$ day$^{-1}$ with a semi-amplitude of 24.2 mmag in $V$. \citet{Kang.07} give two possible interpretations for this variable, that it is a hybrid $\gamma$ Doradus/$\delta$ Scuti type pulsator, or that it is a close semi-detached binary system with a $\delta$ Scuti primary and a late-type giant secondary showing ellipsoidal variations. They claim that the 2MASS color for the object suggests a late G or early K spectral type leading them to conclude that the binary hypothesis is more likely. 

Note that $\gamma$ Doradus pulsators are early F-type main sequence stars that have periods in the range of 0.4-3 days, amplitudes of a few percent and typically oscillate in multiple frequencies \citep{Krisciunas.98, Kaye.99}. The variations in these stars are thought to be due to high-order g-mode pulsations. The instability strip for this type of variable extends from $7200-7550$ K on the zero-age main sequence (ZAMS) and from $6900-7400$ near the end of the main sequence phase \citep{Dupret.04}. $\delta$ Scuti pulsators, on the other hand, are late A or early F-type stars that lie in the intersection of the classical instability strip and the main sequence. These stars have periods less than 0.3 days and amplitudes ranging from a few mmag to almost a full magnitude. The variations in this case are due to p-mode pulsations. The $\delta$ Scuti and $\gamma$ Doradus instability strips overlap so that hybrid stars, exhibiting both modes of pulsation, may exist.

To aid in understanding this interesting object we have conducted a multi-frequency analysis of our $r$ light curve using the Discrete Fourier Transform (DFT) with the CLEAN deconvolution algorithm \citep{Roberts.87}. Figure~\ref{fig:V17DFTCLEAN} shows the DFT power spectrum before and after implementing CLEAN; we also show the power spectrum of the window function for reference. We confirm the high-frequency signal at $f_{1} = 22.796$ day$^{-1}$ and we find four additional low frequencies of $f_{2} = 1.418$ day$^{-1}$, $f_{3} = 2.263$ day$^{-1}$, $f_{4} = 2.054$ day$^{-1}$, and $f_{5} = 0.823$ day$^{-1}$. None of these frequencies are harmonics of each other. The error on all of these measurements is $\sim 0.003$ day$^{-1}$. 

\placefigure{fig:V17DFTCLEAN}

After identifying the frequencies using DFT/CLEAN we fit a series of cosinusoids to the light curve of the form:
\begin{equation}
r = r_{0} + \sum_{i=1}^{7}a_{i}\cos(2\pi f_{i}(t - t_{0}) + \Phi_{i})
\end{equation}
to get the semi-amplitude and phase of each mode. For consistency with \citet{Kang.07} we adopt $t_{0} = 2453000.0$. We find semi-amplitudes of $a_1 = 26.4$ mmag, $a_2 = 13.7$ mmag, $a_3 = 12.8$ mmag, $a_4 = 12.9$ mmag, and $a_5 = 11.2$ mmag, and phases $\Phi_{1} = 2.854$, $\Phi_{2} = 1.132$, $\Phi_{3} = 4.866$, $\Phi_{4} = 4.757$, $\Phi_{5} = 5.477$. Alternatively, if we assume only the two frequencies from \citet{Kang.07} of $f_{1} = 22.796$ day$^{-1}$ and $f_{2} = 2.396$ day$^{-1}$ we find $a_1 = 27.2$ mmag, $a_2 = 8.0$ mmag, $\Phi_{1} = 2.144$, and $\Phi_{2} = 1.996$.

The presence of multiple frequencies in the range $0.8 < f < 2.3$ indicates that the low frequency variations are not due to ellipsoidal variability, and is instead suggestive of $\gamma$ Doradus pulsations. This is not in conflict with the color of the star since the late-G/early-K spectral type estimate by \citet{Kang.07} assumes no reddening. Taking $B-V = 0.6$ from \citet{Kalirai.01} and noting that the reddening to the source must be at least the cluster value of $E(B-V) > 0.23$ given that the star lies below the cluster main sequence on the CMD, we find $(B-V)_{0} < 0.37$ which, for a dwarf star, corresponds to a spectral type earlier than F5 \citep[e.g.][]{SchmidtKaler.82}. We conclude therefore that the star may very well be a hybrid $\gamma$ Doradus/$\delta$ Scuti pulsator. These stars are particularly interesting for astroseismology and to date only a handful of them are known \citep[see][]{King.06}. 

\subsection{The Hot Variable V1148}

As seen from its position on the CMD, star V1148 is an outlier with respect to the general population of variables found in our survey (figure~\ref{fig:PulsandEBonCMD}). The star has $g - r = -0.396 \pm 0.013$, $r - i = -0.247 \pm 0.015$, $r = 19.97 \pm 0.012$, $B-V = -0.071 \pm 0.004$ and $V = 19.770 \pm 0.003$. This places the star roughly 2 magnitudes above the cluster white dwarfs. As seen in figure~\ref{fig:PulsLC} the star shows saw-tooth like variability with a period of $0.157728 \pm 0.000010$ days and an amplitude of $0.07$ mag in $r$. This variable is likely to be a quiescent cataclysmic variable, though we will consider several hypotheses below.

While the position of the star on the CMD is consistent with it being a foreground white dwarf, the period is nearly an order of magnitude longer than the longest seen for white dwarf pulsations \citep[the longest periods being 5-50 minutes for GW Vir stars which are still embedded in planetary nebulae, see][]{Corsico.06}. Such a long period could be possible for a binary system, with the variations resulting from the reflection and/or ellipticity effects, or from an accretion hot-spot on the white dwarf. The sawtooth appearance of the light curve would be most consistent with the accretion hypothesis, though we note that the accretion rate would be quite stable over the course of a month if this hypothesis is correct.

Since the star cannot have $(B-V)_{0} < -0.33$ \citep{SchmidtKaler.82}, the maximum reddening that it could be seen through is $E(B-V) = 0.26$ ($A_{V} = 0.8$). From figure~\ref{fig:WUMAdist}, we can estimate that the star must be closer than $\sim 5$ kpc assuming that the star lies along an average line of sight. We can therefore put a limit on its absolute magnitude of $M_{V} \ga 5.5$. This rules out a main sequence star which must have $M_{V} \la 1.1$ given the color \citep{SchmidtKaler.82}. Even if there were no extinction, if the star were on the main sequence it would have to be at least $\sim 50$ kpc away, placing it well beyond the disk of the galaxy, or $\sim 300$ kpc assuming at least the reddening of the cluster. It is possible that the object could be a hypervelocity B star which has been ejected from the galaxy, these have been found at distances of 50-100 kpc \citep[e.g.][]{Brown.07}; though if this is the case the extinction to the star would have to be anomalously low. It is interesting to note, though, that the light curve shape, period, amplitude and color are all consistent with the star being a main sequence $\beta$-Cephei variable \citep{Sterken.93}. 

It is possible that the source is a sub-dwarf B (sdB) star, though the typical absolute magnitude of an sdB is $M_{V} \sim 4.6$ or brighter \citep{Moehler.97} which is slightly brighter than the limit we set based on the maximum extinction for the source. There is a class of pulsating sdB stars, called PG 1716 stars, with periods comparable to what we see for V1148 \citep{Green.03}. These stars pulsate in high-order g-modes and typically have very low amplitudes ($1$ mmag or less), with the highest observed amplitudes of $\sim 4$ mmag \citep{Kilkenny.06}. If V1148 is a member of this class, then its amplitude is more than an order of magnitude larger than the other members of its class. Such a large amplitude would be unusual for a high order g-mode pulsator since it would have to be pulsating in a very low degree $l$ mode. Alternatively the star might be a short-period sdB in a binary system \citep[e.g.][]{OToole.06}.

Alternatively, the star might be a blend between a white dwarf or an sdB star and a much fainter variable.  If the variable is a background $\delta$-Scuti or SX Phe star, the period and light curve shape suggest that it would be a fundamental mode pulsator. In that case we can estimate its intrinsic color and magnitude using equations~\ref{eqn:DeltaScutiPC} and \ref{eqn:DeltaScutiPL}. This yields $(V-I_{C})_{0} = 0.36$, and $M_{V} = 1.05$, or $B-V \sim 0.3$. Assuming a minimum apparent $B-V$ of $-0.33$ for the white dwarf/sdB star, and an extinction of $A_{V} = 2.0$ mag to the $\delta$-Scuti, the $\delta$-Scuti must have $V \ga 21$, and be $d \ga 40$ kpc away. As seen in figure~\ref{fig:WUMAdist} there does appear to be an unblended SX Phe star at $d \sim 33$ kpc, and it is possible that there could be an even more distant one that happens to lie along the line of sight to a white dwarf. 

The chances for a random blend of this nature though are rather low. Out of the 16431 point sources detected in our field with $g$, $r$ and $i$ measurements and with $r < 24$, there are 128 pairs of objects that are separated by $0.1-1\arcsec$. We do not detect any evidence that the object is anything other than a point source, so if the objects are within two magnitudes of each other then we estimate that they must be separated by $0\farcs1$ at the most. We would expect $\sim 1$ such random pairing out of all point sources in our field. The chances that one of the paired objects is a white dwarf and the other is a distant halo SX Phe star (both fairly rare in our field) is thus very low. 

Finally, we note that the variable star may be a short period rotating K or M dwarf (\S 5.3) that is in a binary system with a white dwarf or sdB star.  Such a pairing is more likely than the chance alignment of a white dwarf with a distant halo SX Phe star.  A period of $0.157$ days would be on the short side for such a star (see figure~\ref{fig:PerVI}), though double this period would also be permissible since the star may have multiple spots. The shape of the light curve is similar to that seen for other variables in the cluster. The spotted cluster stars have amplitudes $A_{r} < 0.2~{\rm mag}$; taking this as the maximum amplitude for a spotted star in the $r$-band we put an upper limit on the difference in magnitude between the companion and the white dwarf of $r_{comp} - r_{WD} < 0.6~{\rm mag}$. The minimum $g-r$ value that a star can have is $g-r = -0.6$ \citep[see][]{Girardi.04}, so to reproduce the color of the system the companion would have to have $g-r < 0.09$ (i.e. F0 or earlier). Therefore, this hypothesis appears to be inconsistent with the observations.

The most likely explanation for this variable is that it is foreground white dwarf with a close M dwarf companion, with the variations due to an accretion hot-spot on the surface of the white dwarf. Understanding the nature of this variable will require spectroscopy to determine $\log(g)$ and $T_{eff}$ of the hot star and to check for evidence of accretion and/or a blend with a cooler star. 

\subsection{V898}

The light curve of the variable V898 (figure~\ref{fig:V898LC}) is quite unique relative to the other variables in the survey.  The light curve shows continuous non-periodic variability with a number of relatively small amplitude outbursts ($0.1 - 0.2$ mag).  The typical duration of these outbursts is 0.5 days with a time-scale between outbursts of 1-2 days.  The variation seen in this source is similar to the flickering seen in many cataclysmic variables \citep[see][]{Duerbeck.96}, confirming this identification would require additional observations including spectroscopy. The source matches to 2mass05523126+3236324 and has $B-V = 0.604$, $V = 16.090$, $g - r = 0.474$, $r - i = 0.208$, $r = 15.876$, $J - K_{S} = 0.311$, $H - K_{S} = 0.110$ and $K_{S} = 14.374$. We estimate that the reddening to the object is $E(B-V) \sim 0.29$. 

\placefigure{fig:V898LC}

\subsection{Detached Eclipsing Binary V1380}

The eclipsing binary V1380 is one of three detached eclipsing binaries that lie near the cluster main sequence. V1380 matches to 2mass05531108+3224434, it has an orbital period of $2.1916 \pm 0.0016$ days, and photometry $B - V = 0.594$, $V = 14.952$, $g-r = 0.42$, $r-i = 0.15$, $r = 14.79$, $J - K_{S} = 0.331$, $H - K_{S} = 0.084$, and $K_{S} = 13.331$. The source is just at our saturation threshhold, and in many of the images it is saturated, so there are only 1111 points in its light curve (a factor of $\sim 4$ fewer than most light curves). If the system is a member of the cluster, its primary star would have $M \sim 1.2 M_{\odot}$ based on its $r$ magnitude and assuming equal mass components. We note, however, that the star does appear to be slightly below the main sequence, so its photometric membership is questionable.

To determine the period of the out of eclipse variations, we remove the in-eclipse points from the light curve and, using L-S, find a period of $0.939 \pm 0.01$ days with a peak-to-peak amplitude of $0.019$ mag. We subtract a sinusoid fit to this signal and find in the residual light curve an additional period of $1.166 \pm 0.01$ days with an amplitude of $0.006$ mag. To verify the detection of both periods, we have also used DFT/CLEAN to search for multiple periods. Figure~\ref{fig:V1380CLEAN} shows the window power spectrum, the raw DFT power spectrum of the light curve, and the power spectrum after using CLEAN. Using DFT/CLEAN we find two strong period detections at $0.941 \pm 0.011$ days and $1.174 \pm 0.013$ days. In figure~\ref{fig:V1380LC} we show the original light curve phased at the orbital period of the system, the out of eclipse points phased at $0.941$ days, the residual light curve after subtracting the above period phased at $1.174$ days, and the full light curve after subtracting a double sinusoid fit to the out of eclipse points. The model is subtracted through the eclipses, though this is technically incorrect. We adopt the DFT/CLEAN periods for the rest of the discussion.

\placefigure{fig:V1380CLEAN}

\placefigure{fig:V1380LC}

The additional periods are most likely not the rotation periods of the two binary components. The time-scale for the synchronization of rotation periods with the orbital period is much shorter than the time-scale for the circularization of the orbit \citep{Zahn.77}. Since the orbit appears to be circular (see below), we expect that it should also be synchronized. We will show that the additional periods are most likely due to one or both of the stars being a $\gamma$-Doradus type pulsator. 

We will now determine the physical masses and radii of the components of this binary system. To do this we will combine radial velocity curves with our light curve.

This binary was among the sources that we obtained spectra for with Hectochelle. The spectroscopic classification procedure discussed in Paper I yields a temperature of $T_{eff} = 6600 \pm 400$ K, and $v\sin i = 21.5 \pm 0.7$ km/s, these values are an average of the two stars. The metallicity is unconstrained so we assume solar metallicity. Only the spectra for the first two nights are used in the classification since the two components were not well separated in the last two nights; classifying spectra from the last two nights yields an artificially high value for $v\sin i$. For the cluster reddening of $E(B-V) = 0.227$ (Paper I), the expected color of the system is $B-V = 0.65 \pm 0.1$ which is consistent with the observed value of $B-V = 0.594$.

We combine the spectra from each of the four usable nights and calculate the cross-correlation as a function of radial velocity (RV) with \emph{xcsao} using the best template from the classification procedure. The cross-correlation plots show two clear peaks, indicating that we do detect the system as a double-lined spectroscopic binary. We then use the TwO-Dimensional CORrelation algorithm \citep[TODCOR;][]{Zucker.94} to determine the velocities for each component. We use the same template for both stars as the nearly equal depth eclipses means that the stars have similar surface temperatures. Table~\ref{tab:V1380RV} lists the radial velocity and light ratio measurements from the spectra. The average light ratio in the RV31 filter is $L_{2}/L_{1} = 0.992 \pm 0.025$.

\placetable{tab:V1380RV}

To combine the radial velocity curves with the light curve we need to improve the period that is derived from the light curve. To do this we fit a zero-eccentricity orbital solution to the radial velocity curves, assuming the period determined from the light curve, and allowing for a shift in phase.  From the phase shift we calculate a single ``observed'' primary eclipse time at the epoch of the RV observations. We then determine eclipse minimum times from the light curve using the Detached Eclipsing Binary Light curve fitter \citep[DEBiL;][]{Devor.05} and fit for the ephemeris using the equation:
\begin{equation}
T_{min} = HJD_{0} + nP.
\end{equation}
Here $T_{min}$ is the observed minimum time, $HJD_{0}$ is the reference primary eclipse epoch, $P$ is the orbital period and $n$ is the eclipse number. We find $HJD_{0} = 2453724.5306 \pm 0.0004$ and $P = 2.19258 \pm 0.00004$ days, where the errors are the standard errors from the the least squares fit. The observed eclipse timing measurements are listed in table~\ref{tab:V1380timing}. We include secondary eclipses in the fit and assume zero eccentricity, which we justify below.

\placetable{tab:V1380timing}

Using the above ephemeris, we fit a zero-eccentricity orbit to the RV curves finding $K_{1} = 118.3 \pm 1.3$ km/s, $K_{2} = 118.3 \pm 1.3$ km/s and $\gamma = 5.3 \pm 0.7$ km/s. The mass ratio is thus $q = 1.00 \pm 0.01$, so the two components have equal mass to within the precision of our observations. This is another example of a twin binary star system \citep{Pinsonneault.06}. The systemic radial velocity $\gamma$ is more than $3\sigma$ below the mean cluster value of $9.4 \pm 1.2$ km/s (Paper I). Figure~\ref{fig:V1380RV} shows the radial velocity observations with the best fit orbit.

\placefigure{fig:V1380RV}

Since the system is well detached, we model the light curve using the JKTEBOP program \citep{Southworth.04a,Southworth.04b} which is based on the Eclipsing Binary Orbit Program \citep[EBOP;][]{Popper.81,Etzel.81,Nelson.72}, but includes more sophisticated minimization and error analysis routines. We assume a linear limb darkening law, adopting $u = 0.564$ for both the primary and secondary stars. This coefficient is appropriate for the the $r$ filter with $T_{eff} = 6600$ K, $\log(g) = 4.5$ and $[M/H] = 0.0$ \citep{Claret.04}. We use a gravity darkening exponent of $1.0$, which is the value expected for a radiative envelope star, though the effect is negligible in this system. The resulting parameters are given in table~\ref{tab:V1380JKTEBOP}, where the listed errors are the $1\sigma$ errors from 100 Monte Carlo simulations. \citet{Southworth.05} have shown that the Monte Carlo routine in JKTEBOP provides robust error estimates, though we note that uncertainties in the limb darkening and reflection laws are not included in the Monte Carlo simulations. In conducting the simulations we set the error on each point in the light curve equal to $3.9$ mmag, which is the RMS of the residual from the best fit. 
The best fit model to the light curve is shown in figure~\ref{fig:V1380JKTEBOPLC}.

The ratio of the radii, $k$, is poorly constrained from the light curve. A better constraint could be obtained by including the spectroscopic constraint on the luminosity ratio, however since the spectra were obtained over a significantly different wavelength range from the light curve, including this constraint is not straightforward and beyond the scope of this paper.

To put a limit on the eccentricity of the system, we have also fit the light curve allowing $e \cos \omega$ and $e \sin \omega$ to vary. We find $e = 0.003 \pm 0.01$, which is consistent with a circular orbit. The error is determined from 100 Monte Carlo simulations. The $3\sigma$ upper bound on the eccentricity is $e < 0.03$. We have also attempted varying the third light contribution to the system, finding $L_{3} = 0.06 \pm 0.14$ which is consistent with zero, though poorly constrained.

\placetable{tab:V1380JKTEBOP}

\placefigure{fig:V1380JKTEBOPLC}

Combining the parameters determined from the RV curves with the parameters determined from the light curve, we find the following masses and radii for the stars: $M_{P} = 1.58 \pm 0.04 M_{\odot}$, $R_{P} = 1.39 \pm 0.07 R_{\odot}$, $M_{S} = 1.58 \pm 0.04 M_{\odot}$, and $R_{S} = 1.38 \pm 0.07 R_{\odot}$. The errors in the masses are dominated by the uncertainties in the velocity semi-amplitudes, while the errors in the radii are dominated by the large uncertainty in $R_{S}/R_{P}$.

In figure~\ref{fig:V1380agemetchi2} we plot the age-metallicity $\chi^{2}$ contours for the eclipsing binary system. To calculate $\chi^{2}$ at each age/metallicity point we use the Yale-Yonsei version 2 isochrones \citep[Y2][]{Demarque.04} and find the pair of masses which minimize
\begin{eqnarray}
\chi^{2} & = & \frac{(K_{P} - K_{P,isoc}(M_{P},P,\sin i))^{2}}{\sigma_{K_{P}}^2} + \frac{(K_{S} - K_{S,isoc}(M_{S},P,\sin i))^{2}}{\sigma_{K_{S}}^{2}} + \frac{(\alpha - \alpha_{isoc}(M_{P},M_{S},P))^{2}}{\sigma_{\alpha}^{2}} + \nonumber \\
 & & \frac{(k - k_{isoc}(M_{P},M_{S}))^{2}}{\sigma_{k}^{2}} + \frac{(T_{ave} - T_{ave,isoc}(M_{P},M_{S}))^{2}}{\sigma_{T}^2}
\end{eqnarray}
where the ``isoc'' subscript denotes values calculated from the isochrones using the trial masses ($M_{P}$ and $M_{S}$) and the measured orbital period ($P$) and $\sin i$, $\alpha = (R_{P} + R_{S})/a$, $k = R_{S}/R_{P}$, and $T_{ave}$ is the average temperature of the two stars. The contours in figure~\ref{fig:V1380agemetchi2} show the $68.3\%$, $95.4\%$ and $99.7\%$ confidence levels. The point shows the values for the cluster using the Y2 isochrones (see Paper I). The binary system is younger and has a lower metallicity than the cluster. 

Figure~\ref{fig:V1380MRTeff} shows the observed masses and radii of the components together with the expected relation for the cluster from the Y2 isochrones and the relation for $[M/H] = -0.25$ and an age of 100 Myr. The stars appear to have radii that are too small given their masses for them to have the metallicity and age of the cluster. 

\placefigure{fig:V1380agemetchi2}

\placefigure{fig:V1380MRTeff}

The individual $r$ magnitudes of the stars are $r_{p} = 15.53$ and $r_{s} = 15.55$. As we stated at the beginning of this subsection, the expected masses for the stars if the system were a member of the cluster would be $M_{p} = M_{s} = 1.18 \pm 0.04 M_{\odot}$, with an uncertainty dominated by the systematic uncertainty in the $M-r$ relation. The observed masses are $\sim 7\sigma$ above the expected masses, this together with the metallicity discrepancy (see figure~\ref{fig:V1380agemetchi2}), and the $\sim 3\sigma$ discrepancy between the systemic velocity and the mean RV of the cluster stars leads us to conclude that the system is most likely not a member of the cluster, and is instead located in the background galactic disk.

Finally, we return to the nature of the non-eclipse variability seen in the light curve. The periods and amplitudes of the non-eclipse variations are consistent with one or both of the stars being a $\gamma$ Doradus pulsator. The spectroscopic temperature is slightly lower than, though consistent with, the instability strip for this type of pulsator.

As seen in figure~\ref{fig:V1380agemetchi2}, the eclipsing binary components appear to be relatively young ($< 700$ Myr at $3 \sigma$, $< 200$ Myr at $1 \sigma$). This is consistent with some indications that the $\gamma$ Doradus phenomenon is restricted to stars with ages $< 250$ Myr \citep{Krisciunas.98}, though there are potentially a few $\gamma$ Doradus variables as old as $1$ Gyr that have been discovered \citep{Pepper.06b}. To our knowledge, V1380 is only the second $\gamma$ Doradus variable found in an eclipsing binary, the other system being VZ CVn \citep{Ibanoglu.07}. These systems are particularly interesting since the geometry allows for the direct determination of the masses and radii of the stars, and with precise photometry it may be possible to use the eclipses to aid in identifying the pulsation modes \citep{Riazi.06}.

\subsection{Detached Eclipsing Binary V1028}

Based on its photometry, the detached eclipsing binary V1028 is a candidate cluster member. V1028 matches to 2mass05523843+3223296, it has an orbital period of $5.496 \pm 0.18$ days, a primary eclipse epoch of $HJD_{0} = 2453725.6977$, and photometry $B - V = 1.591$, $V = 19.428$, $g-r = 1.61$, $r-i = 0.93$, $r = 18.73$, $J - K_{S} = 0.695$, $H - K_{S} = 0.137$, and $K_{S} = 15.289$. If the system is a member of the cluster, its primary would have a mass of $0.6 < M < 0.7 M_{\odot}$ based on the $r$ magnitude of the system. We will now demonstrate that the components of this system likely have masses of $\sim 0.6 M_{\odot}$ and may very well be members of the cluster.

This system shows out of eclipse variations with a peak-to-peak amplitude of 0.028 mag. Since the maximum of this variation does not occur at either of the quadrature points or near the eclipses, and since the system is well detached, the variation is most likely due to spots on one of the stars rather than due to proximity effects. This confirms that at least one of the stars has a rotation period synchronized to the orbital period. The primary and secondary eclipses occur 0.5 apart in phase, so it appears that the system is also circularized. 

Before fitting an eclipse model to the light curve, we first remove the out of eclipse variation using a sinusoid. Assuming the cluster reddening and assuming all the light comes from the primary component, we estimate the temperature of the primary is $\sim 5000$ K from the $J-K_{S}$ and $H-K_{S}$ colors, for which the $r$ quadratic limb darkening coefficients are $a = 0.538$, $b = 0.199$ \citep{Claret.04}. Using JKTEBOP, we fit the light curve assuming zero eccentricity and find the parameters listed in table~\ref{tab:V1028JKTEBOP}. As before we estimate the parameter errors using 100 Monte Carlo simulations taking the error at each observation to be $8.0$ mmag. In this case we refine the period and primary eclipse epoch using JKTEBOP. We find that it is necessary to include a significant amount of third light ($L_{3} = 0.266 \pm 0.006$) in the fit, though it is unclear if this is an artifact of not correctly modelling the spot through the eclipses. The resulting fractional radii for the primary and secondary stars are $R_{P}/a = 0.0439 \pm 0.0003$ and $R_{S}/a = 0.0416 \pm 0.0004$. Figure~\ref{fig:V1028JKTEBOPLC} shows the best fit model to the light curve.

\placetable{tab:V1028JKTEBOP}

\placefigure{fig:V1028JKTEBOPLC}

To estimate the masses of the component stars we use Kepler's third law and the mass-radius relation for low-mass stars. Assuming a power-law mass-radius relation of the form $M = M_{0}(R/R_{\odot})^{b}$, the mass of each component is given by:
\begin{equation}
M_{P,S} = M_{0}\left(\frac{R_{P,S}}{a}\right)^{b}\left(74.471\frac{M_{0}}{M_{\odot}}\left(\frac{P}{1\,{\rm day}}\right)^{2}\left(\left(\frac{R_{P}}{a}\right)^b + \left(\frac{R_{S}}{a}\right)^b\right)\right)^{b/(3-b)}.
\label{eqn:MfromMR}
\end{equation}
For the mass-radius relation we use the double-lined detached eclipsing binary data given in \citet{Ribas.06} to find $(M/M_{\odot}) = 1.04 (R/R_{\odot})^{1.08}$. 

Using eq.~\ref{eqn:MfromMR} we find component masses of $M_{P} \sim 0.61 M_{\odot}$, and $M_{S} \sim 0.57 M_{\odot}$. The mass of the primary is thus close to what we would expect based on the $r$ magnitude of the system if the stars were members of the cluster. The individual magnitudes of the two component stars are $r_{P} \sim 19.5$ and $r_{S} \sim 20.1$, with a third component having $r_{3} \sim 20.4$. The magnitude difference of $0.59 \pm 0.03$ between the primary and secondary stars is quite a bit larger than the value of $0.3$ that one would expect from the estimated masses (see the mass-$r$ magnitude relation in Paper I). However, given that we do not properly model the spots on the stars through the eclipses, and that we do not have radial velocity data to confirm the masses, the discrepancy may simply be an artifact of our modelling. 

We have shown that the stars likely have masses of $\sim 0.6 M_{\odot}$ and may very well be members of the cluster. The system is therefore interesting for follow-up as only $\sim 6$ double-lined detached eclipsing binary systems with main sequence stars smaller than the components in this system are known, and the models for these stars are known to be in error by $5-20\%$ \citep{Ribas.06}. If the system proves to be a member of the cluster, then the additional constraints on its age, metallicity and apparent distance modulus would make the system particularly useful in testing low mass stellar models. The downside though to any follow-up is that with $V \sim 19.4$, a significant amount of large telescope time would be needed to obtain its radial velocity curve. Note that the expected semi-amplitude of the radial velocity curve is $\sim 60~{\rm km/s}$.

\section{Conclusion}

In this paper we have identified 1430 new variable stars in the field of M37. This includes 20 new eclipsing binaries and 31 new short-period pulsating stars. Approximately 500 of the variables are F-M main sequence cluster members; for these stars the variability is due to rotation coupled with significant surface brightness inhomogeneities. This data set is significantly larger than any other set of stellar rotation periods for a cluster that is $\ga 500$ Myr old and will thus provide a unique window on the rotation of lower main sequence stars. We argue that many of the other variable stars are young, rapidly rotating main sequence stars in the galactic disk by showing that the amplitude of these variables is anti-correlated with the Rossby number.

We have used 10 W UMa eclipsing binaries and 4 fundamental mode $\delta$-Scutis to investigate the relation between distance and extinction along the line of sight, and have shown that the extinction apparently cuts off at nearly twice the scale height of the disk. 

Finally, we have analyzed a few particularly interesting variable stars including a previously identified multi-mode pulsator which we argue to be a hybrid $\gamma$-Doradus/$\delta$-Scuti, two possible quiescent cataclysmic variables, a DEB with at least one $\gamma$-Doradus pulsating component, and a low mass DEB that is a possible cluster member. We have combined radial velocity and light curves to obtain a physical solution for the DEB+$\gamma$-Doradus system.

\acknowledgements
This research has made use of the SIMBAD database, operated at CDS, Strasbourg, France. We are grateful to C.~Alcock for providing partial support for this project through his NSF grant (AST-0501681). Funding for M.~Holman came from NASA Origins grant NNG06GH69G. We would like to thank G.~Torres for help with running the eclipsing binary analysis software and for helpful comments, G.~F\"{u}r\'{e}sz and A.~Szentgyorgyi for help in preparing the Hectochelle observations, S.~Meibom for help with the stellar classification, A.~Schwarzenberg-Czerny for discussion of the period finding techniques, J.~Kaluzny for pointing out the CV interpretation of V1148 to us and the staff of the MMT, without whom this work would not have been possible. We would also like to thank the MMT TAC for awarding us a significant amount of telescope time for this project, and the referee, D.~Weldrake, for thoughtful comments that improved the quality of this paper.

\appendix

\section{The VARTOOLS Program}

In conducting the research described in this paper we have developed software to perform a number of common analysis routines on a set of light curves. The program is called {\scshape Vartools} and we release it to the community\footnote{http://www.cfa.harvard.edu/\~{}jhartman/vartools}. The program reads in a light curve, or a set of light curves, in ascii format and performs a series of commands on them; it is designed to run the light curves through a pipeline of routines.

At the time of writing the commands that can be run include the period finding routines discussed in \S 3, routines to fit a Fourier series to light curves, routines to linearly decorrelate light curves against generic signals, and implementations of the SYSREM and TFA de-trending algorithms mentioned in \S 3.1. This is not an exhaustive list; a full description of the program can be found on the website.

\newpage

\clearpage

\clearpage

\clearpage

\clearpage

\clearpage

\clearpage

\clearpage

\clearpage

\clearpage

\clearpage

\clearpage

\clearpage

\clearpage

\begin{figure}[p]
\plotone{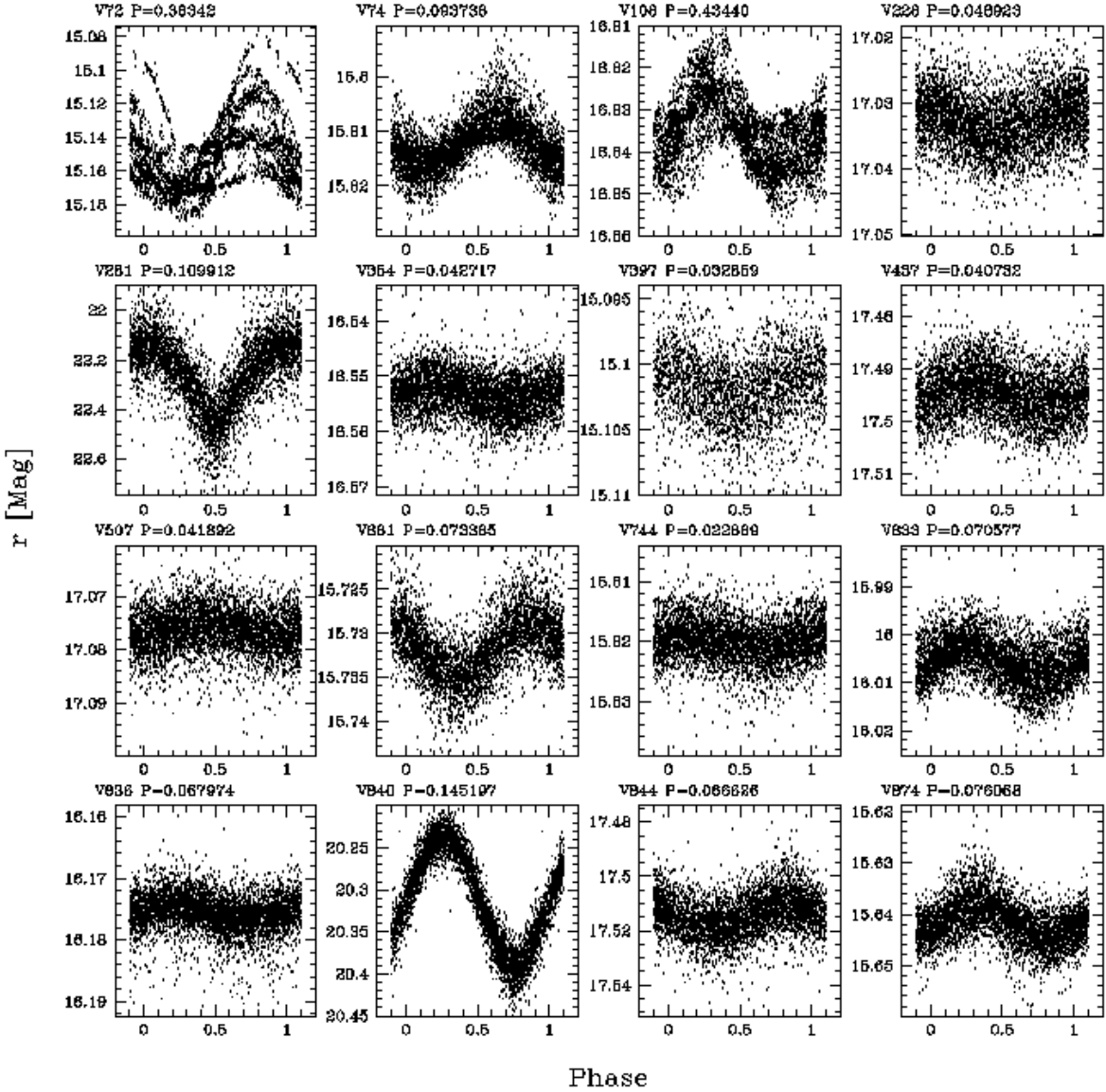}
\caption{Phased light curves for 31 visually selected short-period pulsating variables. Note that several of these stars, including V72, V108, V1244 and V1368, are multi-periodic.}
\label{fig:PulsLC}
\end{figure}

\clearpage

\addtocounter{figure}{-1}
\begin{figure}[p]
\plotone{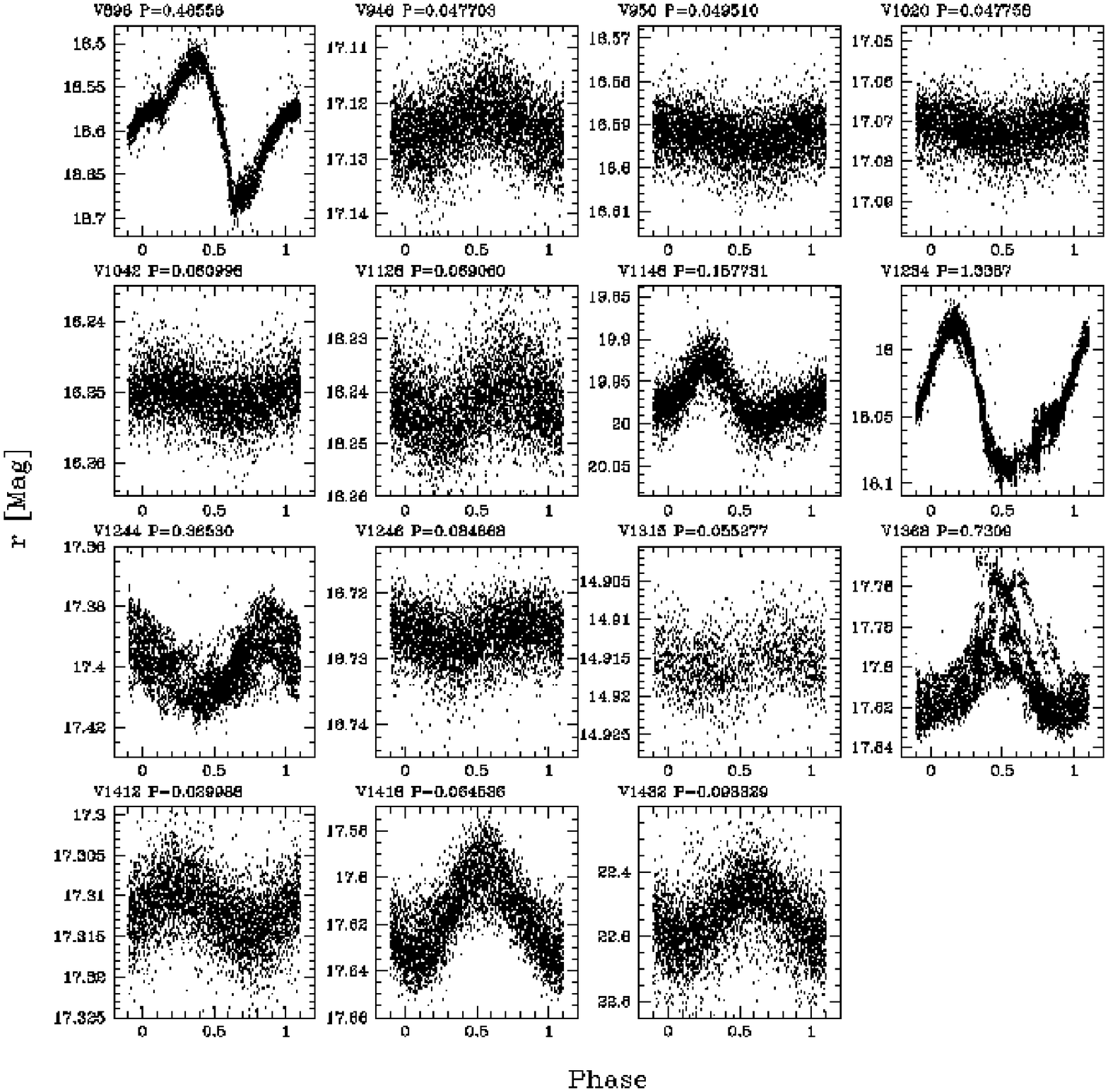}
\caption{{\it Continued}.}
\end{figure}

\clearpage

\begin{figure}[p]
\plotone{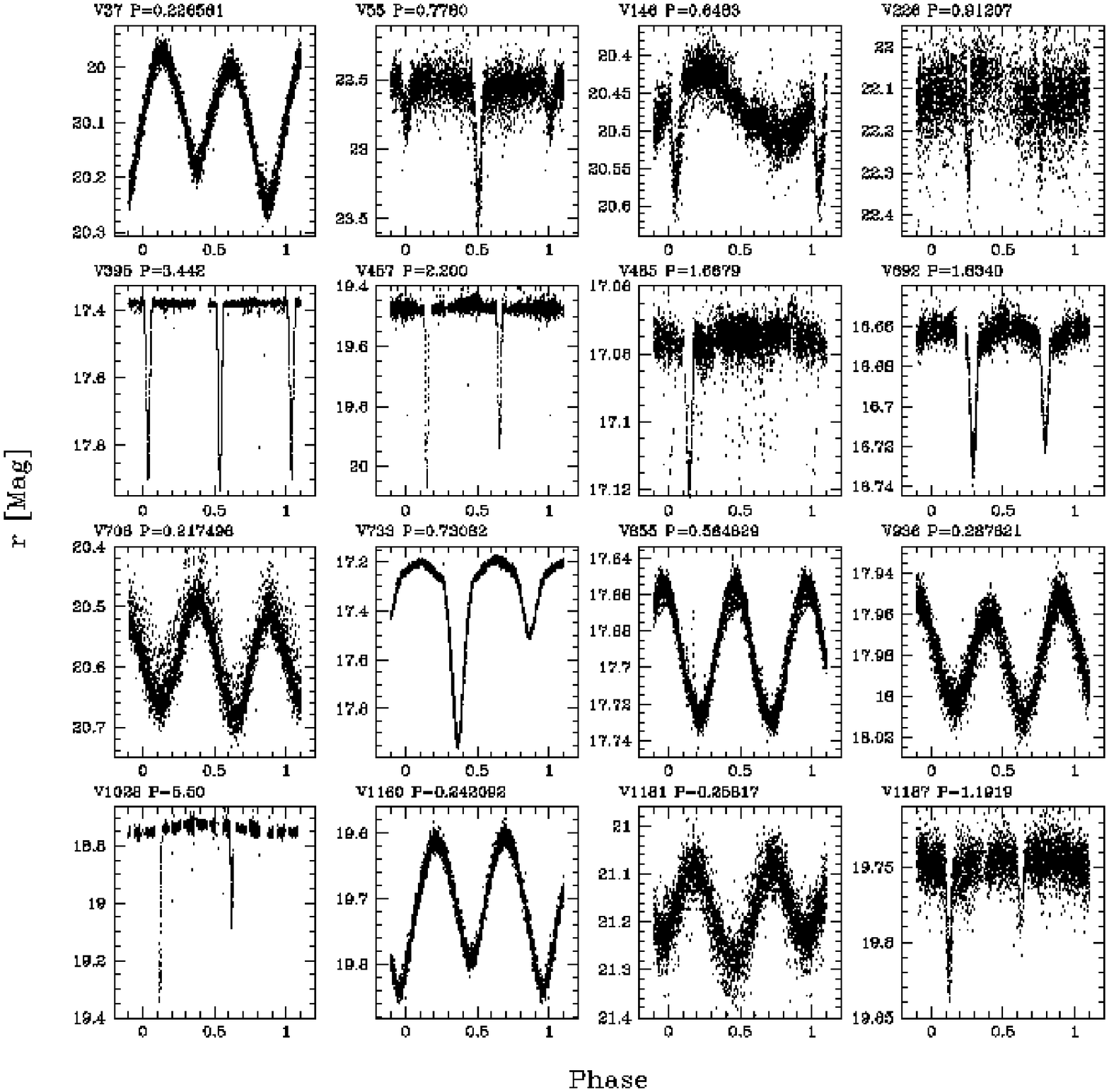}
\caption{Phased light curves for 20 visually selected eclipsing binaries.}
\label{fig:EBLC}
\end{figure}

\clearpage

\addtocounter{figure}{-1}
\begin{figure}[p]
\plotone{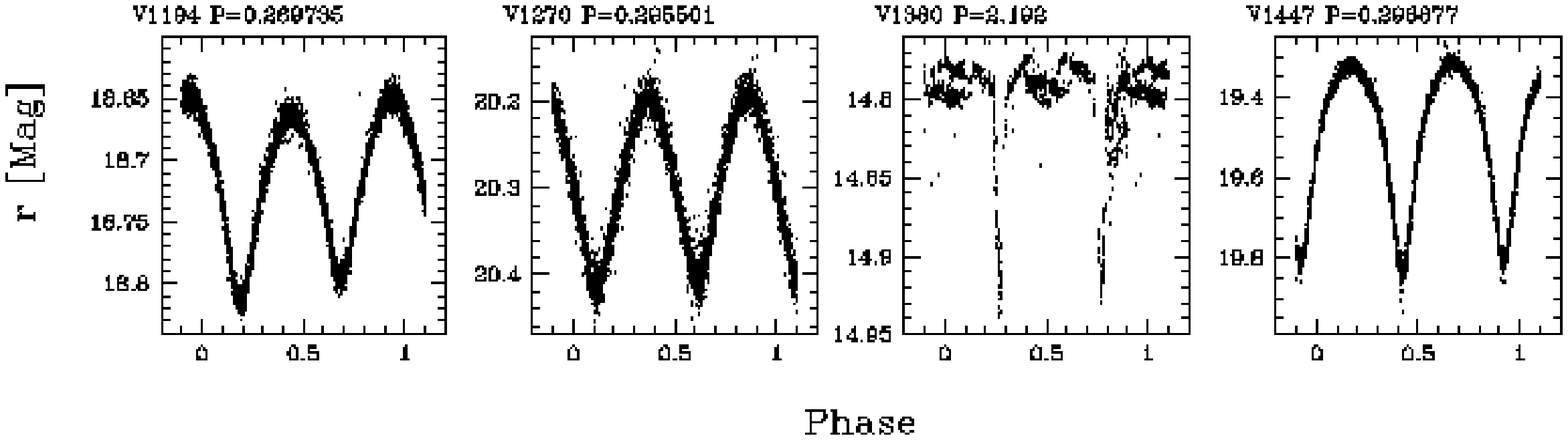}
\caption{{\it Continued}.}
\end{figure}

\clearpage

\begin{figure}[p]
\plotone{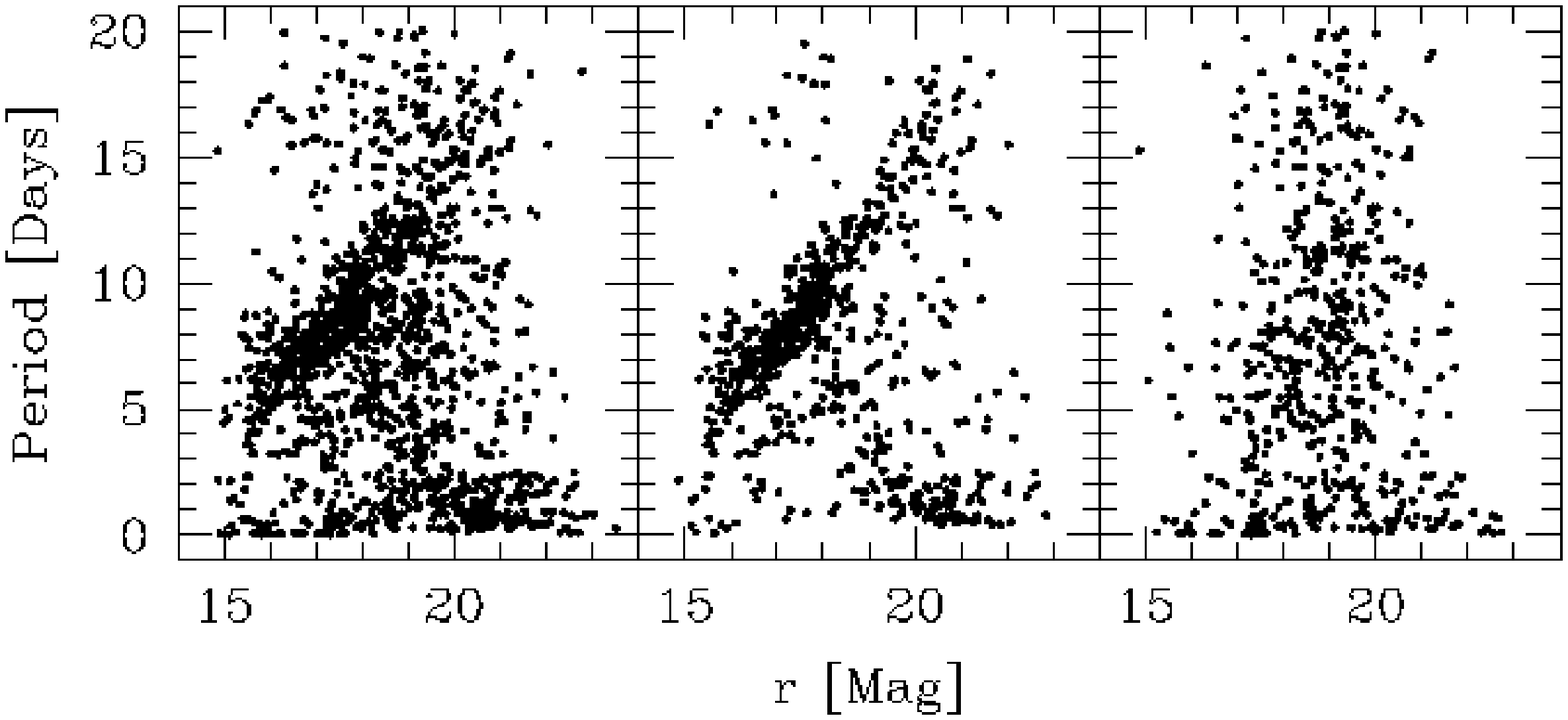}
\caption{Period-$r$ relation for (left) all variables stars, (center) photometrically selected candidate cluster members and (right) non-cluster members. Note that there is a strong correlation between period and magnitude for cluster members while for non-members there is no clear correlation.}
\label{fig:PerMag}
\end{figure}

\clearpage

\begin{figure}[p]
\plotone{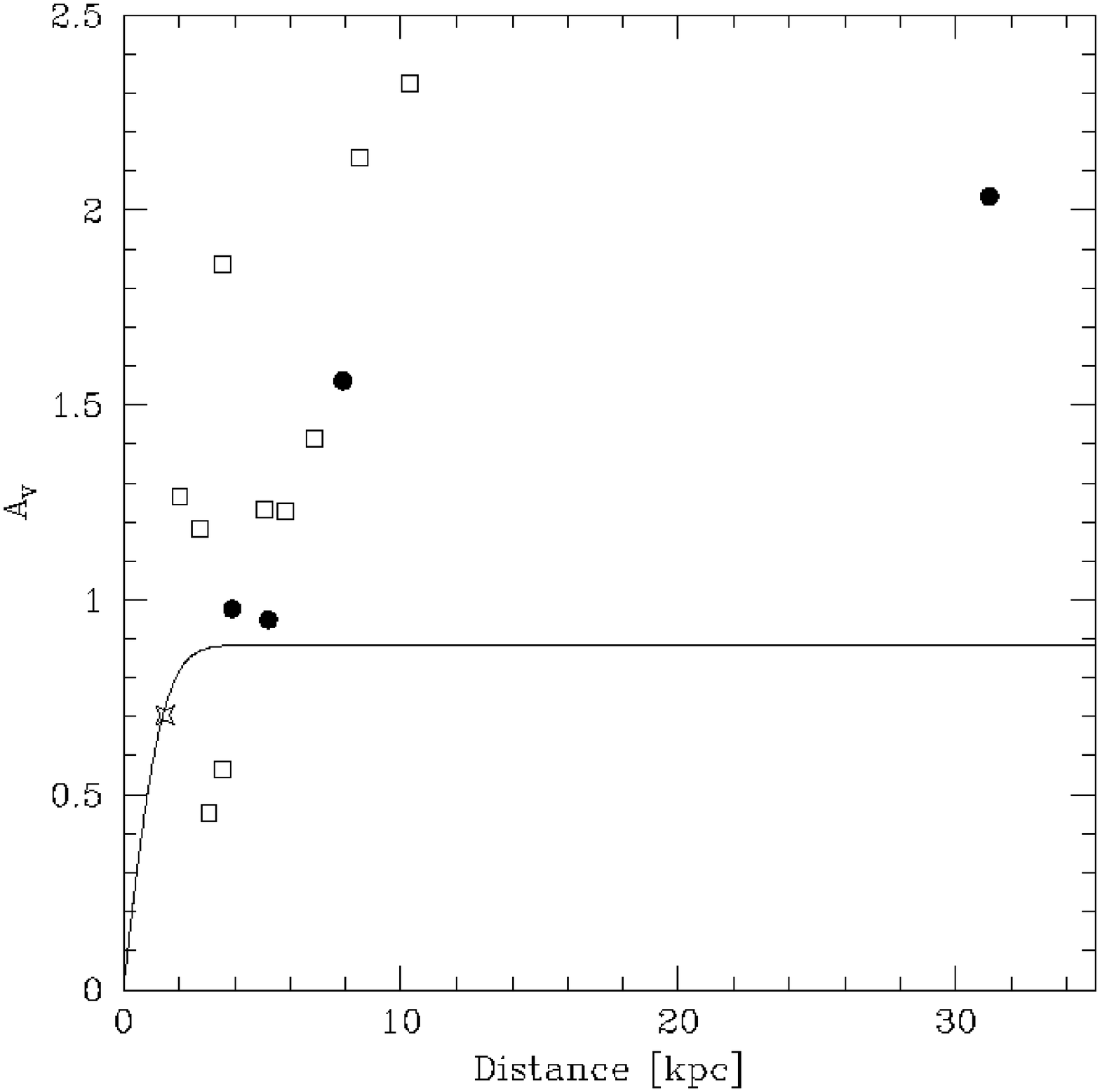}
\caption{$A_{V}$ vs distance for 10 W UMa eclipsing binary systems (open squares), 4 $\delta$-Scuti stars (filled points) and the open cluster (star). The line shows the Besan\c{c}on model \citep{Robin.03} for the galactic latitude/longitude of the field, assuming an interstellar extinction of $0.7$ mag/kpc.}
\label{fig:WUMAdist}
\end{figure}

\clearpage

\begin{figure}[p]
\plotone{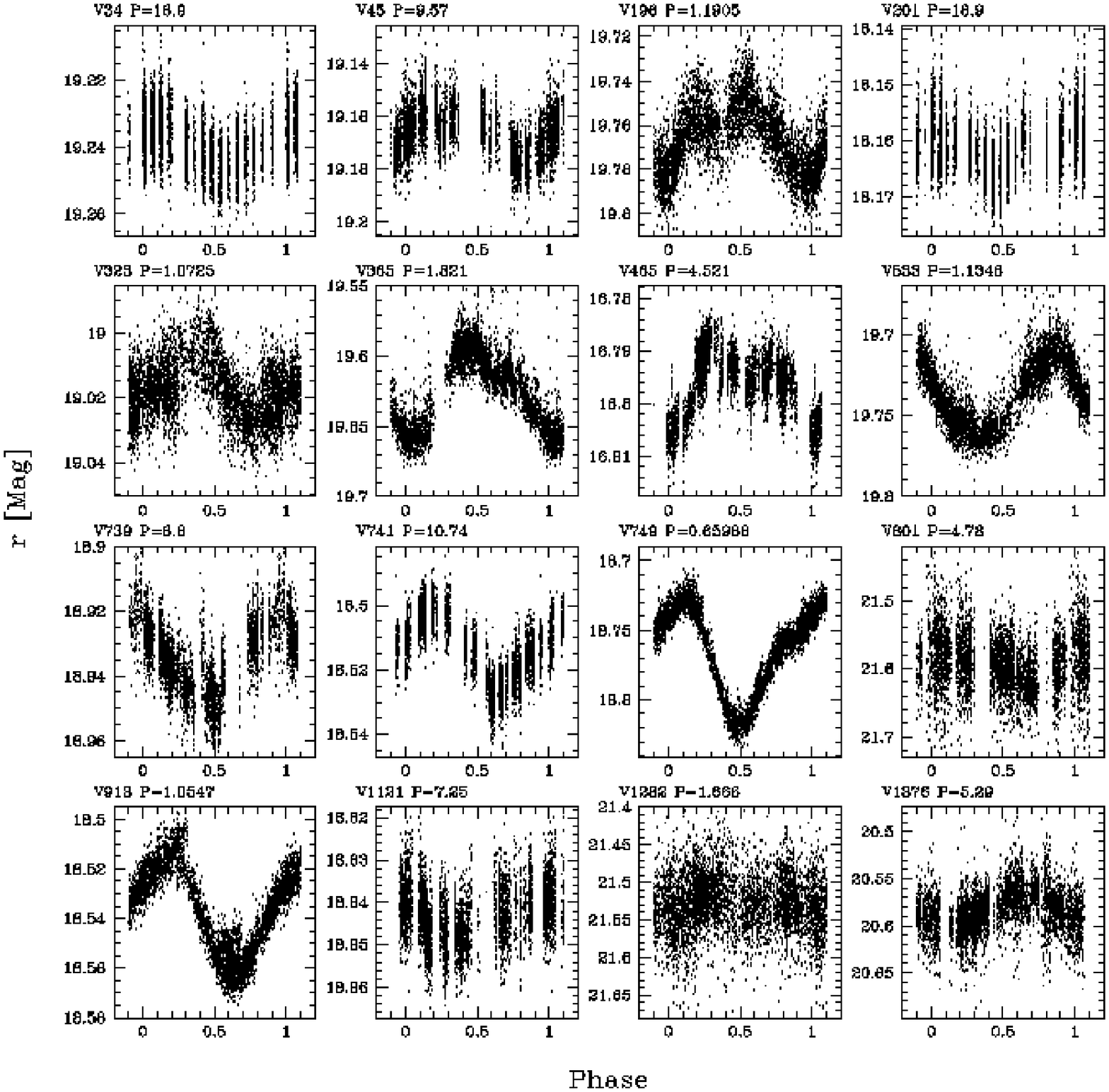}
\caption{Phased light curves for a random sample of 16 variable stars that are not members of the cluster, eclipsing binaries, or short-period pulsators. Many of these variables are likely young, heavily spotted, rapidly rotating stars.}
\label{fig:ExampleRotLC}
\end{figure}

\clearpage

\begin{figure}[p]
\plotone{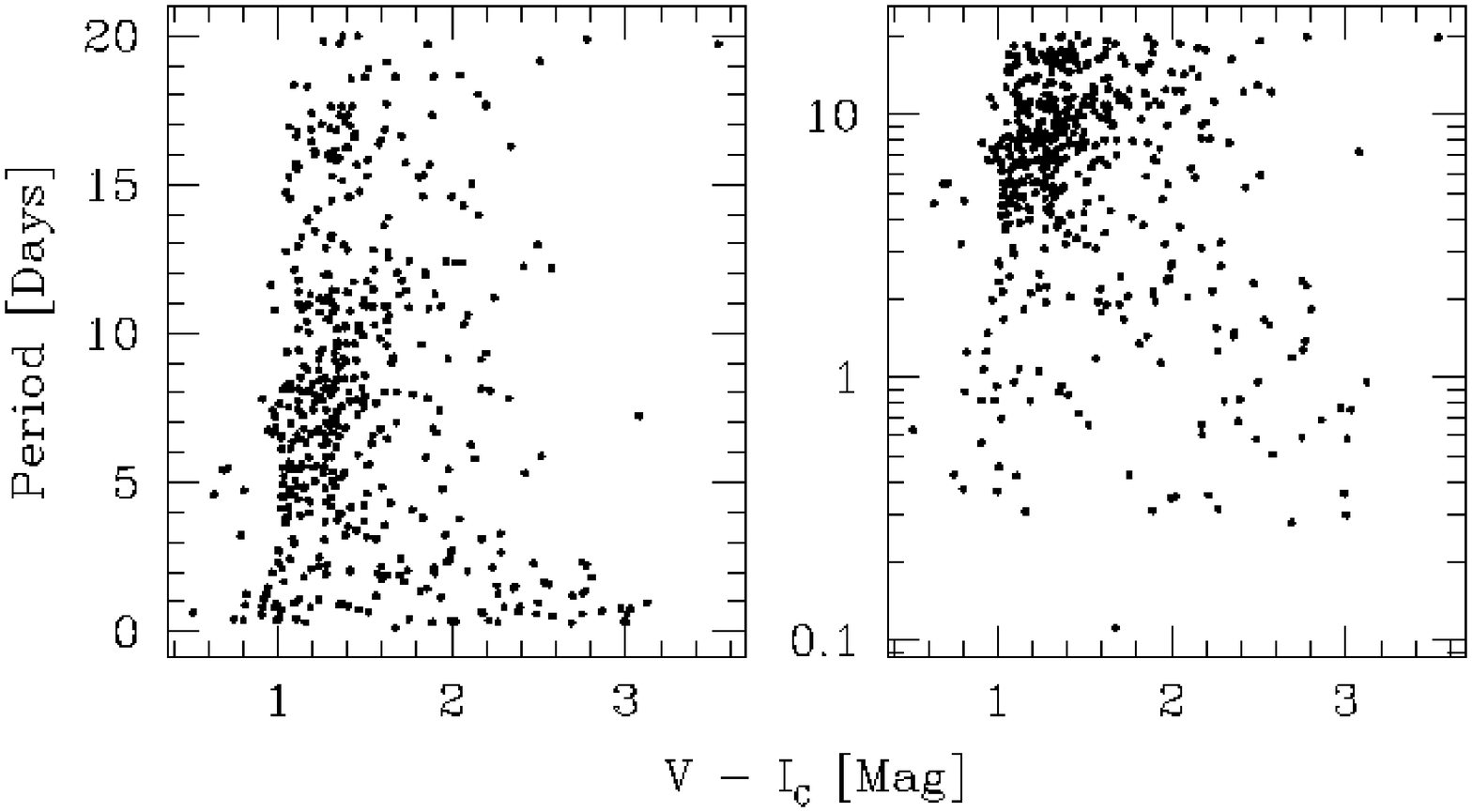}
\caption{Period vs. $V-I_{C}$ for variable stars that are not members of the cluster, not eclipsing binaries, and not $\delta$-Scuti or RR Lyrae type pulsators. The left panel shows the period on a linear scale, the right panel on a logarithmic scale.}
\label{fig:PerVI}
\end{figure}

\clearpage

\begin{figure}[p]
\plotone{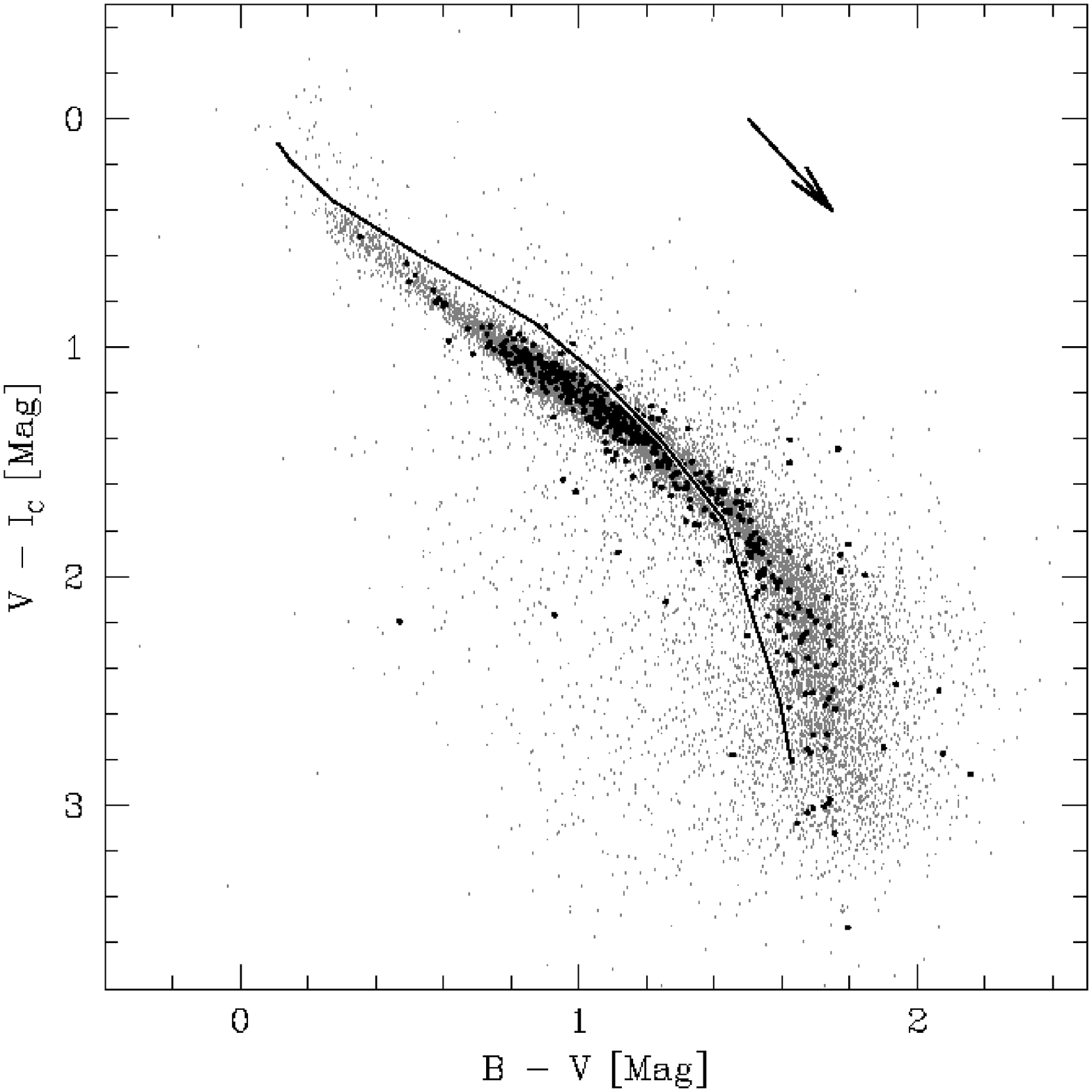}
\caption{$B-V$ vs $V-I_{C}$ color-color plot showing all stars (gray points), field variables that are not eclipsing binaries nor obvious pulsators (dark points), and the fiducial \emph{de-reddened} relation for the cluster (solid line). The arrow shows the direction of the reddening vector. Using this diagram we estimate the reddening to all the stars with $B-V < 1.2$, $V-I_{C} < 1.8$.}
\label{fig:BVVI_notincluster}
\end{figure}

\clearpage

\begin{figure}[p]
\plotone{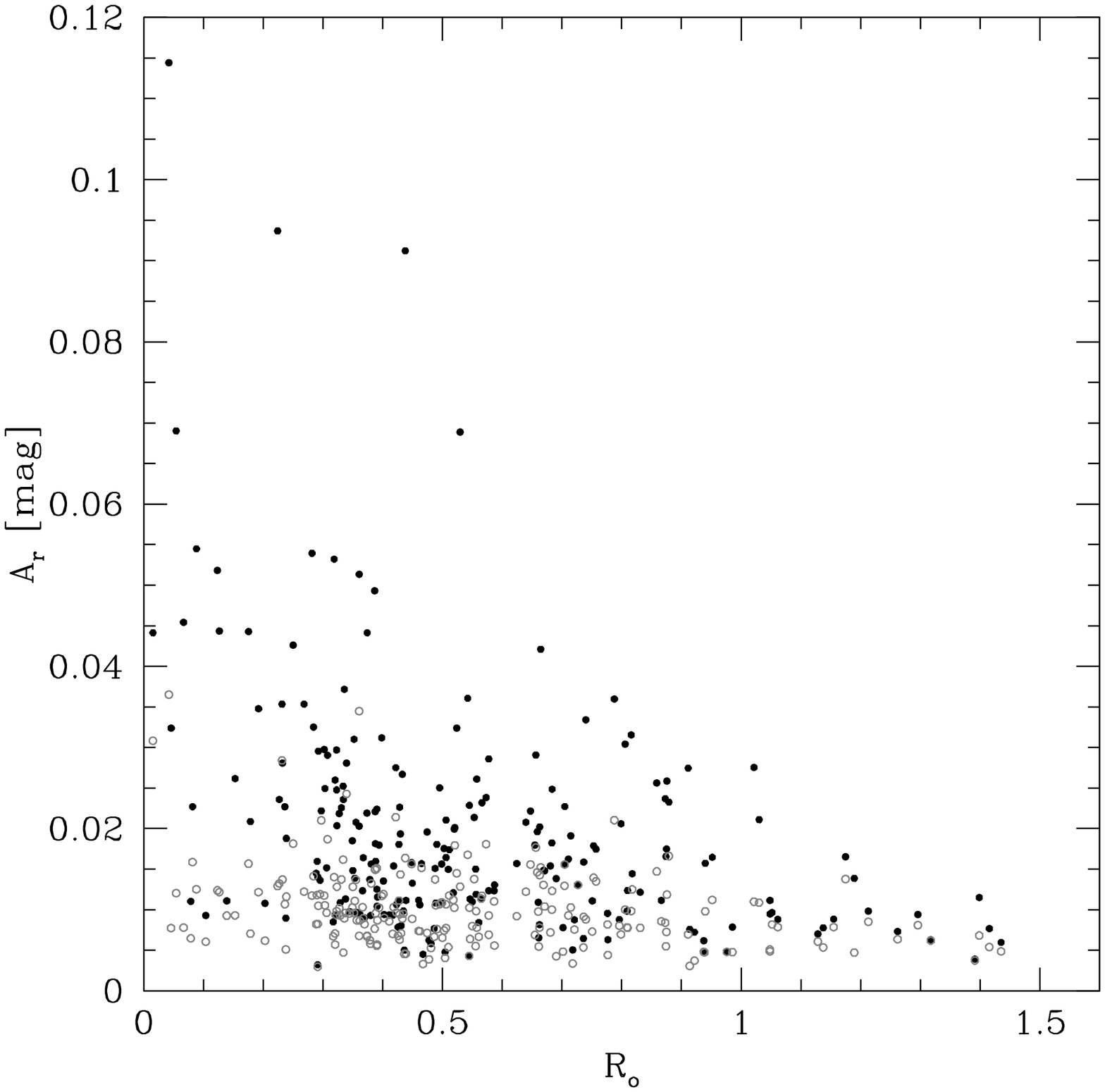}
\caption{The $r$-band amplitude ($A_{r}$) versus the Rossby number ($R_{o}$) for field variables not classified as eclipsing binaries or pulsating stars (see text for a description of the selection). The filled points show the measured values, the open circles show the minimum amplitude that each variable could have had and still have been detected. The anti-correlation between $R_{o}$ and $A_{r}$ is consistent with the hypothesis that these variables are heavily spotted, rotating stars.}
\label{fig:AmplitudeRossby_notincluster}
\end{figure}

\clearpage

\begin{figure}[p]
\plotone{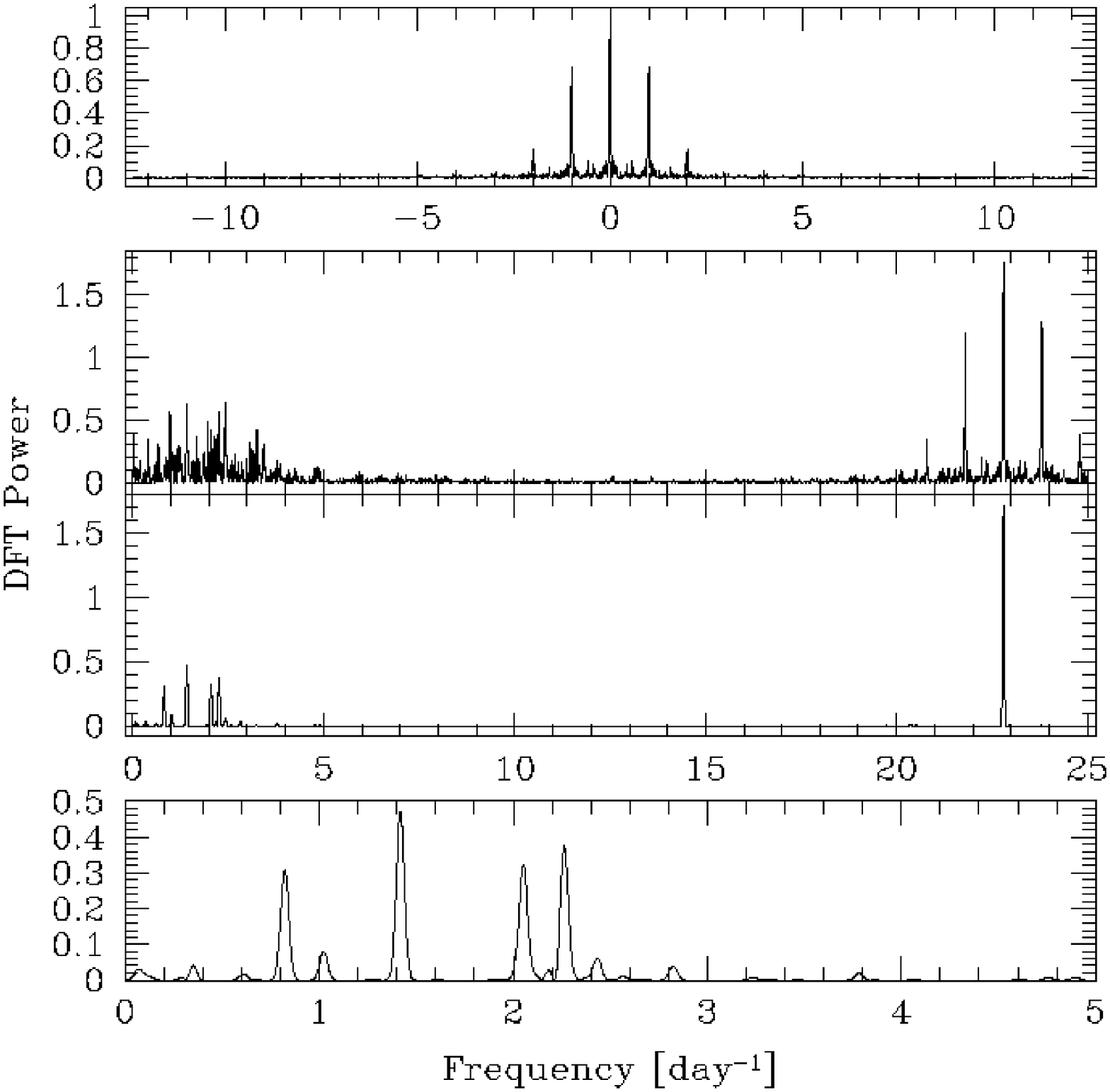}
\caption{The DFT power spectrum of the light curve for V17. The top panel shows the window power spectrum, the second panel shows the raw DFT power spectrum, the third panel shows the power spectrum after implementing the CLEAN deconvolution algorithm, the bottom panel shows the CLEAN power spectrum zoomed in on the low frequency region. The bottom three plots have been scaled by a factor of $10^{4}$. The CLEAN power spectrum reveals five frequencies at $22.796$ days$^{-1}$, $1.418$ day$^{-1}$, $2.263$ day$^{-1}$, $2.054$ day$^{-1}$, and $0.823$ day$^{-1}$.}
\label{fig:V17DFTCLEAN}
\end{figure}

\clearpage

\begin{figure}[p]
\plotone{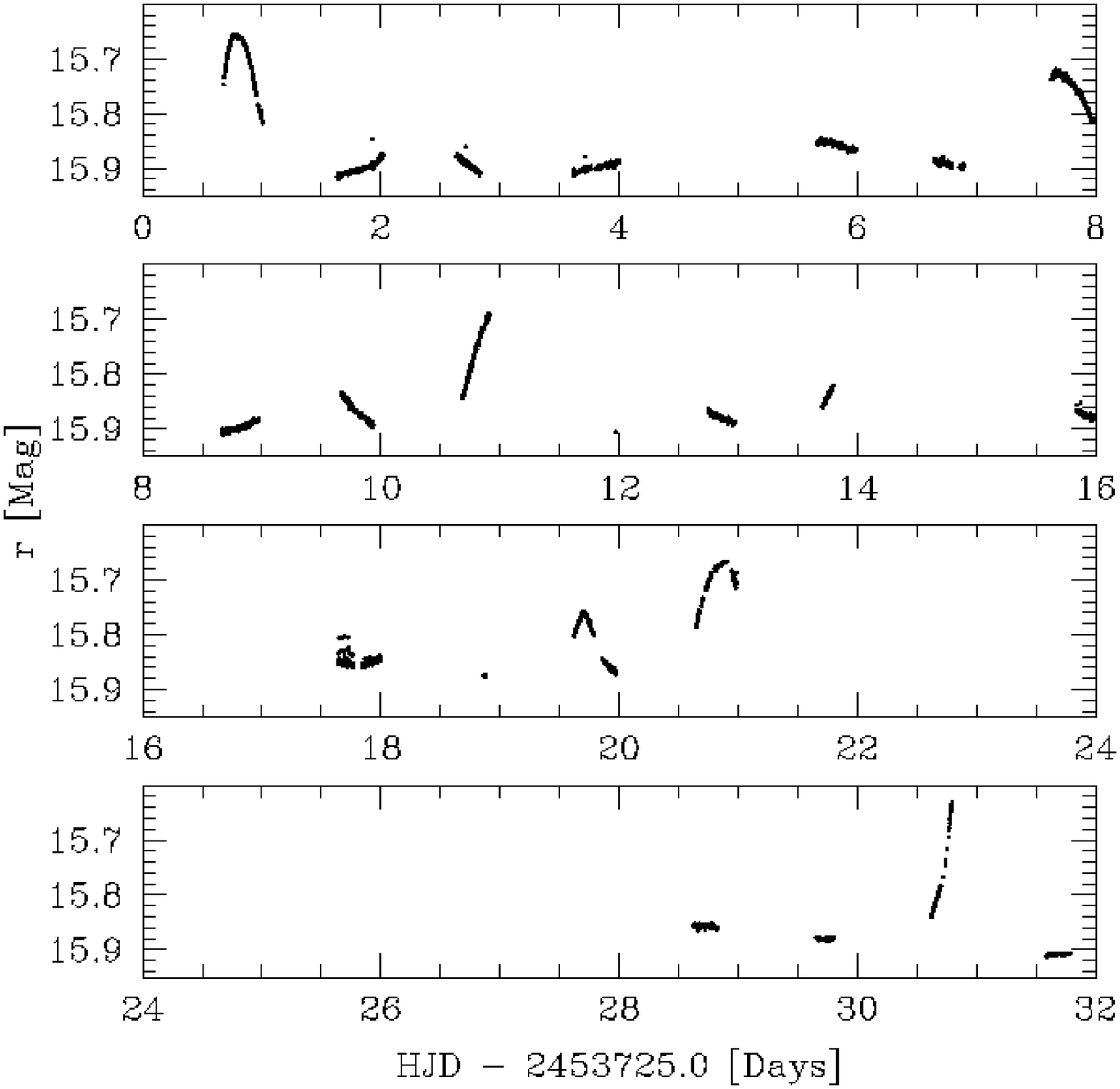}
\caption{Light curve of the variable V898. The light curve is non-periodic, showing continuous variability with repeated outbursts having amplitudes of $0.1 - 0.2$ mag, durations of 0.5 days and a time between outbursts of 1-2 days.}
\label{fig:V898LC}
\end{figure}

\clearpage

\begin{figure}[p]
\plotone{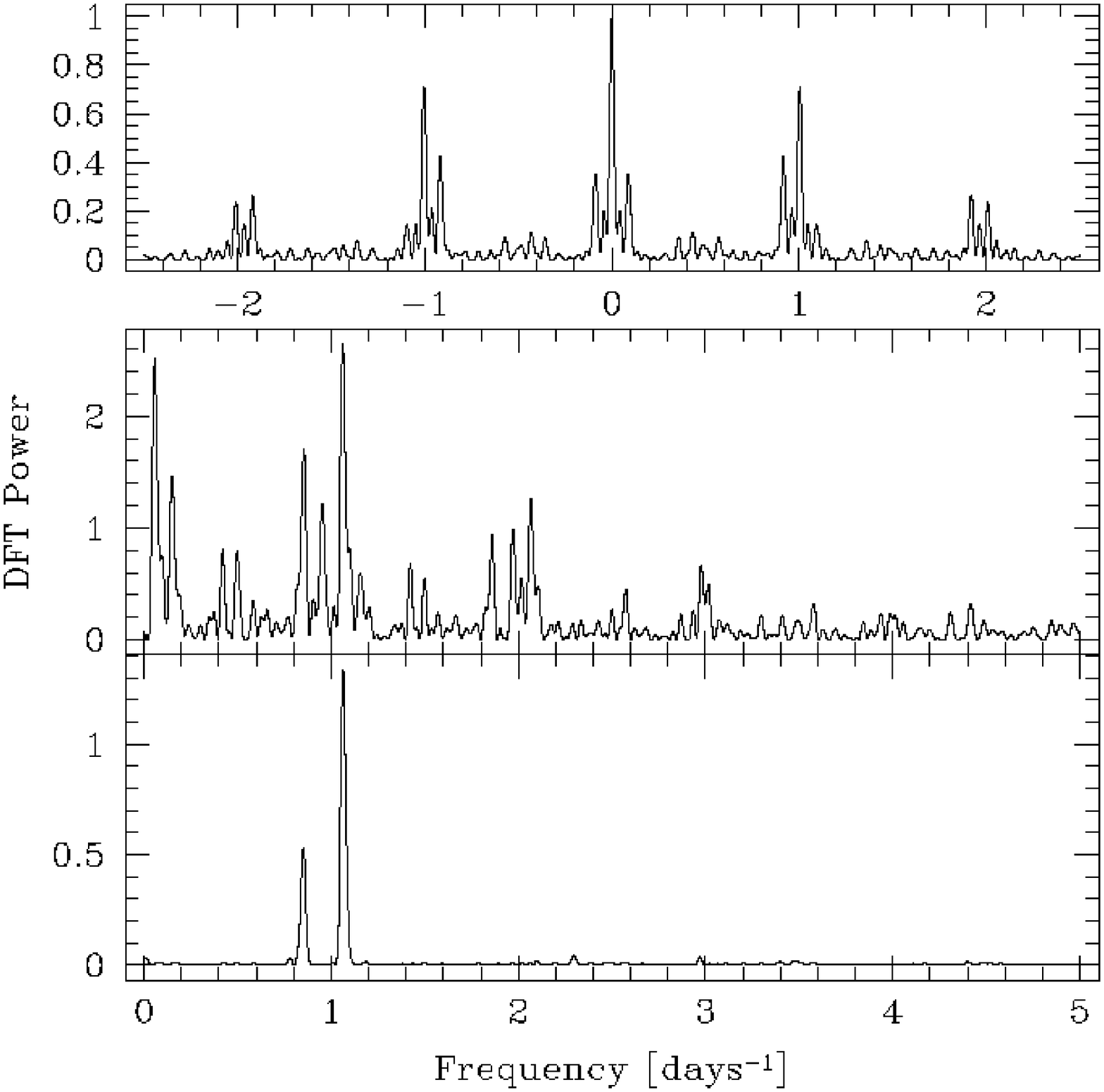}
\caption{The DFT power spectrum of the light curve for the detached eclipsing binary system V1380. The top panel shows the window power spectrum, the middle panel shows the raw DFT power spectrum, the bottom panel shows the power spectrum after 10 iterations of the CLEAN deconvolution algorithm. The bottom two plots have been scaled by a factor of $10^{5}$. The CLEAN power spectrum reveals two strong signals, one at a frequency of $1.063 \pm 0.013$ days$^{-1}$, the other at a frequency of $0.852 \pm 0.013$ days$^{-1}$. These correspond to periods of $0.941 \pm 0.011$ days, and $1.174 \pm 0.013$ days, the same periods found by L-S.}
\label{fig:V1380CLEAN}
\end{figure}

\clearpage

\begin{figure}[p]
\plotone{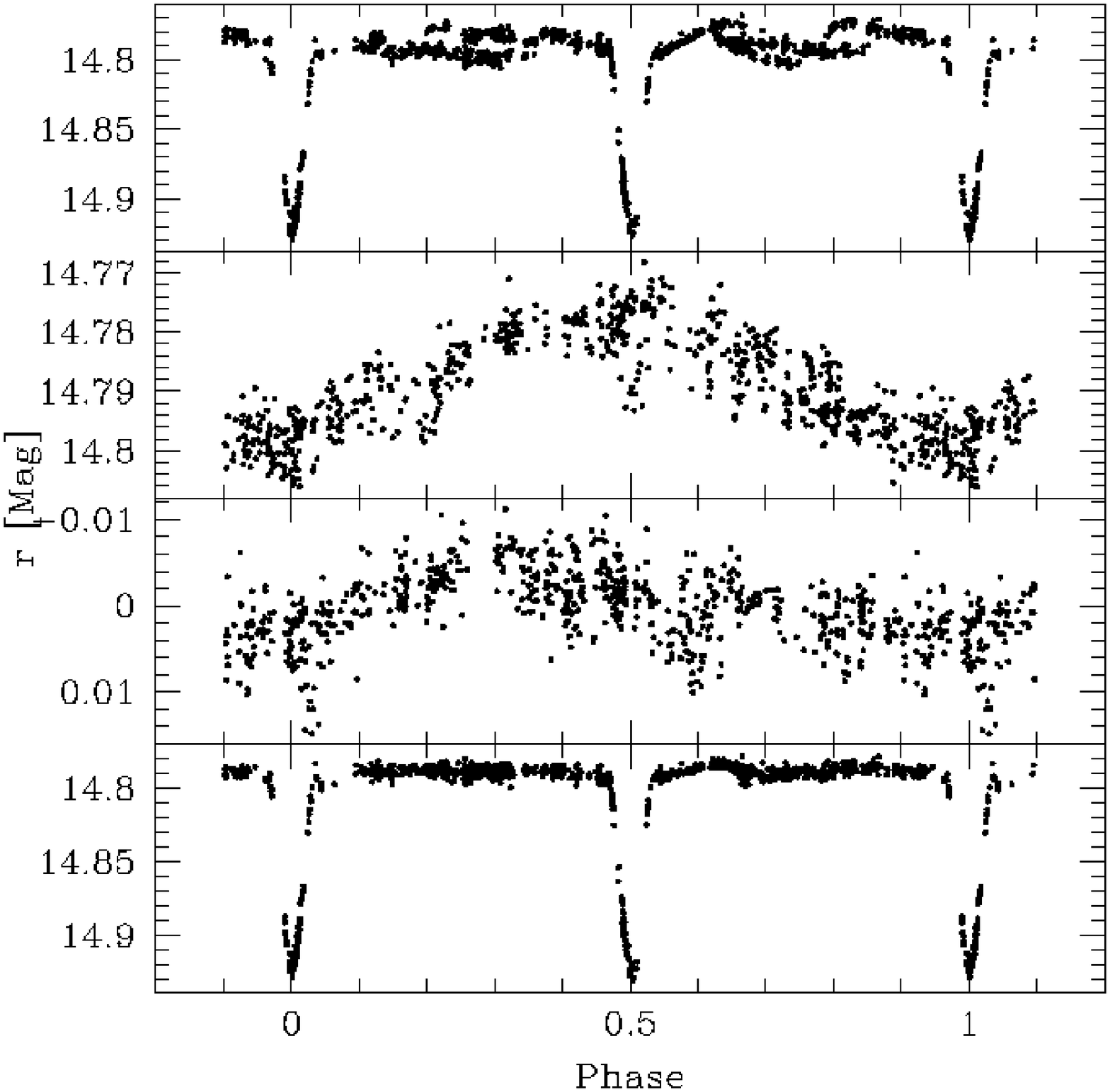}
\caption{Light curve of the detached eclipsing binary system V1380. Top: phased at the orbital period of 2.1916 days. Second from top: out of eclipse points phased at 0.941 days. Second from bottom: out of eclipse points, after subtracting a sinusoid fit to the second from top plot, phased at 1.174 days. Bottom: phased at the orbital period after removing the signals in the middle two panels.}
\label{fig:V1380LC}
\end{figure}

\clearpage

\begin{figure}[p]
\plotone{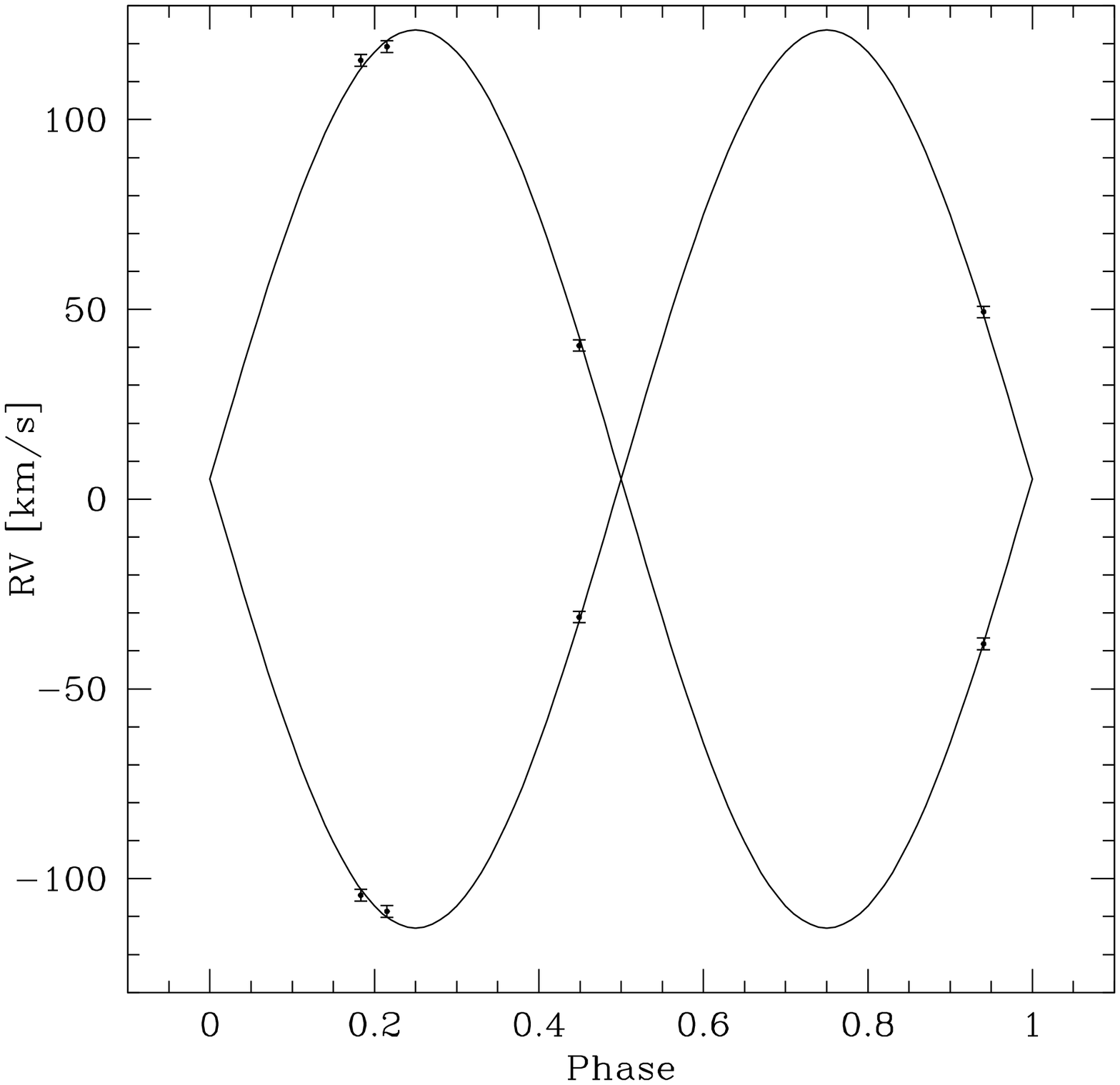}
\caption{Radial velocity curves for the primary and secondary components of the double-lined eclipsing binary system V1380. The solid line shows a zero-eccentricity fit to the observations. We assume an error of 2.2 km/s for each observation. The stars have equal mass to within $1\%$.}
\label{fig:V1380RV}
\end{figure}

\clearpage

\begin{figure}[p]
\plotone{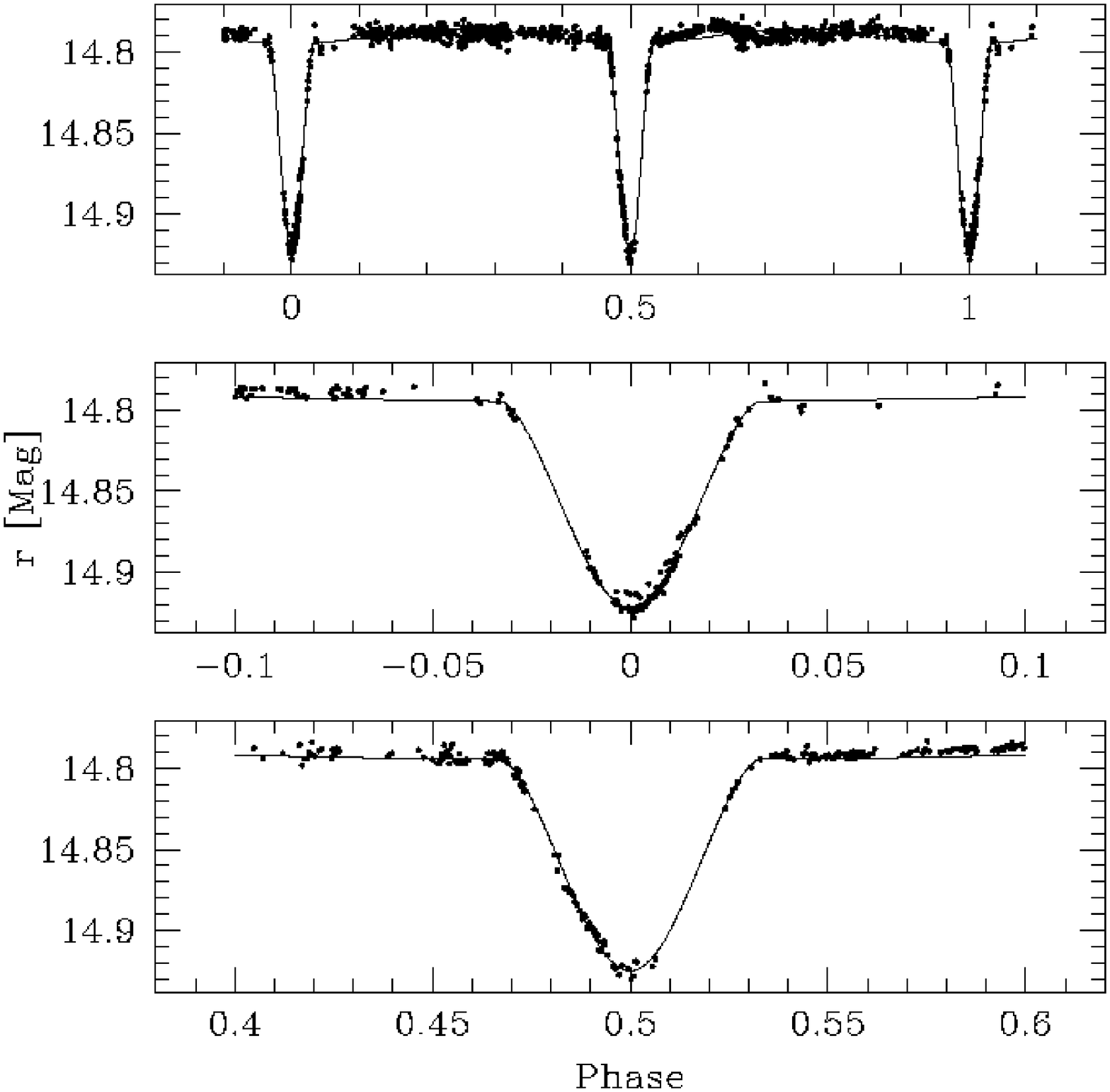}
\caption{Best fit detached eclipsing binary model to the light curve of V1380. The top panel shows the full phased light curve, the bottom panels are zoomed in on the eclipses.}
\label{fig:V1380JKTEBOPLC}
\end{figure}

\clearpage

\begin{figure}[p]
\plotone{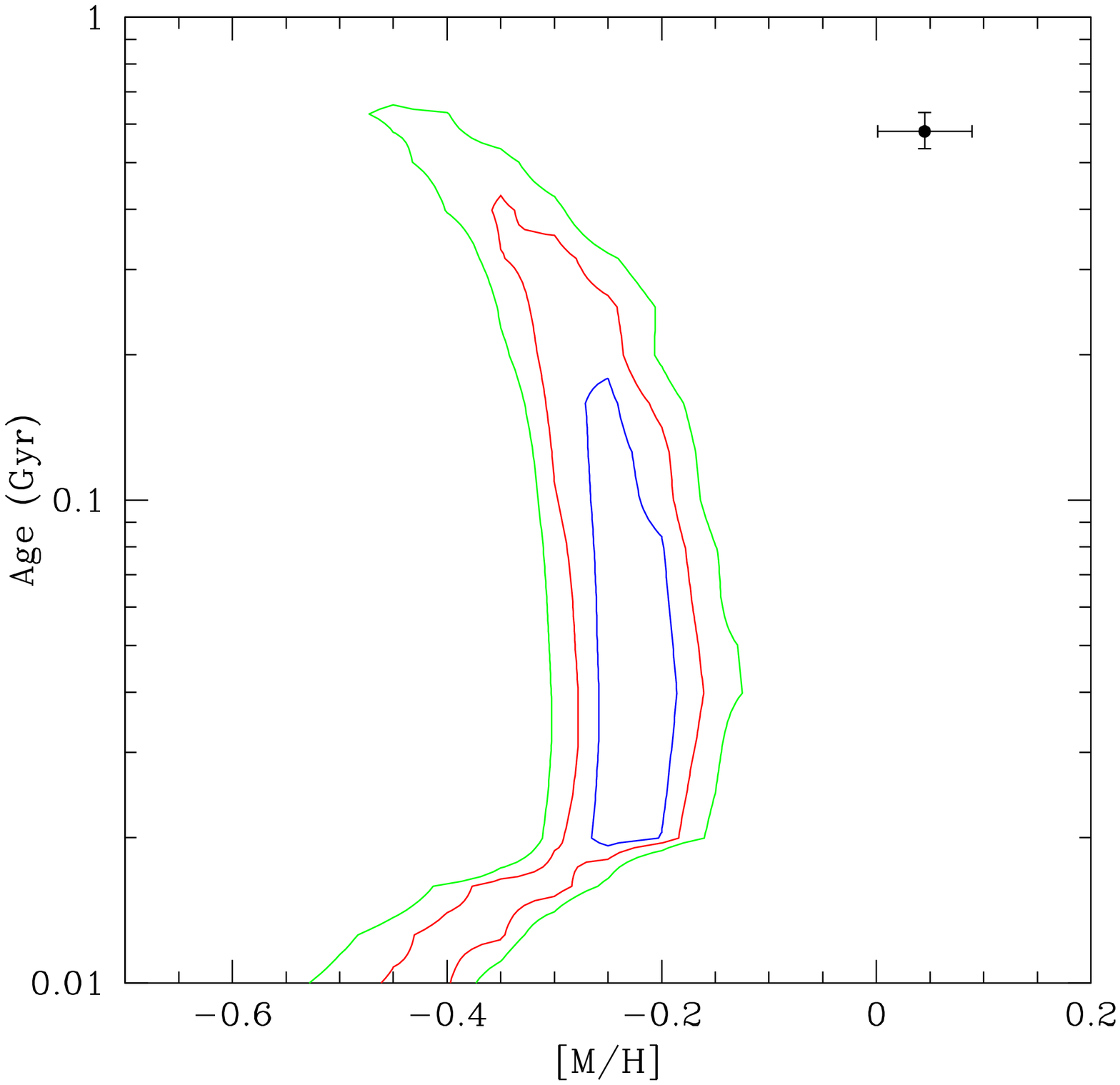}
\caption{Age-metallicity contours for the eclipsing binary system V1380 from comparing the Y2 isochrones with the observed masses/radii/temperatures of the components. The contours show the $68.3\%$, $95.4\%$ and $99.7\%$ confidence levels. The point shows the values for the cluster. The binary system appears to have a lower metallicity and younger age than the cluster.}
\label{fig:V1380agemetchi2}
\end{figure}

\clearpage

\begin{figure}[p]
\plotone{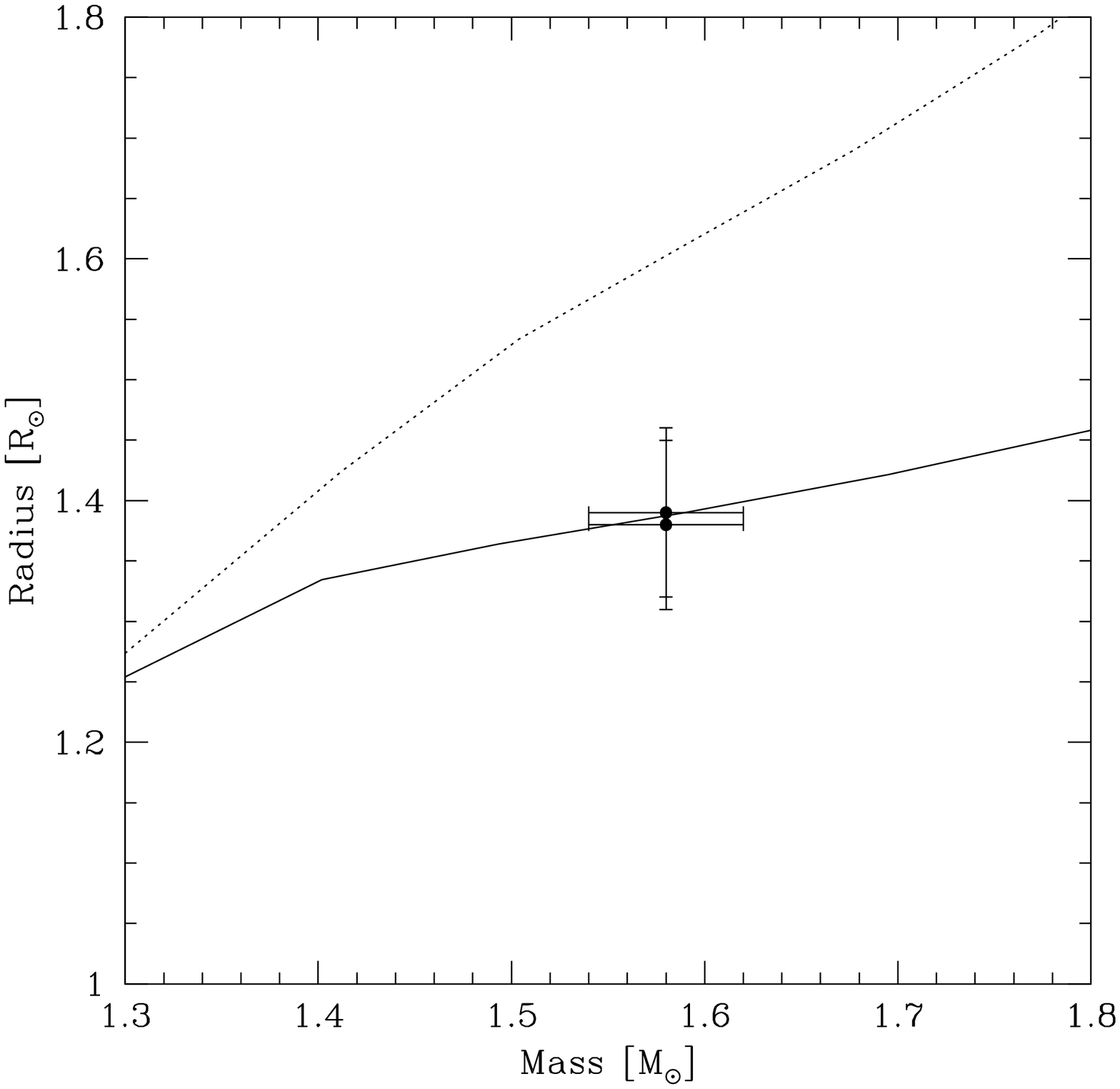}
\caption{The masses and radii of the two components of V1380 together with the expected relations from the Y2 isochrones for the cluster metallicity of $[M/H] = 0.045$ and age of $580$ Myr (dotted line), and for a lower metallicity of $[M/H] = -0.25$ and age of $100$ Myr (solid line). The stars appear to have radii that are too small, given their masses, for them to have the metallicity and age of the cluster.}
\label{fig:V1380MRTeff}
\end{figure}

\clearpage

\begin{figure}[p]
\plotone{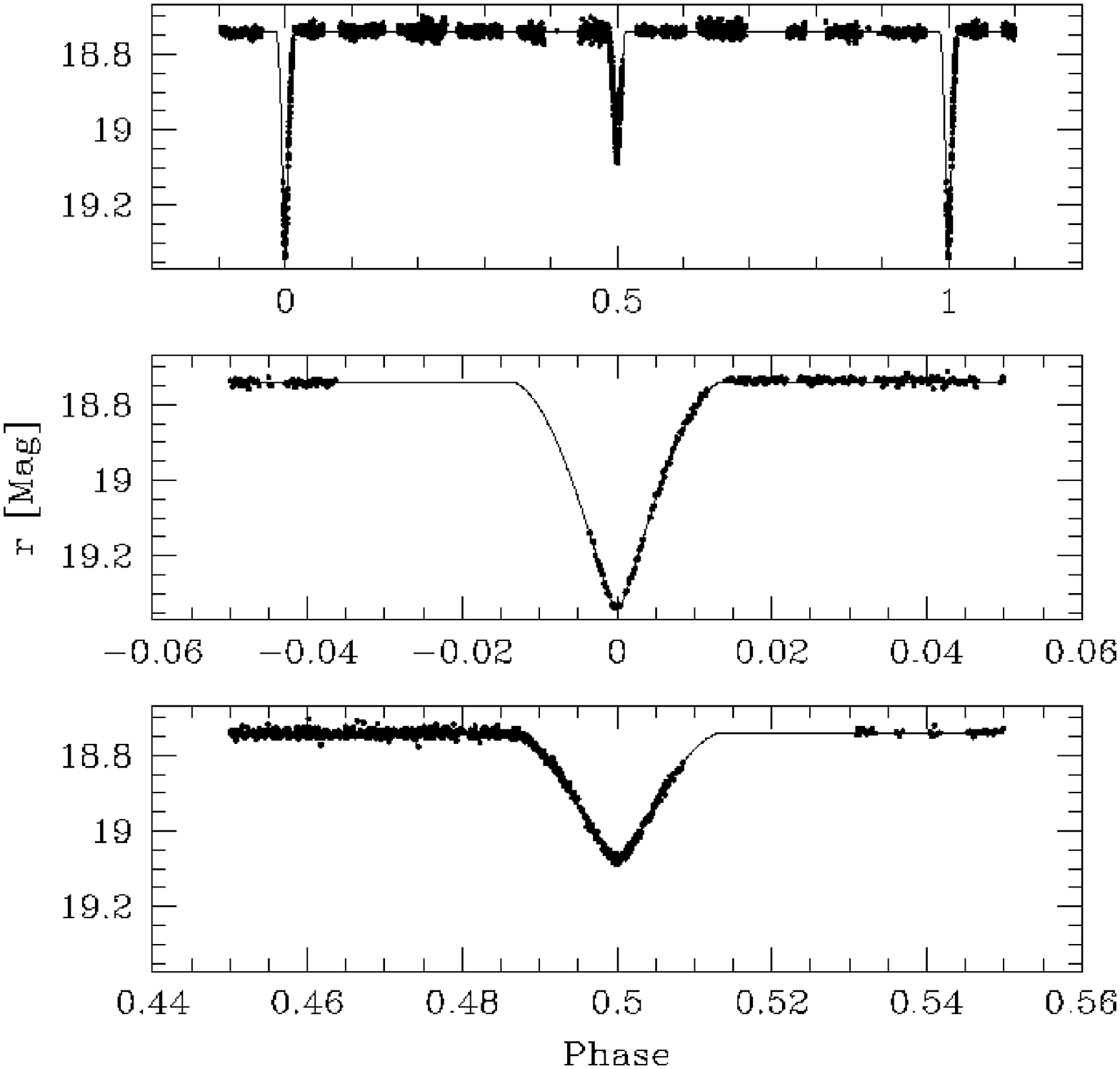}
\caption{Best fit detached eclipsing binary model to the light curve of V1028. The top panel shows the full phased light curve, the bottom panels are zoomed in on the eclipses. In this plot we adopt the zero-point of phase to be the time of primary eclipse.}
\label{fig:V1028JKTEBOPLC}
\end{figure}

\clearpage

\begin{deluxetable}{lrrrrrrrrrrrrrrr}
\tabletypesize{\tiny}
\rotate
\tablewidth{0pc}
\tablecaption{Variable Star Catalog\label{tab:M37_varcatalog}}
\tablehead{
\colhead{ID} & 
\colhead{RA (J2000)} & 
\colhead{DEC (J2000)} & 
\colhead{Chip} & 
\colhead{Period (days)} & 
\colhead{$\sigma_{P}$ (days)} & 
\colhead{$<r>$ (mag)\tablenotemark{a}} & 
\colhead{$RMS_{r}$ (mag)\tablenotemark{b}} & 
\colhead{$g$ (mag)\tablenotemark{c}} & 
\colhead{$r$ (mag)\tablenotemark{c}} & 
\colhead{$i$ (mag)\tablenotemark{c}} & 
\colhead{$B$ (mag)\tablenotemark{d}} & 
\colhead{$V$ (mag)\tablenotemark{d}} & 
\colhead{Type\tablenotemark{e}} & 
\colhead{Selection\tablenotemark{f}}
}
\startdata
     1 & 05:52:20.35 & +32:33:20.4 & $\cdots$ & $\cdots$ & $\cdots$ & $\cdots$ & $\cdots$ & $\cdots$ & $\cdots$ & $\cdots$ & 14.015 & 13.406 & 10 & 000 \\
     2 & 05:52:16.54 & +32:28:15.7 & $\cdots$ & $\cdots$ & $\cdots$ & $\cdots$ & $\cdots$ & $\cdots$ & $\cdots$ & $\cdots$ & 15.341 & 14.653 & 10 & 000 \\
     3 & 05:52:33.01 & +32:32:42.0 & 14 &  0.42248314 &  0.00001000 & 15.7511 &  0.1057 & 16.350 & 15.789 & 15.460 & 16.779 & 16.051 & 10 & 111 \\
     4 & 05:52:53.26 & +32:33:01.5 &  5 &  0.55819369 &  0.00002600 & 15.6407 &  0.0815 & 16.396 & 15.634 & 15.157 & 16.886 & 15.948 & 10 & 111 \\
     5 & 05:53:00.64 & +32:24:50.8 &  8 &  0.27878242 &  0.00000400 & 15.9035 &  0.1140 & 16.374 & 15.915 & 15.433 & 16.769 & 16.184 & 01 & 111 \\
     6 & 05:51:50.53 & +32:32:34.8 & 23 &  0.10983967 &  0.00000100 & 15.9552 &  0.1263 & 16.564 & 15.973 & 15.800 & $\cdots$ & $\cdots$ & 01 & 111 \\
     7 & 05:52:39.09 & +32:36:31.2 & 13 &  0.35773482 &  0.00000400 & 17.3113 &  0.1562 & 18.290 & 17.252 & 17.195 & 18.705 & 17.890 & 10 & 111 \\
     8 & 05:52:34.32 & +32:32:18.7 & $\cdots$ &  0.11950000 & $\cdots$ & $\cdots$ & $\cdots$ & $\cdots$ & $\cdots$ & $\cdots$ & 12.925 & 12.498 & 01 & 000 \\
     9 & 05:52:14.91 & +32:24:41.0 & $\cdots$ &  0.09085950 & $\cdots$ & $\cdots$ & $\cdots$ & $\cdots$ & $\cdots$ & $\cdots$ & 13.668 & 13.298 & 01 & 000 \\
    10 & 05:52:00.49 & +32:36:48.2 & $\cdots$ &  0.07843140 & $\cdots$ & $\cdots$ & $\cdots$ & $\cdots$ & $\cdots$ & $\cdots$ & 15.293 & 14.762 & 01 & 000 \\
    11 & 05:52:03.38 & +32:35:13.7 & 22 &  0.11867118 &  0.00000200 & 15.2542 &  0.0296 & 15.669 & 15.245 & 15.059 & 15.908 & 15.456 & 01 & 110 \\
\enddata
\tablenotetext{a}{The flux averaged $r$ magnitude of the light curve.}
\tablenotetext{b}{The root-mean-square of the light curve.}
\tablenotetext{c}{The magnitude of the source in the photometric catalog (Paper I)}
\tablenotetext{d}{Value from \citet{Kalirai.01}}
\tablenotetext{e}{A two-bit flag. The first bit denotes whether or not the star is an eclipsing binary. The second denotes whether or not it shows pulsations.}
\tablenotetext{f}{A three-bit flag. The first bit denotes if the variable was selected by LS, the second if it was selected by AoV and the third if it was selected by BLS.}
\end{deluxetable}

\clearpage

\begin{deluxetable}{lr}
\tabletypesize{\footnotesize}
\tablewidth{0pc}
\tablecaption{Summary of Variable Star Catalog}
\tablehead{ & \colhead{Number}}
\startdata
Total & 1454 \\
Previously Known & 24 \\
Total Detected in Survey & 1445 \\
Recovered Previously Known & 15 \\
New Discoveries & 1430 \\
Eclipsing Binaries & 31 \\
Pulsating Variables & 44 \\
With $BVgri$ Photometry & 1378 \\
Near $gri$ main sequence & 753 \\
Near $BVgri$ main sequence & 527 \\ 
\enddata
\label{tab:M37_varsummary}
\end{deluxetable}

\clearpage

\begin{deluxetable}{lrrrrrrrrrrrrrrr}
\tabletypesize{\tiny}
\rotate
\tablewidth{0pc}
\tablecaption{Catalog of Suspected Variable Stars\label{tab:M37_varcatalog_suspect}}
\tablehead{
\colhead{ID} & 
\colhead{RA (J2000)} & 
\colhead{DEC (J2000)} & 
\colhead{Chip} & 
\colhead{Period (days)} & 
\colhead{$\sigma_{P}$ (days)} & 
\colhead{$<r>$ (mag)\tablenotemark{a}} & 
\colhead{$RMS_{r}$ (mag)\tablenotemark{b}} & 
\colhead{$g$ (mag)\tablenotemark{c}} & 
\colhead{$r$ (mag)\tablenotemark{c}} & 
\colhead{$i$ (mag)\tablenotemark{c}} & 
\colhead{$B$ (mag)\tablenotemark{d}} & 
\colhead{$V$ (mag)\tablenotemark{d}} & 
\colhead{Type\tablenotemark{e}} & 
\colhead{Selection\tablenotemark{f}}
}
\startdata
     1 & 05:51:20.85 & +32:34:50.5 & 31 &  7.17495773 &  0.02262700 & 18.2823 &  0.0189 & $\cdots$ & 18.217 & $\cdots$ & 20.295 & 18.923 & 00 & 110 \\
     2 & 05:51:29.80 & +32:35:42.0 & 31 & $\cdots$ & $\cdots$ & 14.4660 &  0.1214 & $\cdots$ & 14.466 & $\cdots$ & 14.699 & 14.355 & 00 & 001 \\
     3 & 05:51:33.12 & +32:30:33.8 & 33 &  6.50096134 &  0.01918200 & 19.4794 &  0.0157 & $\cdots$ & 19.479 & 18.653 & 21.571 & 20.077 & 00 & 100 \\
     4 & 05:51:37.12 & +32:40:23.4 & 29 &  4.55114850 &  0.00475400 & 15.8300 &  0.0037 & $\cdots$ & 15.829 & 15.537 & 16.886 & 16.071 & 00 & 100 \\
     5 & 05:51:37.70 & +32:40:40.7 & 29 &  0.85990601 &  0.00037500 & 21.4176 &  0.0440 & $\cdots$ & 21.383 & $\cdots$ & $\cdots$ & $\cdots$ & 00 & 100 \\
     6 & 05:51:42.45 & +32:35:40.4 & 31 & 15.66710882 &  0.11219700 & 18.9369 &  0.0084 & $\cdots$ & 18.939 & 18.418 & 21.483 & 20.447 & 00 & 100 \\
     7 & 05:51:45.62 & +32:27:16.7 & 34 & 11.85243048 &  0.08133600 & 19.5583 &  0.0205 & $\cdots$ & 19.564 & 18.635 & 21.824 & 20.221 & 00 & 100 \\
     8 & 05:51:50.79 & +32:34:01.8 & 23 &  0.67217666 &  0.00018800 & 20.2537 &  0.0371 & $\cdots$ & 20.282 & 19.141 & 22.477 & 21.063 & 00 & 110 \\
     9 & 05:51:51.70 & +32:38:43.0 & 21 &  0.29449917 &  0.00007200 & 23.5355 &  0.2779 & $\cdots$ & 23.506 & 22.150 & $\cdots$ & $\cdots$ & 10 & 010 \\
    10 & 05:51:56.52 & +32:43:44.1 & 19 & 11.73852830 &  0.03931500 & 18.3177 &  0.0081 & $\cdots$ & 18.318 & 19.454 & 20.049 & 18.656 & 00 & 100 \\
    11 & 05:51:56.83 & +32:37:02.5 & 22 &  0.60196052 &  0.00019900 & 20.8365 &  0.0395 & $\cdots$ & 20.824 & 19.529 & 22.803 & 21.452 & 00 & 100 \\
    12 & 05:51:57.09 & +32:33:08.4 & 23 &  0.89557446 &  0.00023500 & 22.4039 &  0.1092 & $\cdots$ & 22.387 & 20.483 & 24.875 & 23.399 & 00 & 100 \\
    13 & 05:51:59.78 & +32:23:33.2 & 26 &  0.93428743 &  0.00049300 & 19.2657 &  0.0101 & $\cdots$ & 19.266 & $\cdots$ & $\cdots$ & $\cdots$ & 00 & 100 \\
    14 & 05:52:00.97 & +32:28:10.0 & 25 &  1.37550745 &  0.00056300 & 17.0477 &  0.0123 & $\cdots$ & 17.049 & 16.885 & $\cdots$ & $\cdots$ & 00 & 100 \\
    15 & 05:52:03.88 & +32:39:10.1 & 21 &  2.85961673 &  0.00233400 & 19.2151 &  0.0211 & $\cdots$ & 19.207 & 18.404 & 21.225 & 19.879 & 00 & 100 \\
    16 & 05:52:09.58 & +32:26:58.4 & 25 & 18.41735750 &  3.59827300 & 22.7522 &  0.1035 & $\cdots$ & 22.770 & 20.951 & 25.276 & 23.538 & 00 & 010 \\
    17 & 05:52:15.33 & +32:35:44.7 & 22 &  0.74276537 &  0.00021900 & 21.0196 &  0.0436 & $\cdots$ & 21.011 & 19.582 & $\cdots$ & $\cdots$ & 00 & 100 \\
    18 & 05:52:15.42 & +32:30:46.7 & 24 &  0.90985227 &  0.00054200 & 21.9206 &  0.1236 & $\cdots$ & 21.956 & 20.277 & $\cdots$ & $\cdots$ & 00 & 100 \\
    19 & 05:52:16.06 & +32:32:44.6 & 23 &  5.25393598 &  0.01656200 & 21.0112 &  0.0550 & $\cdots$ & 21.018 & 20.086 & $\cdots$ & $\cdots$ & 00 & 100 \\
    20 & 05:52:17.60 & +32:30:12.6 & 24 &  8.46159122 &  0.01841700 & 17.9894 &  0.0125 & $\cdots$ & 17.983 & 17.305 & 19.724 & 18.465 & 00 & 100 \\
    21 & 05:52:30.27 & +32:30:49.8 & 15 &  1.42595003 &  0.00214900 & 18.8178 &  0.0092 & $\cdots$ & 18.822 & 18.356 & 19.981 & 19.118 & 00 & 100 \\
    22 & 05:52:34.04 & +32:29:22.6 & 15 &  8.58126897 &  0.01497500 & 17.9452 &  0.0057 & $\cdots$ & 17.949 & 17.591 & 19.033 & 18.214 & 00 & 110 \\
    23 & 05:52:34.50 & +32:40:23.0 & 11 &  2.09687226 &  0.00201000 & 20.2432 &  0.0209 & $\cdots$ & 20.218 & $\cdots$ & $\cdots$ & $\cdots$ & 00 & 100 \\
    24 & 05:52:40.10 & +32:36:10.4 & 13 & 11.89564092 &  0.07649400 & 18.9641 &  0.0087 & $\cdots$ & 18.970 & 18.218 & 21.161 & 19.489 & 00 & 100 \\
    25 & 05:52:40.60 & +32:31:50.1 & 14 &  4.47741049 &  0.00336800 & 19.0914 &  0.0349 & $\cdots$ & 19.088 & 18.364 & $\cdots$ & $\cdots$ & 00 & 110 \\
    26 & 05:52:50.60 & +32:34:48.9 &  4 &  5.20926065 &  0.19755000 & 15.2892 &  0.0151 & $\cdots$ & 15.290 & 14.945 & $\cdots$ & $\cdots$ & 00 & 110 \\
    27 & 05:52:54.61 & +32:39:28.6 &  3 & 17.23385097 &  0.08937600 & 15.8058 &  0.0052 & $\cdots$ & 15.808 & 15.450 & $\cdots$ & $\cdots$ & 00 & 100 \\
    28 & 05:53:09.81 & +32:21:34.5 &  9 &  2.59599006 &  0.00113400 & 15.6733 &  0.0148 & $\cdots$ & 15.655 & $\cdots$ & 16.310 & 15.699 & 00 & 111 \\
    29 & 05:53:11.99 & +32:23:53.9 &  8 &  0.58837547 &  0.00003100 & 18.2538 &  0.0971 & $\cdots$ & 18.298 & 17.203 & $\cdots$ & $\cdots$ & 10 & 111 \\
\enddata
\tablenotetext{a}{The flux averaged $r$ magnitude of the light curve.}
\tablenotetext{b}{The root-mean-square of the light curve.}
\tablenotetext{c}{The magnitude of the source in the photometric catalog (Paper I). Note that when the $i$ magnitude is missing no color term was applied in transforming from the instrumental $r$ and $i$ magnitudes to the Sloan 2.5 m system.}
\tablenotetext{d}{Value from \citet{Kalirai.01}}
\tablenotetext{e}{A two-bit flag. The first bit denotes whether or not the star is an eclipsing binary. The second denotes whether or not it shows pulsations.}
\tablenotetext{f}{A three-bit flag. The first bit denotes if the variable was selected by LS, the second if it was selected by AoV and the third if it was selected by BLS.}
\end{deluxetable}

\begin{deluxetable}{lrrrrrr}
\tabletypesize{\footnotesize}
\tablewidth{0pc}
\tablecaption{Extinction and Distances for Fundamental Mode $\delta$-Scuti Stars}
\tablehead{\colhead{ID} & \colhead{$V$} & \colhead{$B-V$} & \colhead{$V-I_{C}$} & \colhead{Period (days)} & \colhead{$A_V$} & \colhead{Distance (kpc)}}
\startdata
    V5 & 16.184 & 0.585 & 1.194 &  0.278782 & 1.56 & 7.90 \\
    V6 & 16.177 & 0.691 & 0.806 &  0.109840 & 0.95 & 5.22 \\
   V11 & 15.456 & 0.452 & 0.826 &  0.118671 & 0.98 & 3.92 \\
  V840 & 20.695 & 1.711 & 1.377 &  0.145197 & 2.04 & 31.21 \\
\enddata
\label{tab:DeltaScutidist}
\end{deluxetable}

\clearpage

\begin{deluxetable}{lrrrrrr}
\tabletypesize{\footnotesize}
\tablewidth{0pc}
\tablecaption{Extinction and Distances for W UMa Eclipsing Binaries}
\tablehead{\colhead{ID} & \colhead{$V$} & \colhead{$B-V$} & \colhead{$V-I_{C}$} & \colhead{Period (days)} & \colhead{$A_V$} & \colhead{Distance (kpc)}}
\startdata
    V3 & 16.051 & 0.728 & 1.026 &  0.422483 & 1.18 & 2.78 \\
    V4 & 15.948 & 0.938 & 1.234 &   0.55819 & 1.27 & 2.04 \\
    V7 & 17.890 & 0.815 & 1.117 &  0.357735 & 1.23 & 5.10 \\
   V20 & 17.628 & 0.892 & 1.008 &  0.289456 & 0.45 & 3.07 \\
   V24 & 20.506 & 1.275 & 1.518 &  0.252674 & 1.23 & 5.84 \\
  V855 & 18.156 & 1.227 & 1.617 &  0.564829 & 1.86 & 3.56 \\
 V1160 & 20.234 & 1.134 & 1.435 &  0.242092 & 1.42 & 6.88 \\
 V1181 & 21.847 & 1.402 & 1.871 &   0.25817 & 2.32 & 10.30 \\
 V1194 & 19.030 & 1.136 & 1.267 &  0.269735 & 0.56 & 3.58 \\
 V1447 & 19.930 & 1.066 & 1.535 &  0.296877 & 2.13 & 8.52 \\
\enddata
\label{tab:WUMAdist}
\end{deluxetable}

\clearpage

\begin{deluxetable}{rrrr}
\tabletypesize{\footnotesize}
\tablewidth{0pc}
\tablecaption{Radial velocity and light ratio measurements from the spectra of V1380.}
\tablehead{\colhead{HJD} & \colhead{Primary Velocity [km/s]} & \colhead{Secondary Velocity [km/s]} & \colhead{Light Ratio [$L_{2}/L_{1}$]}}
\startdata
 2454154.6776 & -104.36 & 115.64 & 0.991 \\
 2454165.7104 & -108.62 & 119.24 & 0.955 \\
 2454170.6082 & -31.08 & 40.47 & 0.995 \\
 2454171.6861 & 49.34 & -38.14 & 1.025 \\
\enddata
\label{tab:V1380RV}
\end{deluxetable}

\clearpage

\begin{deluxetable}{lrr}
\tabletypesize{\footnotesize}
\tablewidth{0pc}
\tablecaption{Eclipse time measurements for V1380. The last measurement given is inferred from a fit to the radial velocity data.}
\tablehead{\colhead{Eclipse Number} & \colhead{Observed Minimum [HJD]} & \colhead{Error [HJD]}}
\startdata
 0.5   & 2453725.6289 & 0.0008 \\
 1.0 & 2453726.7223 & 0.0004 \\
 1.5   & 2453727.8188 & 0.0003 \\
 2.0 & 2453728.9160 & 0.0004 \\
 7.5 & 2453740.9756 & 0.0003 \\
 195 & 2454152.0803 & 0.0037  \\
\enddata
\label{tab:V1380timing}
\end{deluxetable}

\clearpage

\begin{deluxetable}{lrr}
\tabletypesize{\footnotesize}
\tablewidth{0pc}
\tablecaption{Parameters for Eclipsing Binary V1380.}
\tablehead{\colhead{Parameter} & \colhead{Value} & \colhead{Error}}
\startdata
$P_{orb}$ & 2.19258 days & 0.00004 days\\
$HJD_{0}$ & 2453724.5306 & 0.0004\\
$K_{1}$ & 118.3 km/s & 1.3 km/s\\
$K_{2}$ & 118.3 km/s & 1.3 km/s\\
$\gamma$ & 5.3 km/s & 0.7 km/s\\
$J_{S}/J_{P}$ & 1.02 & 0.02 \\
$(R_{P} + R_{S})/a$ & 0.266 & 0.002 \\
$R_{S}/R_{P}$ & 0.99 & 0.1 \\
$i$ & $80.00^{\circ}$ & $0.09^{\circ}$ \\
$a$ & $10.41$ $R_{\odot}$ & $0.23$ $R_{\odot}$ \\
$M_{1}$ & $1.58$ $M_{\odot}$ & $0.04$ $M_{\odot}$ \\
$M_{2}$ & $1.58$ $M_{\odot}$ & $0.04$ $M_{\odot}$ \\
$R_{1}$ & $1.39$ $R_{\odot}$ & $0.07$ $R_{\odot}$ \\
$R_{2}$ & $1.38$ $R_{\odot}$ & $0.07$ $R_{\odot}$ \\
\enddata
\label{tab:V1380JKTEBOP}
\end{deluxetable}

\clearpage

\begin{deluxetable}{lrr}
\tabletypesize{\footnotesize}
\tablewidth{0pc}
\tablecaption{Light Curve Parameters for V1028.}
\tablehead{\colhead{Parameter} & \colhead{Value} & \colhead{Error}}
\startdata
$P_{orb} = P_{rot}$ & 5.496 days & 0.18 days\\
$HJD_{0}$ & 2453725.6989 & 0.0001\\
$J_{S}/J_{P}$ & 0.649 & 0.004\\
$(R_{P} + R_{S})/a$ & 0.0855 & 0.0003\\
$R_{S}/R_{P}$ & 0.95 & 0.01\\
$i$ & $89.73^{\circ}$ & $0.05^{\circ}$\\
$L_{3}$ & 0.266 & 0.006\\
\enddata
\label{tab:V1028JKTEBOP}
\end{deluxetable}

\end{document}